\documentstyle[12pt]{article}
\def\hybrid{\topmargin -20pt	\oddsidemargin 0pt
	\headheight 0pt	\headsep 0pt
	\textwidth 6.25in	
	\textheight 9.5in	
	\marginparwidth .875in
	\parskip 5pt plus 1pt	\jot = 1.5ex}

\hybrid

\begin{document}
\def\x{\times}
\def\beq{\begin{equation}}
\def\eeq{\end{equation}}
\def\beqa{\begin{eqnarray}}
\def\eeqa{\end{eqnarray}}
\def\D{ {\cal D}}
\def\L{ {\cal L}}
\def\C{ {\cal C}}
\def\N{ {\cal N}}
\def\calE{{\cal E}}
\def\lin{{\rm lin}}
\def\Tr{{\rm Tr}}
\def\mxth{\mathsurround=0pt }
\def\xversim#1#2{\lower2.pt\vbox{\baselineskip0pt \lineskip-.5pt
x  \ialign{$\mxth#1\hfil##\hfil$\crcr#2\crcr\sim\crcr}}}
\def\simgr{\mathrel{\mathpalette\xversim >}}
\def\simle{\mathrel{\mathpalette\xversim <}}

\def\a{\alpha}
\def\b{\beta}
\def\dota{ {\dot{\alpha}} }
\def\lag{Lagrangian}
\def\Kahler{K\"{a}hler}
\def\kahler{K\"{a}hler}
\def\A{ {\cal A}}
\def\C{ {\cal C}}
\def\D{ {\cal D}}
\def\F{{\cal F}}
\def\L{ {\cal L}}
\def\R{ {\cal R}}
\def\x{ \times }
\def\beq{\begin{equation}}
\def\eeq{\end{equation}}
\def\beqa{\begin{eqnarray}}
\def\eeqa{\end{eqnarray}}

\sloppy
\newcommand{\be}{\begin{equation}}
\newcommand{\eq}{\end{equation}}
\newcommand{\ov}{\overline}
\newcommand{\un}{\underline}
\newcommand{\p}{\partial}
\newcommand{\la}{\langle}
\newcommand{\ra}{\rangle}
\newcommand{\bl}{\boldmath}
\newcommand{\ds}{\displaystyle}
\newcommand{\nl}{\newline}
\newcommand{\Nzahl}{{\bf N}  }
\newcommand{\zzahl}{ {\bf Z} }
\newcommand{\Zzahl}{ {\bf Z} }
\newcommand{\Qzahl}{ {\bf Q}  }
\newcommand{\Rzahl}{ {\bf R} }
\newcommand{\Czahl}{ {\bf C} }
\newcommand{\wt}{\widetilde}
\newcommand{\wh}{\widehat}
\newcommand{\fs}[1]{\mbox{\scriptsize \bf #1}}
\newcommand{\ft}[1]{\mbox{\tiny \bf #1}}
\newtheorem{satz}{Satz}[section]
\newenvironment{Satz}{\begin{satz} \sf}{\end{satz}}
\newtheorem{definition}{Definition}[section]
\newenvironment{Definition}{\begin{definition} \rm}{\end{definition}}
\newtheorem{bem}{Bemerkung}
\newenvironment{Bem}{\begin{bem} \rm}{\end{bem}}
\newtheorem{bsp}{Beispiel}
\newenvironment{Bsp}{\begin{bsp} \rm}{\end{bsp}}
\renewcommand{\arraystretch}{1.5}


\renewcommand{\thesection}{\arabic{section}}
\renewcommand{\theequation}{\thesection.\arabic{equation}}

\parindent0em

\def\S4{\frac{SO(4,2)}{SO(4) \otimes SO(2)}}
\def\P3{\frac{SO(3,2)}{SO(3) \otimes SO(2)}}
\def\MGd{\frac{SO(r,p)}{SO(r) \otimes SO(p)}}
\def\SOd{\frac{SO(r,2)}{SO(r) \otimes SO(2)}}
\def\SO2{\frac{SO(2,2)}{SO(2) \otimes SO(2)}}
\def\SUm{\frac{SU(n,m)}{SU(n) \otimes SU(m) \otimes U(1)}}
\def\SUS{\frac{SU(n,1)}{SU(n) \otimes U(1)}}
\def\SK{\frac{SU(2,1)}{SU(2) \otimes U(1)}}
\def\SU{\frac{ SU(1,1)}{U(1)}}

\begin{titlepage}
\begin{center}

\hfill HUB-IEP-94/50\\
\hfill hep-th/9412209\\

\vskip .1in

{\bf THRESHOLD CORRECTIONS AND SYMMETRY ENHANCEMENT IN STRING
COMPACTIFICATIONS}

\vskip .2in

{\bf Gabriel Lopes Cardoso,
Dieter L\"{u}st }  \\
\vskip .1in

{\em  Humboldt Universit\"at zu Berlin\\
Institut f\"ur Physik\\
D-10115 Berlin, Germany}\footnote{e-mail addresses:
GCARDOSO@QFT2.PHYSIK.HU-BERLIN.DE, LUEST@QFT1.PHYSIK.HU-BERLIN.DE}

\vskip .1in
and
\vskip .1in

{\bf Thomas Mohaupt}
\vskip .1in

{\em
DESY-IfH Zeuthen\\
Platanenallee 6\\
D-15738 Zeuthen, Germany}\footnote{e-mail address: MOHAUPT@HADES.IFH.DE}

\end{center}

\vskip .2in

\begin{center} {\bf ABSTRACT } \end{center}
\begin{quotation}\noindent
We present the computation of threshold functions for Abelian
orbifold compactifications. Specifically, starting from the
massive, moduli-dependent string spectrum after compactification, we derive
the threshold functions as
target space duality invariant free energies (sum over  massive string states).
In particular we work out the dependence on the continuous
Wilson line moduli fields. In addition we concentrate on the
physically interesting effect that at certain critical points in the
orbifold moduli spaces additional massless states appear in the
string spectrum leading to logarithmic singularities in the
threshold functions. We discuss this effect for the gauge coupling
threshold corrections; here the appearance of additional massless
states is directly related to the Higgs effect in string theory.
In addition the singularities in the threshold functions are
relevant for the loop corrections to the gravitational coupling
constants.
\end{quotation}
\end{titlepage}
\vfill
\eject

\newpage

\section{Introduction}

Four-dimensional string models possibly provide a consistent description of
all known  interactions including gravity. At energy scales small
compared to the typical string scale $M_{string}$,
which is directly related to the
Planck mass $M_{\rm Planck}=O(10^{19}~{\rm GeV})$, it is convenient
to use an effective Lagrangian
for the light string modes $\phi_L$ where the effect of the heavy string modes
$\phi_H$
is integrated out. Based on several motivations, the main area of research in
this context is on four-dimensional string vacua with  $N=1$ space-time
supersymmetry, implying that the effective string
interactions are of the form of an $N=1$
supergravity-matter action. Studying
four-dimensional string vacua with space-time supersymmetry, it turns out that
infinitely many of them are connected by continuous deformations of
the underlying two-dimensional conformal field theory, parametrized by coupling
constants called {\sl moduli}, collectively denoted by $T_i$.
 At the level of the effective supergravity
action, and neglecting non-perturbative effects, moduli are described
by massless scalar fields with flat potential, whose vacuum
expectation values parametrize the continuous deformations. The moduli
play a very important role in the effective action,
since various coupling constants among the matter fields like the tree-level
Yukawa couplings \cite{HamVa,DiFrMaSh}
or the loop corrections to the gauge and gravitational
coupling constants \cite{Vkap}-\cite{AntTay}
are moduli dependent functions. Specifically,
the one-loop moduli dependence of the gauge and gravitational
coupling constants
arises from $\sigma$-model and K\"{a}hler
anomalies
\cite{Louis,CO1,DFKZ,CO2}
and from
integrating out the infinite tower
of massive string modes with moduli dependent masses.

The moduli spaces ${\cal M}$ of four-dimensional string theories have a very
interesting and rich structure. First, as it is true for many known
compactification schemes of the ten-dimensional heterotic string, the
underlying superconformal field theory is invariant under
the target space duality symmetries (see \cite{GPR}) which act
on the the moduli as discrete reparametrizations. These target space duality
transformations act in general non-trivially on the
infinite massive spectrum $\phi_H$, in the
sense that states with different quantum numbers (e.g. discrete internal
momentum and winding numbers) are mapped onto each other.
Consequently, integrating out the massive spectrum $\phi_H$
implies that certain
low energy couplings are given in terms of automorphic functions of the
corresponding duality group. This observation can provide very useful
informations about the structure  of the effective low-energy supergravity
action, in particular when combined with some analyticity arguments of
holomorphic coupling functions \cite{FLST}.

A second very interesting feature of the string moduli spaces comes from
the fact that at certain critical points $P_c$ in the moduli spaces a finite
number of additional
massless states may appear in the
moduli-dependent string spectrum, which are otherwise
massive at generic points in ${\cal M}$. We
will call these states $\phi_H'$. Very often, at $P_c$ these
fields correspond to additional holomorphic spin one currents on the
world sheet. Then the gauge symmetry of the four-dimensional string
is enlarged
which is nothing else than the stringy version of the well known Higgs
effect. The role of the Higgs fields is now taken by moduli fields.
The field-theoretical formulation of the stringy Higgs effect, i.e. the
correct coupling of the relevant moduli to the gauge bosons, was
investigated in \cite{DHS,IbaLuLer} for the case of  the standard ${\bf Z}_3$
orbifold \cite{DHVW1,DHVW2}.

The appearance of additional massless fields at some critical points $P_c$
in ${\cal M}$ implies that the description of the string compactification
by an low-energy effective action contains discontinuities, since
near the critical points the fields $\phi_H'$ should be kept as light
degrees of freedom
and should not be integrated out from the spectrum. In other words,
when integrating over all massive fields including $\phi_H'$ the
effective action may acquire singularities at the critical points $P_c$.
Let us make this more clear by considering as an example
the one loop running of a gauge coupling constant in the low-energy
field theory. The one-loop
running coupling constant at a scale $p$ is given by (neglecting
K\"{a}hler and $\sigma$-model anomalies)
\be
{1\over g^2(p^2)}={1\over g^2(M^2_{string})}+{b\over 16\pi^2}
\log{M_{string}^2\over p^2}+{|\Delta(T_i)|^2\over
16\pi^2},
\label{rgg}
\eq
where  $b$ is the one-loop $\beta$-function
coefficient of the light modes $\phi_L$ and $\Delta(T_i)$ is  the
moduli-dependent threshold correction of the heavy string modes.
$\Delta(T_i)$
can be regarded as the suitably regularized free energy
\cite{VafaOog,FerKouLuZw}
of all massive modes, $\Delta(T_i)\propto\log\det M^2_{\phi_H}(T_i)$,
where the $M_{\phi_H}(T_i)$'s are the moduli-dependent masses of
the heavy modes. Clearly, if  this sum contains also states $\phi_H'$ which
become massless at $P_c$, $\Delta(T_i)$ possesses a singularity at $P_c$.
 As we will discuss in detail, the masses of $\phi_H'$ are
generically of the form $M_{\phi_H'}(T_i)\propto (T_i-P_c)$ (here, $T_i$ is one
specific `critical' modulus). Thus we see that $\Delta(T_i)$ exhibits a
logarithmic singularity of the form
\be
\Delta\rightarrow n\log(T_i-P_c),
\eq
where $n$ accounts for the degeneracy of
states which become massless at $P_c$.
In order to get finite threshold corrections
$\tilde\Delta(T_i)$ over the whole moduli space it is
useful to separate in eq.(\ref{rgg})
the contribution of the states $\phi_H'$ from
the states $\phi_H$ which are always massive. Then eq.(\ref{rgg})
can be written as
\be
{1\over g^2(p^2)}={1\over g^2(M^2_{string})}+{b\over 16\pi^2}\log{M_{string}^2
\over p^2}+
{b'\over 16\pi^2}\log|T_i-P_c|^2+{|\tilde\Delta(T_i)|^2\over
16\pi^2}.
\label{rgga}
\eq
Here $b'$ is the contribution of the states $\phi_H'$ to the $\beta$-function
coefficient, and $\tilde\Delta(T_i)$ does not contain the states $\phi_H'$.
Thus the logarithmic singularity in the moduli $T_i$
is nothing else than the threshold effect of $\phi_H'$ with (intermediate) mass
scale $M_{\phi_H'}(T_i)\propto(T_i-P_c)$.

This discussion was entirely based on field theoretical arguments.
As already discussed in \cite{CveLu}, in string theory these
threshold functions are given in terms of automorphic functions  of the
underlying target space duality group
with the appropriate singularity structure. In fact, since in string theory
there is generically an infinite number of states which may become
massless at duality equivalent points in ${\cal M}$ (at one particular point
in ${\cal M}$ only a finite number of states can become massless) the
relevant automorphic functions possess singularities at an infinite number
of points being related by discrete duality transformations.
In this paper we will calculate explicitly these divergent threshold
functions which are related to the discontinuities in the string
spectrum and to
the stringy Higgs effect. To be specific we concentrate on Abelian orbifold
compactifications of the ten-dimensional heterotic string. We will
discuss the dependence of the threshold corrections as functions
of the moduli associated with the six-dimensional orbifold, denoted by
$T$ and $U$, as well as of the
socalled Wilson line moduli \cite{IbaNilQue}
which take values in the heterotic gauge group.
The orbifold moduli $T$, $U$ (metric and antisymmetric tensor)
will be relevant for
the discussion of the Higgs effect in the compactification sector; here
the relevant automorphic functions will be given in terms of the
absolute modular invariant function $j$. On the other hand, the Wilson line
moduli are responsible for the Higgs effect in the heterotic gauge group.
This is
of rather phenomenological importance since the Wilson line Higgs field
may break some GUT gauge group to the gauge group of the standard model.
Moreover it may be even possible to identify some of the
Wilson line moduli with the supersymmetric standard model Higgs fields $H_1$
and $H_2$.

Our paper is organized as follows. In sections 2 and 3
we discuss the structure
of the gauge groups in orbifold compactifications as a function of
the various moduli fields. In section 4 we compute the masses
of the generically massive fields. Some of these masses become zero at
certain fixed points in the orbifold moduli spaces.  Here we will use
the results of our previous paper \cite{CLM} where we have determined
the moduli spaces plus the target space duality transformations in the
presence of Wilson line moduli. In sections
5, 6 and 7 we will apply the results of
section 4 to
compute the target space free energies as (infinite) sums over the
massive string states. Finally we will explain the relation of these
free energies to the string threshold corrections. We will display the
dependence of the threshold corrections in terms of $T$ and $U$ as well as
in terms of the generic Wilson line moduli. A
discussion of this is provided in section 8.

\section{Gauge groups in orbifold compactifications with
continuous Wilson lines}

\setcounter{equation}{0}

In this section we will start to
discuss the moduli dependence of the
gauge group of an orbifold compactification.
We will concentrate on the so called gauge sector here,
whereas the compactification sector will be studied in the next
section.
Our aim is
to determine the unbroken gauge group in the presence of the most
general continuous Wilson lines compatible with a given twist.
For a certain class of twists the
minimal and therefore generic gauge groups are easily determined, and
we describe the method for determining them.  The results are listed
in tables given at the end of this paper.
We also point out under which circumstances
this method fails to give the minimal gauge groups, in which case it only
yields a lower bound on these minimal
gauge groups, that is not saturated. For these cases,
in which a more detailed analysis is necessary, we outline
how one has to proceed in order to determine the generic gauge groups.

\subsection{Wilson lines in the Narain model}

Since many properties of an orbifold model can be understood
easily in terms of the underlying Narain model \cite{Nar,NSW},
let us first
recall how the gauge group of a toroidal compactification
depends on the Wilson line moduli \cite{Moh1}.
We will concentrate
on the so called gauge sector which is generated by the sixteen
extra left--moving worldsheet bosons ${\bf X}(z) := X^{I}_{L}(z)$,
$I=1,\ldots,16$.

There are sixteen chiral conserved currents $\p X^{I}_{L}(z)$
on the worldsheet.  These can be combined with the right--moving
ground state, which is an $N=4$ space time vector supermultiplet,
to give a $U(1)^{16}$ $N=4$ gauge theory. For special values
of the moduli
the Narain lattice $\Gamma = \Gamma_{22,6}$ contains vectors of the
form
\be    {\bf P} = \left( {\bf p}_{L}; {\bf p}_{R} \right) =
\left( {\bf v}, 0_{6} ; 0_{6} \right) \in \Gamma,\;\;\;
{\bf p}_{L}^{2} =  {\bf v}^2  = 2.
\label{ExtraGBGS}
\eq
Then there are extra conserved currents
$\exp(i\, {\bf v} \cdot {\bf X}(z))$,
leading to vertex operators for massless
charged gauge bosons with
charges ${\bf v} = (v^{I})$, thus extending the gauge group to a
rank 16 reductive non--abelian Lie group
\be
G^{(16)} = G^{(l)} \otimes U(1)^{16-l}
\eq
where $G^{(l)}$ is semi--simple and has rank $l$.

As discussed in \cite{Moh1} these
extended symmetries can be described in the following way.
Every vector ${\bf P} \in \Gamma$ of the Narain lattice
can be written as
\be
{\bf P} =
q^{I} {\bf l}_{I} + n^{i} \ov{\bf k}_{i} + m_{i} {\bf k}^{i},
\eq
where the integers $q^{I}$, $n^{i}$ and $m_{i}$ are the
charge, winding and momentum quantum numbers,
$I=1,\ldots,16$, $i=1,\ldots,6$.
The standard basis vectors are \cite{Gin}
\be
{\bf l}_{I} = \left( {\bf e}_{I}, -\frac{1}{2}
({\bf e}_{I} \cdot {\bf A}_{i}) {\bf e}^{i};
-\frac{1}{2}({\bf e}_{I} \cdot {\bf A}_{i}) {\bf e}^{i} \right),
\label{lvector}
\eq
\be
\ov{\bf k}_{i} = \left( {\bf A}_{i}, (4 G_{ij} + D_{ij} )
\frac{1}{2} {\bf e}^{i}; D_{ij}\frac{1}{2} {\bf e}^{i} \right)
\eq
and
\be
{\bf k}^{i} = \left( 0_{16}, \frac{1}{2}{\bf e}^{i};
\frac{1}{2}{\bf e}^{i} \right),
\label{kvector}
\eq
where
\be
D_{ij} = 2 \left( B_{ij} - G_{ij} - \frac{1}{4} {\bf A}_{i}
\cdot {\bf A}_{j} \right).
\eq
This basis is a function of the moduli
\be
G_{ij} = G_{ji} \in M(6,6,{\bf R}),\;\;\;
B_{ij} = - B_{ji} \in M(6,6,{\bf R}),\;\;\;
{\bf A}_{i} \in {\bf R}^{16}
\eq
of the Narain model, namely the metric and the axionic
background field and the Wilson lines.
${\bf e}_{I}$ are basis vectors of a selfdual
sixteen dimensional lattice $\Gamma_{16}$
(the $E_{8} \otimes E_{8}$ root lattice
or the $SO(32)$ root lattice extended by the spinor weights
of one chirality),
and the ${\bf e}^{i}$
are a basis of the dual $\Lambda^{*}$ of the compactification
lattice.

Vectors of the form (\ref{ExtraGBGS}) have quantum numbers
such that $q^{I}C^{(16)}_{IJ} q^{J}$ $:= q^{T} C^{(16)} q$
$=2$, where $C^{(16)}$ is the lattice metric of $\Gamma_{16}$
and $n^{i}m_{i}$ $:= n^{T}m$ $=0$.
\footnote{More generally all possible extra massless
states correspond to Narain vectors with quantum
numbers satisfying
$q^T C^{(16)} q + 2 n^T m = 2$. This is a consequence of the
mass formula as we will recall in section \ref{ModOrbUnt}.
A second subclass of this set will be the subject of the
next section.}
The Wilson lines
${\bf A}_{i}$ must be chosen such that
\be
{\bf v} \cdot {\bf A}_{i} \in {\bf Z}
\eq
where ${\bf v} = q^{I} {\bf e}_{I}$. Then
\be
{\bf P} = q^{I} {\bf l}_{I} + ({\bf v} \cdot {\bf A}_{i})\,
{\bf k}^{i} = \left( {\bf v}, 0_6 ; 0_6 \right)
\eq
is a Narain vector
with ${\bf v}^{2}=q^T C^{(16)}q = 2$. If one sets for example ${\bf A}_{i}=0$,
then
all roots of the lattice $\Gamma_{16}$ are in the Narain lattice
and therefore the generic gauge group $U(1)^{16}$ is extended
to $E_{8} \otimes E_{8}$ or $SO(32)$, depending on the choice
of $\Gamma_{16}$.
Other solutions, which have as gauge groups
all
possible maximal rank regular reductive subgroups of
$E_{8} \otimes E_{8}$ and  $SO(32)$ were
constructed in \cite{Moh1}.

Finally note that a small deformation $\delta {\bf A}_{i}$
of the Wilson lines, if it destroys some of the conditions
${\bf v} \cdot {\bf A}_{i} \in {\bf Z}$, acts
on the lattice as a deformation
\be
({\bf v},0_{6};0_{6}) \rightarrow ({\bf v}, {\bf w}; {\bf w})
\eq
with
${\bf w}$
$=-\frac{1}{2} ({\bf v} \cdot \delta {\bf A}_{i}) {\bf e}^{i}$,
which makes the corresponding state acquire
a mass $\frac{\alpha'}{2} M^{2} =$ ${\bf w}^{2}$ in a smooth
way. This is a version of the stringy Higgs effect \cite{IbaLuLer}.

\subsection{Definition of the orbifold}

We can now proceed to extend this to the untwisted sector of
an orbifold compactification. First of all one has to select
a Narain
lattice with some symmetry that can be modded out. Our reference
lattice will be the one with vanishing Wilson lines, which therefore
factorises as $\Gamma_{22;6} = \Gamma_{16} \oplus \Gamma_{6;6}$.
For definiteness, $\Gamma_{16}$ will be the $E_{8} \otimes E_{8}$
lattice. Then, we have to specify the twist action  on
$\Gamma_{16}$ and $\Gamma_{6;6}$. Whereas the twist action
on  $\Gamma_{6;6}$ is defined by choosing one of the 18 twists
$\theta$ of the compactification lattice $\Lambda$
that lead to $N=1$ space time supersymmetry \cite{ErlKle},
the twist on $\Gamma_{16}$ will be
a Weyl twist (inner automorphism) $\theta'$ of $E_{8} \otimes E_{8}$.
The total twist $\Theta = (\theta',\theta,\theta)$
of the Narain lattice is constrained
by world sheet modular invariance \cite{NSV2}.
The level matching conditions worked out in \cite{NSV2}
restrict the eigenvalues of $\Theta$. In this paper we will not
present a classification of ${\bf Z}_{N}$ Weyl orbifolds.
Instead we will take one of the $E_{8}$ as a hidden sector and
assume that the Weyl twist in this sector has been chosen
in such a way that it cancels the WS modular anomalies of the
internal twist and of the Weyl twist in the first $E_{8}$.
The orbifold model defined this way still has some
Wilson line moduli left. In order to be compatible
with the twist the Wilson lines must satisfy \cite{IbaNilQue,Moh2}
\be
\theta_{I}^{J} A_{Jj} = A_{Ii} \theta^{i}_{j}
\label{ConstraintWL}
\eq
where $A_{Ij}$ is a matrix containing the components of
the Wilson lines and $\theta_{I}^{J}$, $\theta^{i}_{j}$
are the matrices of the gauge and of the internal twist
with respect to the lattice bases ${\bf e}_{I}$ and
${\bf e}^{i}$ of $\Gamma_{16}$ and $\Lambda^{*}$
respectively.
Wilson line moduli do exist if an eigenvalue appears both
in the gauge twist and in the internal twist. More precisely,
if a complex conjugated pair of  eigenvalues
(a real eigenvalue) appears $d$ times in the gauge twist and
$d'$ times in the internal twist, this then leads to
$2 dd'$ ($dd'$) real moduli \cite{CLM,Moh2}.

\subsection{Minimal gauge groups in the presence of Wilson lines}

Let us now work out the gauge groups for Weyl twists of a $E_{8}$
in the presence of generic continuous Wilson lines.
The basic idea is the following.  All Weyl twists of $E_{8}$ are
induced by twists that have a nontrivial action on some
sublattice ${\cal N}$. This sublattice can be chosen to be
the root lattice of a regular semi--simple subalgebra.
The twist action on the sublattice can then be described
by a so called Carter diagram, which can be thought of as a
generalization of the well known Dynkin diagram \cite{Car}. In fact
most Weyl twists of $E_{8}$ are induced by Coxeter twists of
regular subalgebras and in this case the Carter diagram is
identical with the Dynkin diagram of this subalgebra. To get
all inequivalent Weyl twist one has to add a few twists
of subalgebras, which are not Coxeter twists. These are then
described by Carter diagrams that are not Dynkin diagrams.
For more details on Carter diagrams and their relation
to Weyl twists see \cite{SchWar} and \cite{HolMyh}.

Given the sublattice ${\cal N}$ on which the twist acts non--trivially
one has to look for the  largest complementary sublattice
${\cal I}$ on which it acts trivially. That means that ${\cal I}$ is
defined by
\be
E_{8} \supset {\cal N} \oplus {\cal I}
\eq
together with
\be
\left( E_{8} \supset {\cal N} \oplus {\cal I}\mbox{   and   }
{\cal I}' \supset {\cal I}  \right)
\Longrightarrow {\cal I}' = {\cal I}.
\eq
If we now decompose $E_{8}$ into conjugacy classes with
respect to ${\cal N} \oplus {\cal I}$,
a familiar procedure used in covariant lattice
models, we get schematically that
\be
E_{8} = ({\cal N},0) + (0,{\cal I})
+ \sum_{i} (W_{i}({\cal N}),W_{i}({\cal I}))
\eq
This means that there are three types of lattice vectors,
namely those belonging to the sublattices ${\cal N}$ and ${\cal I}$
and those which have a non--vanishing projection
onto both ${\cal N}$ and ${\cal I}$.
For a review of lattice techniques see \cite{LSW}.

Using this decomposition the effect of switching on
continuous Wilson lines becomes quite obvious. If we
assume for the moment that that all eigenvalues of the
gauge twist also appear in the internal twist, which is
true for all ${\bf Z}_{3}$, ${\bf Z}_{4}$, ${\bf Z}_{6}'$
and ${\bf Z}_{7}$ orbifolds, then the Wilson lines will
take arbitrary values in $\la {\cal N} \ra_{\fs{R}}$.
Thus for any generic choice of the Wilson lines only states
corresponding to lattice vectors in $(0,{\cal I})$ are massless.
Note that these states are automatically twist invariant.
Therefore the gauge group of the orbifold contains at least
a semi--simple group corresponding to the roots of the
lattice ${\cal I}$. However, the rank of this group is not a priori
guaranteed to be $8 - \dim({\cal N})$. This follows from the
fact that the subgroup, to which $E_{8}$ is broken, need not
to be semi--simple but only reductive, that is there can
be $U(1)$ factors around.
On the other hand, when considering the action of the
twist on the Cartan subalgebra, one knows that
$\dim({\cal N})$ Cartan generators are not invariant under the twist
and are therefore projected out, whereas $8 - \dim({\cal N})$
are invariant under the twist. Therefore, the rank of the unbroken gauge
group is $8 - \dim({\cal N})$ and the gauge group itself is given by
\be
G_{\cal I} \otimes U(1)^{8 - \dim({\cal N})
- \mbox{\scriptsize rk}(G_{\cal I})}
\label{giu1}
\eq
where $G_{\cal I}$ is the semi-simple Lie group associated
to the roots of the lattice ${\cal I}$.

Decompositions of the form ${\cal N} \oplus {\cal I}$
can easily be found
using the formalism of extended Dynkin diagrams \cite{Cah}. There
are, however, cases where the decomposition is not unique. This
is not in contradiction with ${\cal I}$ being maximal, because
$\supset$ is a partial ordering relation, only. It is well
known that it happens in a few number of cases that
the twist of the full algebra does not only depend
on the isomorphic type of the subalgebra that is twisted, but
also on the precise embedding \cite{Car}. This is easily illustrated by the
following example, namely
by decomposing $E_{8}$ into cosets with respect to
$A_{1}^{8}$ and then taking the $A_{1}^{4}$ Coxeter twist.
{From} the decompostion one sees that, depending on the choice of
the
$A_{1}$'s, one has that ${\cal I}=D_{4}$ or ${\cal I}=A_{1}^{4}$.
These twists are known as $A_{1}^{4\,I}$ and $A_{1}^{4\,II}$,
respectively.

Fortunately, we can use the results of the classification
of conjugacy classes of the Weyl group, which is
equivalent to the classification of  Weyl twist
modulo conjugation, for determining the minimal gauge groups.
In his work \cite{Car} Carter gives, for all twists,
the decompositions of the
root system, which specify the group $G_{\cal I}$. Then,
according to (\ref{giu1}),
all one has
to do is to add some $U(1)$ factors, if necessary.
We have listed all the minimal gauge groups
that can appear in the context of $N=1$ supersymmetric
${\bf Z}_{N}$ orbifold compactifications.

Note, however, that we have so far assumed that the Wilson lines
are really allowed to take values in all of
$\la {\cal N} \ra_{\fs{R}}$. But this is not the case if
an eigenvalue of the gauge twist does not appear in the internal
twist. The simplest example for this is provided
by the $A_{3}$ Coxeter twist which has, as a lattice twist,
the eigenvalues $\omega^{i}$, $i = 1,2,3$ where
$\omega= \exp(2 \pi i / 4)$. This can be combined with
a ${\bf Z}_{8}$ twist of the internal space
with eigenvalues $\Omega^{k}$,
$k=1,2,3,5,6,7$, where $\Omega = \exp(2 \pi i / 8)$.
Since $\theta$ does not have the eigenvalue $-1$, there
are no Wilson lines taking values in the $-1$ eigenspace
of $\theta'$. Therefore, the minimal gauge group is larger then
the expected $SO(10)$. A detailed analysis shows that
the unbroken gauge group is the non--simply laced group
$SO(11)$. Note that this is not in contradiction with the
twist being defined through a Weyl twist of $E_{8}$,
because an inner automorphism of the full group may be
an outer one of a subgroup. Therefore, breaking to
non--regular subgroups is possible, if Wilson lines are turned on.

In these cases, which include many of the ${\bf Z}_{6}$
and ${\bf Z}_{8}$,
and most of the ${\bf Z}_{12}$ and ${\bf Z}_{12}'$
orbifolds,
our list only provides a lower bound on the
gauge group that is not saturated. Note that the analysis to be
performed in order
to get the minimal gauge group is, in these cases, the same as the one
one has to use in order to get intermediate gauge groups,
that is gauge groups that are neither minimal nor maximal.
A combination of counting and embedding
arguments as used in \cite{Moh2} is in many cases sufficent
to determine the gauge group.

The maximal gauge groups which appear for vanishing Wilson lines
have been described in \cite{HolMyh}. Here the gauge group of the
torus modes is $E_{8}$ and therefore all the three classes of
vectors in the decomposition are present. In order to determine
the gauge group of the orbifold one has to form twist invariant
combinations of the states corresponding to lattice vectors
in $({\cal N},0)$ and $(W_{i}({\cal N}),W_{i}({\cal I}))$,
because these vectors
transform non--trivially under the twist. For convenience
we have included the results of \cite{HolMyh} in our table.

The table is organized as follows. We list all twists that have
order $2,3,4,6,7,8$ or 12 and can therefore appear in the
context of $N=1$ orbifolds. We quote the conjugacy class of the
twist and its name from \cite{Car}.
So, a Coxeter twist in the subalgebra $X$ is called $X$, whereas
the non--Coxeter twists, if they are needed, are called
$X(a_{i})$. If the twist depends on the embedding of the
subalgebra, then the inequivalente choices are labeled by
$X^{I}, X^{II},\ldots$.
Next, we list the order of the twist.
An asteriks is put on those numbers, where the order of the
Lie algebra twist is twice the order of the corresponding lattice
twist, following \cite{HolMyh}. We also list the non--trivial eigenvalues
of the twist, because they specify the structure of the untwisted
moduli space \cite{CLM}. Actualy, we list the corresponding powers
of the $N$-th primitive root of unity where $N$ is the order of the
twist. The eigenvalues of the Coxeter twists have been calculated
using their relation to the ranks of the inequivalent Casimir
operators, whereas the eigenvalues of non--Coxeter twists are
quoted from \cite{SchWar}.
Then we give the minimal gauge groups, by taking
the semi--simple part from \cite{Car} and adding $U(1)$s if
needed. Finally, we also quote the maximal
gauge groups from \cite{HolMyh}.

\section{Extended gauge groups from the compactification sector
\label{ExGGCompS} }

\setcounter{equation}{0}

In this section we will discuss the enhancement of the gauge group
occuring in
the compactification (also called internal)
sector of Narain compactifications and
Narain orbifolds.  We will also take into account the effect of continuous
Wilson lines on this enhancement.

In the toroidal case,
the generic gauge symmetries come from the twelve
conserved chiral world sheet currents $\p X^{i}_{L}(z)$
and $\ov{\p} X^{i}_{R}(\ov{z})$
corresponding to the left- and right--moving parts of the
six internal coordinates. Thus, the generic gauge group
coming from the compactification sector is
$U(1)^{6}_{L} \otimes U(1)^{6}_{R}$.
The six gauge bosons of the $U(1)^{6}_{R}$ are graviphotons,
that is they are part of an $N=4$ gravitational supermultiplet,
whereas the other six gauge bosons belong to $N=4$ vector multiplets.

In order to break the extended $N=4$ supersymmetry to $N=1$ one
must break the $U(1)^{6}_{R}$ completely. This is done by every
twist that doesn't have 1 as an eigenvalue. To preserve
the minimal $N=1$ supersymmetry, the twist must be in the
subgroup $SU(3) \subset SO(6)$ \cite{DHVW2}. This leads to the
classification {\cite{ErlKle} of $N=1$ twists.

Since the twist acts in a left - right symmetric way
the $U(1)^{6}_{L}$ is automatically also broken completely.
There are, however, similarly to the gauge
sector,
special values of the moduli for which
one has extra conserved currents. Let us discuss this
for the untwisted model first.
The additional massless gauge bosons  are related to Narain
vectors
${\bf P} = q^{I} {\bf l}_{I} + n^{i} \ov{\bf k}_{i}+  m_{i} {\bf k}^{i}$
with quantum numbers $q^{I}=0$, $n^{i}m_{i}$
$:= n^{T}m = 1$.
In order to extend the generic gauge group $U(1)^{6}_{L}$
to $G^{(l)} \otimes U(1)^{6-l}$, where
$G^{(l)}$ is a rank $l \leq 6$ semi--simple simply laced Lie group
with Cartan matrix $C_{ij}$, $i,j = 1,\ldots,l$,
the moduli must satisfy the relations \cite{Moh1}
\be
{\bf A}_{i} \cdot {\bf A}_{j} + 4\, G_{ij} = C_{ij}
\label{aiaj}
\eq
and
\be
4 \, B_{ij} = C_{ij} \mbox{   modulo   } 2.
\label{bijcij}
\eq
This implies that $D_{ij} \in {\bf Z}$ and moreover the
vectors
\be
{\bf P}^{(i)} =
\left( {\bf p}^{(i)}_{L}; {\bf p}^{(i)}_{R} \right)
= \ov{k}_{i} - D_{ij} k^{j} =
\left( {\bf A}_{i}, 2 {\bf e}_{i}; 0_{6} \right)
\eq
are in the Narain lattice. The vectors
${\bf e}_{i} = G_{ij} {\bf e}^{j}$
are basis vectors of the compactification lattice $\Lambda$.
Since $C_{ij}$ is a Cartan matrix,
$({\bf p}^{(i)}_{L})^{2} = 2$ and therefore the ${\bf p}^{(i)}_{L}$
are a set of simple roots of $G^{(l)}$.

For vanishing Wilson lines the conditions
(\ref{aiaj}), (\ref{bijcij})
reduce to the more special
conditions
for points of extended gauge symmetry
that are known from \cite{Gin}.
Note that an extended gauge group is
compatible with, at least,  small deformations of the Wilson
lines, because their effect can be compensated by tuning the
metric.

In the orbifold case,
the moduli are further restricted by the condition of compatibility
with a given
twist.
These conditions are, in the absence of discrete
background fields, given by (\ref{ConstraintWL}) and
\be
D_{ij} \theta^{j}_{\;k} = \theta_{i}^{\;j}D_{jk}
\eq
where
\be
D_{ij} = 2 \left( B_{ij} - G_{ij} - \frac{1}{4} {\bf A}_{i} \cdot
{\bf A}_{j} \right)
\eq
and $\theta_{i}^{\;j}$, $\theta^{i}_{\;j}$ are the matrices of
the internal twist with respect to lattice bases of the
compactification lattice $\Lambda$ and its dual \cite{Moh2}.

As an example, let us discuss the
compactification on a 2-torus $T_2$.
There are two rank two
semi--simple simply laced Lie algebras, namely
$A_{1} \oplus A_{1}$ and $A_{2}$ with corresponding gauge groups
$SU(2)^{2}$ and $SU(3)$.  These enhanced gauge groups occur at
special points in the moduli space.  Let us in the following
first consider the case of vanishing Wilson lines.
The special points of enhanced gauge symmetries are determined
by equations (\ref{aiaj}) and (\ref{bijcij}).
For the maximal enhancement of the gauge group
to $SU(2)^2$ or $SU(3)$, these equations
do completely fix the moduli $G_{ij}$ and $B_{ij}$,
up to discrete transformations. Introducing
complex moduli $T$ and $U$,
\be
T = 2 \left( \sqrt{G} - i B_{12} \right),\;\;\;
U = \frac{1}{G_{11}} \left( \sqrt{G} - i G_{12} \right),
\eq
where $G_{ij}, B_{ij}$, $i,j = 1,2$ are the real moduli
of the 2 - torus, these conditions can be
expressed as
\be U = T = 1
\label{ut1}
\eq
and
\be U=T= e^{i \pi /6}
\label{utro}
\eq
for a maximal enhancement to $A_{1} \oplus A_{1}$ and $A_{2}$,
respectively \cite{IbaLuLer}.  Note that in these equations
we have restricted ourselves to the critical points of the
standard fundamental domain.
If one allows for an abelian factor, then one can
also have the gauge group $SU(2) \otimes U(1)$.  Inspection
of equations (\ref{aiaj}) and (\ref{bijcij}) shows that, in this case, not
all of the
moduli are fixed by these equations, but that
two real moduli parameters are left unfixed.  These two parameters
can be taken
to be one of the two radii of the $T_2$ as well as
the angle between these two radii.
In terms of the
complex moduli $T$ and $U$, this means that the extended gauge
group $SU(2) \otimes U(1)$ occurs for points in moduli space
for which $U=T$.
Note that this
complex subspace $U=T$ contains the
two points (\ref{ut1}) and (\ref{utro}) of maximally extended symmetry.
For generic values $U \not=T$, the toroidal gauge group is given by
$U(1)^{2}$.

Let us now discuss the enhancement of the gauge group in the context of
orbifold compactifications for which the underlying internal 6-torus
decomposes into a $T_6 = T_2 \oplus T_4$.  For concreteness, let us
study the effect on this enhancement of a ${\bf Z}_2$ acting on the
2-torus $T_2$.  This is the situation encountered in a ${\bf Z}_4$-orbifold,
for instance.  The
${\bf Z}_{2}$ twist, given by the reflection $-I_{2}$,
doesn't put any additional constraints on
the four real moduli $G_{ij}$ and $B_{ij}$ and on
their complex version $U$ and $T$. Therefore, the
moduli space associated with the $T_2$ is the same
as in the toroidal case. As is well known from
one--dimensional compactifications, the enhanced ($SU(2)$) and
the generic ($U(1)$) gauge groups get broken to
$SU(2) \rightarrow U(1)$ and
$U(1) \rightarrow \emptyset$ by the ${\bf Z}_2$-twist, respectively.
Thus, in the case of the ${\bf Z}_2$-twist acting on the internal $T_2$,
 the gauge groups
for the points
$U=T=1$, $\exp(i \pi /6) \not= U=T \not=1$ and
$U \not= T$ in moduli space are given by
$U(1)^{k}$ with $k=2,1,0$, respectively.

Something more interesting happens at the $SU(3)$ symmetric
point $U=T=\exp(i \pi /6)$
in the orbifold case. The ${\bf Z}_2$-twist on $T_2$ can be decomposed
into
\be
-I_{2} = W_{1} C^{-1} D
\eq
where $W_{1}$, $C$ and $D$ are the first Weyl reflection, the Coxeter
twist and the diagram automorphism of $A_{2}$, respectively.
Therefore the twist acts as an outer automorphism and breaks
the $SU(3)$ to the maximal non--regular subgroup $SU(2)$.
Thus, we have found a bosonic realization of the conformal
embedding $SU(2)_{k=4} \subset SU(3)_{k=1}$ and
expect that the $SU(2)$ is realized at the higher level $k=4$
in order to have central charge $c=2$.
Indeed, a direct calculation of the OPE of the twist--invariant
combinations of conserved currents shows that there
is a $SU(2)$ current algebra at level $k=4$.

Note that this phenomenon of rank reduction and simultaneous increase
of the level
is also quite generically present in the gauge sector
because, as mentioned in the previous section,
a Weyl twist of the $E_{8}$ will
often act as an outer automorphism of a subalgebra left
unbroken by Wilson lines.
Take as an example the $SU(3)_{3}$
model obtained in \cite{FonIbaQue} through switching on Wilson line
moduli. By inspection of the vertex operators given in
\cite{FonIbaQue} one easily sees that the algebra of the
corresponding deformed untwisted model is $SO(8)_{1}$,
like in some of the models described in \cite{Moh2}.
Therefore all these models  should be bosonic realizations of the
conformal embedding $SU(3)_{3} \subset SO(8)_{1}$.

The points of extended gauge symmetry are
fixed points under some transformation belonging to
the modular group $SO(2,2,{\bf Z})$. There is a
remarkable relation between the order of that transformation
and the number of extra massless gauge bosons.
Namely, we will now show, for a
two--dimensional torus compactification
and for its ${\bf Z}_{2}$ and ${\bf Z}_{3}$ orbifolds,
that
\be
\mbox{order of fixed point} = (\mbox{order of twist})
\times (\mbox{number of extra gauge bosons})
\label{orfixp}
\eq
We will present here the case of the two--dimensional
torus and its
${\bf Z}_{2}$ orbifold. The ${\bf Z}_{3}$ orbifold will
be discussed
at the end of this section.

The modular group of both the torus compactification and
its ${\bf Z}_{2}$ orbifold is, in the absence of Wilson lines,
the group $SO(2,2,{\bf Z})$. This group has four generators, which
we can take to be ${\cal S}, {\cal T}, {\cal D}_{2}, {\cal R}$
as defined in
\cite{GPR}. ${\cal S}$ and ${\cal T}$ generate the
subgroup $SL(2,{\bf Z}) \subset SO(2,2,{\bf Z})$, which is the
subgroup of orientation preserving basis changes in $\Lambda$,
whereas ${\cal R}$ is the reflection of the first coordinate
and ${\cal D}_{2}$ is the factorized duality in the second
coordinate. Note that there is a second $SL(2,{\bf Z})$
subgroup
which commutes with the first one. It is generated by
${\cal S}'$ $={\cal D}_{2}{\cal S}{\cal D}_{2}$ and
${\cal T}'$ $={\cal D}_{2}{\cal T}{\cal D}_{2}$.
Whereas ${\cal T}'$ is the axionic shift symmetry,
${\cal S}'$ is almost the full duality ${\cal D}$
$={\cal D}_{1}{\cal D}_{2}$, namely ${\cal S}'$
$={\cal S}{\cal D}$.
Note that ${\cal D}_{1}$, which is the factorized
duality transformation of the first coordinate,
is not an independent generator of the group
because ${\cal P}$ $:= {\cal R} {\cal S}$ permutes
the two coordinates and
therefore ${\cal D}_{1}= {\cal P}{\cal D}_{2}{\cal P}$.
See \cite{GPR} for a detailed discussion.
The explicit matrices given there specify the linear action
of the modular group $SO(2,2,{\bf Z})$ on the
quantum numbers.

The group $SO(2,2,{\bf Z})$
acts non--faithfully and fractionally linear
on the moduli. The faithfull transformation
group $PSO(2,2,{\bf Z})$ is known to be generated by the
transformations \cite{DijVerVer}
\be
\wt{\cal S}: (U,T) \rightarrow (\frac{1}{U},T) \;\;\;
\wt{\cal T}: (U,T) \rightarrow (U + i,T)
\eq
\be
\wt{\cal R}: (U,T) \rightarrow (\ov{U},\ov{T}),\;\;\;
\wt{\cal M}: (U,T) \rightarrow (T,U).
\eq
Note that as a consequence of our definition of the $U$ modulus
with $G_{11}$ in the denominator, we had to rearrange
the generators as:
$\wt{\cal S} = {\cal S}$, $\wt{\cal T} = {\cal S}{\cal T}{\cal S}$,
$\wt{\cal R} = {\cal R}$ and
$\wt{\cal M} = {\cal R}{\cal S}{\cal D}_{2}$, in order
to achieve that the transformations take their standard form.
The well known subgroups $SL(2,{\bf Z})_{U}$ and $SL(2,{\bf Z})_{T}$
are generated by $\wt{\cal S}, \wt{\cal T}$ and
$\wt{\cal S}' = \wt{\cal M}\wt{\cal S}\wt{\cal M}$,
$\wt{\cal T}' = \wt{\cal M}\wt{\cal T}\wt{\cal M}$
respectively.

We can now prove the statement given above.
The extended gauge groups $SU(2) \otimes U(1)$, $SU(2)^{2}$
and $SU(3)$ appear at the points $U=T \not= 1, e^{i \pi/6}$,
$U=T=1$
and $U=T=e^{i \pi/6}$. These are fixed points of the
transformations $\wt{\cal M}$, $\wt{\cal M}\wt{\cal S}$
and $\wt{\cal M}\wt{\cal T} \wt{\cal S}$ which have
the orders $2$, $4$ and $6$ respectively. By formally
taking the twist to be the identity here, we see that equation (\ref{orfixp})
holds, because the orders of the fixed point transformations
are equal to the numbers of extra massless gauge bosons
appearing at these points. This can moreover be extended
to the point at infinity, which is a fixed point of order
$\infty$ under the translation $\wt{\cal M}\wt{\cal T}$.
Since the limit $U,T \rightarrow \infty$ describes
the decompactification of the torus, infinitely many Kaluza--
Klein states become massless there, as predicted by the
order of the fixed point \cite{CveLu}.

In the ${\bf Z}_{2}$ orbifold the critical points are the same,
but since one must form twist invariant combinations
the numbers of extra massless gauge bosons is divided
by the order of the twist. Therefore equation (\ref{orfixp})
holds as well. The case of the ${\bf Z}_{3}$ orbifold will be
discussed below.

Let us now switch on Wilson lines
in the ${\bf Z}_{2}$ orbifold model discussed above.
As already argued above in terms
of the real moduli, any extended gauge group can be
preserved by an appropriate tuning of the moduli of the
compactification sector. All we have to add here is the prescription
of this tuning in terms of the complex moduli. We will do
this for the case of two complex Wilson line moduli $B,C$
associated with the internal 2-torus $T_2$.
The $B,C$ moduli are \cite{CLM}
\[
B =  \frac{1}{G_{11}} \left(
{\cal{A}}_{11} \sqrt{G} - {\A}_{21} G_{12} + {\A}_{22} G_{11}+
i (-{\A}_{11} G_{12} + {\A}_{12} G_{11} - {\A}_{21} \sqrt{G})\right)
\]
\be
C = \frac{1}{G_{11}} \left(
{\cal{A}}_{11} \sqrt{G} +  {\A}_{21} G_{12} - {\A}_{22} G_{11} +
i (-{\A}_{11} G_{12} + {\A}_{12} G_{11} +{\A}_{21} \sqrt{G}) \right).
\eq
The $T$ modulus now reads
\be
T= 2\left(\sqrt{G} (1+ \frac{1}{4}
\frac{ {\A}^{\mu}_1
{\A}_{\mu 1}}{G_{11}})
-i(B_{12}+\frac{1}{4}
{\A}^{\mu}_1 {\A}_{\mu 1} \frac{G_{12}}{G_{11}}
- \frac{1}{4} {\cal{A}}^{\mu}_1 {\cal{A}}_{\mu 2} )
 \right)
\eq
whereas the $U$ modulus is not modified.
Here ${\A}_{\mu i}$ denotes the $\mu$-th component of the
$i$-th Wilson line with respect to an orthonormal frame,
$\mu,i=1,2$.

States which become massless
for generic $U=T$ (where the orbifold gauge group is
$SU(2)\otimes U(1) \rightarrow U(1)$)
stay massless when Wilson lines
are turned on. Thus, no tuning of the complex moduli $U$ and $T$
is necessary.  In order to have
the maximally extended gauge group
$SU(2)^{2} \rightarrow U(1)^{2}$,
the original condition $T=U=1$ is, in the
presence of Wilson lines, replaced by
\be
T = U = \sqrt{ 1 + \frac{BC}{2} }
\eq
whereas in order to have $SU(3) \rightarrow SU(2)$, the new condition is
given by
\be
T = U = \frac{i}{2} + \sqrt{ \frac{3}{4} + \frac{BC}{2} }.
\eq
This follows from the discussion of the zeros
of the mass formula, which will be presented in section
\ref{AutFunNulOrb}.

Consider, as another example, a ${\bf Z}_{3}$ orbifold
defined by the Coxeter twist of the $A_{2}$ root lattice.
Here, the moduli have to be restricted in order to be compatible
with the twist. In terms of complex moduli one has to set
$U = \frac{1}{2} ( \sqrt{3} + i ) = e^{i \pi /6}$ and $C = 0$,
whereas $T$ and $B =: \sqrt{3} {\cal A}$ remain moduli.
Clearly, the $SU(3)$ point ($U=T= e^{i \pi /6}$)
is in the ${\bf Z}_{3}$ moduli
space, and at this special point in moduli space
the toroidal gauge group
$SU(3)$ is broken to $U(1)^{2}$. For generic $T$, the toroidal
gauge group is $U(1)^{2}$, whereas there is no leftover gauge group in
the orbifold case.
Note again that the product of
twist order and number of extra massless states in the orbifold model
equals the order of the fixed point with respect to the modular
group of the untwisted model, namely  $3 \cdot 2=6$.
No tuning of $T$ is required
to preserve the $SU(3) \rightarrow U(1)^{2}$ gauge group in
the presence of Wilson lines.
The condition for having an extended gauge
symmetry is that $T= e^{i \pi /6}$, whereas
$B =: \sqrt{3}{\cal A}$ is arbitrary.

Since the $U$ and the $C$ modulus are frozen to  discrete values,
the modular group of the ${\bf Z}_{3}$ orbifold
is smaller than the one of the untwisted model.
In the case of vanishing Wilson lines the modular group
is obviously
$SU(1,1,{\bf Z})$ $\simeq$ $SL(2,{\bf Z})_{T}$ with
generators $\wt{\cal S}'$ and $\wt{\cal T}'$ as defined above.
The point $T= e^{i \pi /6}$ of extended $U(1)^{2}$ symmetry is
a fixed point under $\wt{\cal T}'\wt{\cal S}'$, which is a
transformation of order 3. This reduction of the order, as
compared to the modular group of the untwisted model,
is due to the fact that the transformation $\wt{\cal M}$,
which is of order 2, is not in the reduced modular
group $SU(1,1,{\bf Z})$.

\section{Mass formulae for $SO(p+2,2)$ and $SU(m+1,1)$ cosets}

\setcounter{equation}{0}

In this chapter we show that in the case of a factorizing
2--torus $T_2$,
the moduli dependent part of
the mass formula for the untwisted sector of an
$N=1$ orbifold can be written as $|{\cal M}|^{2}$/$Y$.
${\cal M}$ is a holomorphic function of the moduli and depends
on the quantum numbers.  $Y$ is a real analytic function of
the moduli, only, and is
related to the K\"ahler potential by
$K = - \log Y$.

\subsection{Torus compactifications and the $SO(22,6)$ coset}

Let us first recall that the mass formula for the
heterotic string compactified on a torus is
\be
\frac{\alpha'}{2} M^{2} = N_{L} + N_{R} + \frac{1}{2}
({\bf p}_{L}^{2} + {\bf p}_{R}^{2}) -1
\eq
and that physical states must also satisfy the level matching
condition
\be
\frac{\alpha'}{2} M^{2}_{L} :=
N_{L} + \frac{1}{2} {\bf p}_{L}^{2} -1 \stackrel{!}{=}
N_{R} + \frac{1}{2} {\bf p}_{R}^{2}
=: \frac{\alpha'}{2} M^{2}_{R}.
\eq
Here, we have absorbed the normal ordering constant of the NS
sector into the definition of the right moving number operator
and restricted ourselves to states surviving the GSO projection.
Thus $N_{R}$ has an integer valued spectrum in both the
NS and the R sector.

Substituting the second equation into the mass formula yields
\be
\frac{\alpha'}{2} M^{2} = {\bf p}_{R}^{2} + 2 N_{R}.
\eq
Our aim is to
make the moduli dependence of the mass explicit.

The first step is to express ${\bf p}_{R}$ in terms of the quantum numbers
$q^{I}, n^{i}, m_{j}$ and the real moduli
$G_{ij}, B_{ij}, {\bf A}_{i}$. This can be done by expanding
the Narain vector
${\bf P} = ({\bf p}_{L}; {\bf p}_{R}) \in \Gamma$ in terms of the lattice
basis ${\bf l}_{I}$, $\ov{\bf k}_{i}$, ${\bf k}^{j}$
(\ref{lvector}) - (\ref{kvector}),
\be
{\bf P} = q^{I} {\bf l}_{I} + n^{i}\ov{\bf k}_{i} + m_{j} {\bf k}^{j}
\eq
and then projecting onto the right handed part by multiplying with
the vectors ${\bf e}_{\mu}^{(R)}$ = $(0_{16}, 0_{6}, {\bf e}_{\mu})$,
where the ${\bf e}_{\mu}$ are an orthonormal basis of ${\bf R}^{6}$
\be
{\bf p}_{R} = (q^{I},n^{i},m_{j}) \left( \begin{array}{c}
 {\bf l}_{I} \cdot {\bf e}_{\mu}^{(R)} \\
 \ov{\bf k}_{i} \cdot {\bf e}_{\mu}^{(R)} \\
  {\bf k}^{j} \cdot {\bf e}_{\mu}^{(R)} \\
 \end{array} \right)
  {\bf e}_{\mu}^{(R)}.
\eq
The norm squared takes then the form
\be
{\bf p}_{R}^{2} = {\bf v}^{T} \Phi \Phi^{T} {\bf v}
\eq
where ${\bf v}$ is the vector of quantum numbers,
\be
{\bf v}^{T} = \left( q^{I},n^{i},m_{j} \right) \in M(1,28,{\bf Z})
\simeq {\bf Z}^{28}.
\eq
The matrix $\Phi$ contains the moduli
\be
\Phi =
\left( \begin{array}{c}
 {\bf l}_{I} \cdot {\bf e}_{\mu}^{(R)} \\
 \ov{\bf k}_{i} \cdot {\bf e}_{\mu}^{(R)} \\
  {\bf k}^{j} \cdot {\bf e}_{\mu}^{(R)} \\
 \end{array} \right) =
\frac{1}{2} \left( \begin{array}{c}
- AE^{*} \\ DE^{*} \\ E^{*} \\
\end{array} \right) \in M(28,6,{\bf R})
\eq
where
\be
A = (A_{Ii}) = ({\bf e}_{I} \cdot {\bf A}_{i}) \in M(16,6,{\bf R})
\eq
and
\be
D = (D_{ij}) =
2 \left( B_{ij} - G_{ij} - \frac{1}{4} {\bf A}_{i} \cdot
{\bf A}_{j} \right) \in M(6,6,{\bf R})
\eq
are the moduli matrices and
\be
E^{*} = (E^{i}_{\;\;\nu}) = ({\bf e}^{i} \cdot {\bf e}_{\nu})
\eq
is a 6--bein whose appearence reflects the fact that the
${\bf e}_{\mu}^{(R)}$
can be rotated by an $SO(6)$ transformation.
${\bf e}^{i}$ are a basis of the dual $\Lambda^{*}$ of the
compactification lattice $\Lambda$.

The matrix $\Phi$ satisfies the equation
\be
\Phi^{T} H_{22,6}  \, \Phi = -I_{6},
\label{O226LB}
\eq
where $H_{22,6}$ is the standard pseudo--euclidean
lattice metric of $\Gamma = \Gamma_{22;6}$:
\be
H_{22,6} = \left( \begin{array}{ccc}
C^{-1}_{(16)} & 0 & 0 \\
0 & 0 & I_{6} \\ 0 & I_{6} & 0 \\
\end{array} \right),
\eq
where $C_{(16)}$ is the Cartan matrix of $E_{8} \otimes E_{8}$.
By a coordinate transformation $\Phi \rightarrow \wt{\Phi}$
equation (\ref{O226LB}) can be brought to the form
\be
\wt{\Phi}^{T} \eta_{22,6} \wt{\Phi} = - I_{6},
\eq
with the standard pseudo--euclidean metric of
type $(+)^{22}(-)^{6}$
\be
\eta_{22,6} = \left( \begin{array}{cc}
I_{22} & 0 \\ 0 & - I_{6} \\
\end{array} \right).
\eq
This is the standard form of the constraint equation which
defines the coset space $SO(22,6)/(SO(22) \otimes SO(6))$
in terms of homogeneous coordinates $\wt{\Phi}$ \cite{Gil}.
Therefore $\Phi$ is a modified  homogenous coset coordinate.
It has the advantage that not only
the deformation group $SO(22,6)$ acts linearly on it,
as usual for homogenous
coordinates, but that the subgroup
of modular transformations acts by integer valued matrices \cite{CLM}.
In the following the pseudo--euclidean lattice metric of
an integer lattice $\Gamma_{m;n}$ and the standard metric
of type $(+)^{m}(-)^{n}$ will be denoted by
$H_{m,n}$ and $\eta_{m,n}$ respectively.

\subsection{Orbifold compactifications and symmetric K\"ahler
spaces}

In the following subsections we will derive mass formulae
and parametrizations for the untwisted moduli of orbifold
compactifications. This will be done in four steps. In the
first step we will recall how one can derive an explicit
real parametrization of orbifold moduli spaces by solving
the constraint equations, imposed by the compatibility requirement with
a given twist, on the moduli $G_{ij}, B_{ij}, {\bf A}_{i}$
of the Narain model.
This solution can then be used in a second step to locally
factorize the moduli space into spaces corresponding
to distinct eigenvalues of the twist. The third step is
then to find appropriate complex coordinates on each factor
space which make explicit its K\"ahler structure. This can be
done by using the relations between the the real moduli and
homogenous coset coordinates. Finally, one can solve the
constraint equations for an independent set of complex moduli.
Plugging these results into the mass formula allows one
to write its moduli dependent part as the ratio of a
holomorphic and a real analytic piece, where the latter one
is related to the K\"ahler potential.

\subsubsection{Untwisted orbifold moduli}

We will now implement the first step of the four described above. Let us
recall that it was shown in \cite{Moh2}, based on earlier
results of \cite{IbaNilQue,EJL}, that the continuous parts
of the background fields  $G_{ij}, B_{ij}, {\bf A}_{i}$
must satisfy the equations
\be
D_{ij} \theta^{j}_{\;k} = \theta_{i}^{\;j} D_{jk},\;\;\;
A_{Ij} \theta^{j}_{\;k} = \theta_{I}^{\;J} A_{Jk}
\label{constraintDA}
\eq
where $\theta_{i}^{\;j}$, $\theta^{j}_{\;k}$ and $\theta_{I}^{\;J}$
are the matrices of the internal twist $\theta$ and of the
gauge twist $\theta'$ with respect to the lattice bases
${\bf e}_{i}$, ${\bf e}^{i}$ and ${\bf e}_{I}$ of the lattices $\Lambda$,
$\Lambda^{*}$ and $\Gamma_{16}$.
Although $\theta$ and $\theta'$ are orthogonal {\em maps}
their matrices with respect to non--orthonormal lattice
bases are not. Thus, it is convenient to express equations
(\ref{constraintDA})
in terms of orthonormal bases ${\bf e}_{\mu}$ and
${\bf e}_{M}$ of ${\bf R}^{6}$ and ${\bf R}^{16}$, before solving
them \cite{EJL,Moh2}. The transformations
between the lattice and orthonormal frames is given by
n--bein matrices
$E^{*}= (E^{i}_{\;\nu}) := ({\bf e}^{i} \cdot {\bf e}_{\nu})$,
$E = (E_{i\nu}) := ({\bf e}_{i} \cdot {\bf e}_{\nu})$,
${\cal E} = ({\cal E}_{I M}) = ({\bf e}_{I} \cdot {\bf e}_{M})$,
etc. Note that
orthonormal bases are "selfdual", ${\bf e}_{M} = {\bf e}^{M}$,
${\bf e}_{\mu} = {\bf e}^{\mu}$,
whereas lattice bases are (generically) not:
${\bf e}_{I} \not= {\bf e}^{I}$, ${\bf e}_{i} \not= {\bf e}^{i}$.
Therefore
$E_{i\mu} = E_{i}^{\;\mu} \not= E^{i}_{\;\mu}$. A useful relation
to be used later is $E^{*}= E^{T,-1}$.

One complication is that the metric moduli
drop out of the $D$--matrix when written with respect to an orthonormal
frame, because ${\bf e}_{\mu} \cdot {\bf e}_{\nu} = \delta_{\mu \nu}$
and therefore (see below for details of the transformation)
\be
D_{\mu \nu} = 2\left( B_{\mu \nu} - \delta_{\mu \nu} - \frac{1}{4}
{\bf A}_{\mu} \cdot {\bf A}_{\nu} \right)
\eq
Instead, they are now contained in the 6--bein $E$ which is
sensitive to deformations of $\Lambda$.
To study the effect of lattice deformations let us fix
a reference lattice $\ov{\Lambda}$ and introduce a deformation
map $S$ (or better a family of deformation maps)
which maps it to $\Lambda$
\be
S: \ov{\Lambda} \rightarrow \Lambda: \ov{\bf e}_{i} \rightarrow
{\bf e}_{i} =
S_{i}^{\;j} \ov{\bf e}_{j}
\Longrightarrow
\ov{G}_{ij}= \ov{\bf e}_{i} \cdot \ov{\bf e}_{j} \rightarrow
G_{ij} = {\bf e}_{i} \cdot {\bf e}_{j}
\eq
This is compatible with the twist
$\theta: \ov{\bf e}_{i} \rightarrow \theta_{i}^{\;j}
\ov{\bf e}_{j}$ if
\be
S_{i}^{\;j} \theta_{j}^{\;k} = \theta_{i}^{\;j} S_{j}^{\;k}.
\label{constraintS}
\eq
In order to transform equations (\ref{constraintDA}) and
(\ref{constraintS}) into the orthonormal frame we will have to work
out some formulae. Consider therefore the deformation
described in terms of the ${\bf e}_{\mu}$
\be
S: {\bf e}_{\mu} \rightarrow {\bf e}'_{\mu} =
S_{\mu}^{\;\nu} {\bf e}_{\nu}
\Longrightarrow {\bf e}_{\mu} \cdot {\bf e}_{\nu} = \delta_{\mu\nu}
\rightarrow {\bf e}'_{\mu} \cdot {\bf e}'_{\nu} =: G_{\mu\nu}
\not= \delta_{\mu\nu}
\eq
Thus, a compactification on $\Lambda$ with background metric
$\delta_{\mu\nu}$ can be reinterpreted as a compactification
on a fixed lattice $\ov{\Lambda}$ in a deformed background
$G_{\mu\nu}$.
Noting that
\be
S_{\mu}^{\;\nu} = {\bf e}'_{\mu} \cdot {\bf e}^{\nu}
= {\bf e}'_{\mu} \cdot {\bf e}_{\nu}
=: S_{\mu\nu} \mbox{   and   }
S^{\mu}_{\;\nu} := {\bf e}'^{\mu} \cdot {\bf e}_{\nu}
= G^{\mu \nu} S_{\mu \nu}
\eq
and defining $S^{*} = (S^{\mu}_{\;\nu})$,
$S = (S_{\mu}^{\;\nu})$ we have that $S^{*} = S^{T,-1}$.

Next we introduce a fixed (moduli independent) coordinate
transformation which connects the orthonormal basis ${\bf e}_{\mu}$
to the lattice basis $\ov{\bf e}_{i}$ of the reference lattice
$\ov{\Lambda}$:
\be
\ov{\bf e}_{i} = T_{i}^{\;\mu} {\bf e}_{\mu}
\Rightarrow
T_{i}^{\;\mu} = {\bf e}_{i} \cdot {\bf e}^{\mu}
\eq
This matrix and its inverse
$T_{\mu}^{\;i} = {\bf e}_{\mu} \cdot {\bf e}^{\;i}$
connect the deformation matrices by
\be
S_{\mu}^{\;\nu} = T_{\mu}^{\;i} S_{i}^{\;j} T_{j}^{\;\nu}
\eq
Therefore, the matrix $T_{i}^{\;\mu}$ also connects
the deformed orthonormal basis ${\bf e}'_{\mu}$ to the deformed
lattice basis ${\bf e}_{i}$:
\be
{\bf e}_{i} = T_{i}^{\;\mu} {\bf e}'_{\mu}
\Rightarrow
T_{i}^{\;\mu} = {\bf e}_{i} \cdot {\bf e}'^{\mu}
= \ov{\bf e}_{i} \cdot {\bf e}^{\mu}.
\eq

The relations
between the four n--bein matrices
$E_{i}^{\;\nu},T_{i}^{\;\mu},S_{i}^{\;j},S_{\mu}^{\;\nu}$
and the
four bases $\ov{\bf e}_{i}$, ${\bf e}_{i}$, ${\bf e}_{\mu}$,
${\bf e}'_{\mu}$
are summarized by
\be
E_{i}^{\;\nu}= S_{i}^{\;j} T_{j}^{\;\nu} = T_{i}^{\;\mu}
S_{\mu}^{\;\nu}
\label{estts}
\eq
which shows that $S_{i}^{\;j}$ and $S_{\mu}^{\;\nu}$
relate the undeformed lattice basis $\ov{\bf e}_{i}$ and the
orthonormal basis ${\bf e}_{\mu}$ to their
images ${\bf e}_{i}$, $\ov{\bf e}_{\mu}$ under the deformation $S$,
whereas $T$ describes a fixed coordinate transformation
relating $\ov{\bf e}_{i}$ to ${\bf e}_{\mu}$ and
${\bf e}_{i}$ to ${\bf e}'_{\mu}$.
$E$ is a family of coordinate transformations relating the
moving frame ${\bf e}_{i}$ to the fixed orthonormal frame
${\bf e}_{\mu}$.
Equation (\ref{estts}) shows that $E$
can be factorised into a moduli--dependent piece
and a moduli--independent one in two different ways.

We can now use the n--bein matrices and transform the equations
(\ref{constraintDA}), (\ref{constraintS}) into the
orthonormal bases.
In order to do this we have to introduce the transformed moduli matrices
\be
D_{\mu \nu} = E_{\mu}^{\;i}D_{ij} E^{j}_{\;\nu} =
T_{\mu}^{\;i}\ov{D}_{ij} T^{j}_{\;\nu},\;\;\;
A_{M \nu} = {\cal E}_{M}^{\;I} A_{Ij} E^{j}_{\;\nu} =
{\cal E}_{M}^{\;I} \ov{A}_{Ij} T^{j}_{\;\nu},\;\;\;
S_{\mu}^{\;\nu} = T_{\mu}^{\;i} S_{i}^{\;j} T_{j}^{\;\nu}
\eq
and the transformed twist matrices
\be
\theta_{\mu}^{\;\nu} = E_{\mu}^{\;i} \theta_{i}^{\;j} E_{j}^{\;\nu} =
T_{\mu}^{\;i} \theta_{i}^{\;j} T_{j}^{\;\nu},\;\;\;
\theta^{\mu}_{\;\nu} = E^{\mu}_{\;i} \theta^{i}_{\;j} E^{j}_{\;\nu} =
T^{\mu}_{\;i} \theta^{i}_{\;j} T^{j}_{\;\nu},\;\;\;
\theta_{M}^{\;N} = {\cal E}_{M}^{\;I} \theta_{I}^{\;J}  {\cal E}_{J}^{\;N}.
\eq
Note that these relations are consistent thanks to (\ref{constraintS}).
Since ${\bf e}_{\mu}$, ${\bf e}_{M}$ are orthonormal bases it follows
that
\be
\theta_{\mu}^{\;\nu} = \theta^{\mu}_{\;\nu} =: \theta_{\mu\nu}
\mbox{   and   } \theta_{M}^{\;N} =: \theta_{MN}
\eq
are orthogonal matrices.

Using the formulae derived above we find that equations
(\ref{constraintDA}), (\ref{constraintS}) are equivalent to
\be
\theta_{MN} A_{N\nu} = A_{M\mu} \theta_{\mu \nu},\;\;\;
\theta_{\mu\nu} D_{\nu\rho} = D_{\mu \nu} \theta_{\nu \rho},\;\;\;
\theta_{\mu\nu} S_{\nu\rho} = S_{\mu \nu} \theta_{\nu \rho}.
\label{constraintsONB}
\eq
The orthonormal bases can be chosen such that the twists
$\theta,\theta'$
take their real standard forms, with nonvanishing $2\times 2$
blocks along the diagonal. A slight modification will turn out
to be useful in order to display the coset structure. Namely,
we will group together degenerate eigenvalues into bigger blocks.
For definiteness let us consider the gauge twist $\theta'$
and assume that it has complex eigenvalues $e^{\pm i \psi_{j}}$
and real eigenvalues $-1$ and $1$ with multiplicities
$m_{j}$, $p$ and $q$. Then there exists a basis $e_{M}$
such that
\be
(\theta_{MN}) = \bigoplus_{j} R_{j} \oplus  -I_{p} \oplus
I_{q}
\eq
with
\be
R_{j} = \left( \begin{array}{cc}
\cos( \psi_{j} ) I_{m_{j}} & - \sin( \psi_{j} ) I_{m_{j}} \\
\sin( \psi_{j} ) I_{m_{j}} & \cos( \psi_{j} ) I_{m_{j}} \\
\end{array} \right)
\in O(2 m_{j}, {\bf R})
\eq
The internal twist $\theta$ can be brought into the same form with
multiplicities $n_{j}$, $r$ and $s$.

In these coordinates the equations (\ref{constraintsONB})
can be easily solved. The result for the Wilson lines
matrix is
\be
(A_{M\nu}) = \bigoplus_{j} A^{(j)} \oplus  A^{(-1)}
\oplus  A^{(+1)},
\label{Adecomp}
\eq
with
\be
A^{(j)} = \left( \begin{array}{cc}
A^{(j)}_{1} & A^{(j)}_{2} \\
- A^{(j)}_{2} & A^{(j)}_{1} \\
\end{array} \right)  \in M(2 m_{j}, 2 n_{j}, {\bf R}), \;\;\;
A^{(-1)} \in M(p,r,{\bf R}),\;\;\;
A^{(+1)} \in M(q,s,{\bf R}).
\eq
and for the $D$- matrix the result is
\be
(D_{\mu \nu}) = \bigoplus_{j} D^{(j)} \oplus  D^{(-1)}
\oplus  D^{(+1)},
\eq
with
\be
D^{(j)} = \left( \begin{array}{cc}
D^{(j)}_{1} & D^{(j)}_{2} \\
- D^{(j)}_{2} & D^{(j)}_{1} \\
\end{array} \right)  \in M(2 n_{j}, 2 n_{j}, {\bf R}), \;\;\;
D^{(-1)} \in M(r,r,{\bf R}),\;\;\;
D^{(+1)} \in M(s,s,{\bf R}).
\label{Dblock}
\eq
$S$ has the same structure as $D$. Note, however, that
only the symmetric positive part (in the polar decomposition) of
$S$ is physically relevant, because the orthogonal part describes
a pure rotation of $\Lambda$. We use here that, when referring to an
orthonormal frame, the matrix of the positive symmetric
(orthogonal) part of an invertible map is a positive symmetric
(orthogonal) matrix. The irrelevance of rotations will also be
manifest by the fact that
the masses only depend on the metric through
\be
G_{ij} = S_{i}^{\;k} \ov{G}_{kl} S^{\;l}_{j} =
T_{i}^{\;\mu} G_{\mu\nu} T^{\nu}_{\;j} =
E_{i}^{\;\mu} E_{\mu j}
\eq
or
\be
G_{\mu\nu} = S_{\mu}^{\;\rho} \delta_{\rho\sigma} S_{\nu}^{\;\sigma}.
\eq

\subsubsection{Factorization of the moduli space}

We can now perform the second step, namely use the constrained
form (\ref{Adecomp}) - (\ref{Dblock})
of the moduli in order to factorize the untwisted
orbifold moduli space in factors corresponding to the various
distinct eigenvalues. In order to display its coset structure
we must work with the homogenous coset coordinate. Therefore
we start by replacing the lattice indices appearing in
$\Phi$ by orthonormal indices. This defines a new homogenous
coordinate $\wh{\Phi}$ on the $SO(22,6)$ coset. In order to
keep the metric moduli inside $\wh{\Phi}$ we do not
use the 6--bein $E^{i}_{\;\nu}$ but only its
moduli--independent part $T^{i}_{\;\nu}$ to define the new
coset coordinate, namely
\[
\Phi =  \frac{1}{2} \left( \begin{array}{c}
- A_{Ij} E^{j}_{\;\nu} \\ D_{ij} E^{j}_{\;\nu} \\ E^{i}_{\;\nu} \\
\end{array} \right) =
\frac{1}{2} \left( \begin{array}{ccc}
{\cal E}_{I}^{\; M} & 0 & 0\\
0 & E_{i}^{\;\mu} & 0 \\
0 & 0 & E^{i}_{\;\mu} \\
\end{array} \right)
\left( \begin{array}{c}
- A_{M \nu} \\ D_{\mu \nu} \\ \delta^{\mu}_{\;\nu} \\
\end{array} \right)
\]
\be
= \frac{1}{2} \left( \begin{array}{ccc}
{\cal E}_{I}^{\; M} & 0 & 0 \\
0 & T_{i}^{\;\mu} & 0 \\
0 & 0 & T^{i}_{\;\mu} \\
\end{array} \right)
\left( \begin{array}{c}
- A_{M \nu} \\ S_{\mu}^{\;\rho}D_{\rho \nu} \\ S^{\mu}_{\;\nu} \\
\end{array} \right) =
\left( \begin{array}{ccc}
{\cal E}_{I}^{\; M} & 0 & 0 \\
0 & T_{i}^{\;\mu} & 0 \\
0 & 0 & T^{i}_{\;\mu} \\
\end{array} \right) \wh{\Phi}
\eq
The new coset coordinate satisfies the coset equation
\be
\wh{\Phi}^{T} \wh{H}_{22,6} \wh{\Phi} = -I_{6}
\label{Hdach}
\eq
in which the pseudo--euclidean lattice metric
$H_{22,6}$ is replaced by
\be
\wh{H}_{22,6} = \left( \begin{array}{ccc}
I_{16} & 0 & 0 \\
0 & 0 & I_{6} \\
0 & I_{6} & 0 \\
\end{array} \right)
\eq
because of the coordinate transformation. Inside the mass formula
we absorbe the transformation matrix
into the component vector ${\bf v}$
\be
\wh{\bf v}^{T} = \left(q^{I}, n^{i}, m_{j} \right)
\left( \begin{array}{ccc}
{\cal E}_{I}^{\;M} & 0 & 0 \\
0 & T_{i}^{\;\mu} & 0 \\
0 & 0 & T^{j}_{\;\nu} \\
\end{array} \right).
\eq
such that
\be
{\bf v}^{T} \phi = \wh{\bf v}^{T} \wh{\phi}.
\eq

Now we can use the block--diagonal form of the $A$, $D$ and $S$
matrices and permute the rows and columns of $\wh{\Phi}$
such that it becomes block--diagonal
\be
\wh{\Phi} \rightarrow
\bigoplus_{j} \wh{\phi}^{(j)} \oplus \wh{\phi}^{(-1)}
\oplus \wh{\phi}^{(+1)}
\eq
with
\be
\wh{\phi}^{(j)} = \frac{1}{2} \left( \begin{array}{c}
-A^{(j)} \\ S^{(j)} D^{(j)} \\ (S^{(j)})^{T,-1} \\
\end{array} \right) \in M(2m_{j} + 2 n_{j}, 2 n_{j},{\bf R})
\eq
and a similar expression for $\wh{\phi}^{(-1)} \in M(p+r,r,{\bf R})$
and $\wh{\phi}^{(+1)} \in M(q+s,s,{\bf R})$.

In the coset equation (\ref{Hdach})
this permutation results in replacing
\be
\wh{H}_{22,6} \rightarrow
\bigoplus_{j} \wh{H}_{2m_{j} + 2 n_{j}, 2 n_{j}}
\oplus \wh{H}_{p+r,r}
\oplus \wh{H}_{q+s,s},
\eq
which makes manifest the factorization into factors
corresponding to the distinct eigenvalues of the twist.

Again we can keep the mass formula form--invariant by absorbing
this permutation into the components $\wh{\bf v}$. Note, however,
that the underlying lattice will not have a corresponding
decomposition, because a generic lattice vector has nonvanishing
projections onto more than one eigenspace.
Therefore the mass formula will not factorize for all states,
but only for those having quantum numbers which live in only one
of the eigenspaces.

Before we proceed to discuss the irreducible factors
in the decomposition, let us recall from \cite{DHVW2} that
$N=1$ space time supersymmetry requires $\theta$ to
have no eigenvalue $+1$ and the eigenvalue $-1$ can only
have multiplicities 0 or 2. Therefore $s=0$ and either
$r=0$ or $r=2$.
This implies that $\wh{\Phi}^{(+1)}$ does not appear.
More generally, if some eigenvalue does only appear in $\theta$
($\theta'$) but not in $\theta'$ ($\theta$) this gives
rise to vanishing rows in the Wilson line
matrix  and therefore also in the rearranged $\wh{\Phi}$.
They correspond to directions of the Narain lattice which possess
no deformations.

\subsubsection{Real eigenvalues and the $SO(p+2,2)$ coset
\label{SOpP22} }

We will now study the subspace which corresponds to the real
twist eigenvalue $-1$ and is parametrized by $\wh{\phi}^{(-1)}$. For
the later discussion of modular symmetries it is convenient
to go one step back and to introduce a lattice basis in this
sector. More precisely (see \cite{CLM})
we can find a sublattice $\Gamma_{p+2,2}$
of the Narain lattice $\Gamma_{22,6}$ on which the
twist $\Theta$ acts as $-I_{p+4}$. Note, however, that this
lattice is only a sublattice but not a direct factor, which
means that there is no decomposition
$\Gamma_{22,6} = \Gamma_{p+2,2} \oplus \cdots$, but only
a sublattice $\Gamma_{p+2,2} \oplus \cdots \subset \Gamma_{22,6}$.
We will now consider those states
that have non--vanishing quantum numbers lying in $\Gamma_{p+2,2}$
only.

The moduli dependent part of the mass formula for such states now
takes the form
\be
{\bf p}_{R}^{2} = {\bf v}^{T} \phi \phi^{T} {\bf v}.
\eq
with quantum numbers
\be
{\bf v}^{T} = \left(q^{1},\ldots,q^{p},
n^{1},n^{2},m_{1},m_{2}
\right)
\eq
which are lattice coordinates for $\Gamma_{p+2,2}$.
For vanishing Wilson lines $\Gamma_{p+2,2}$ factorizes as
\be
\Gamma_{p+2,2} = \Gamma_{p} \oplus \Gamma_{2,2}
\eq
where $\Gamma_{p}$ is a sublattice but not a factor of
the $E_{8} \otimes E_{8}$ or $SO(32)$ lattice $\Gamma_{16}$.
$\Gamma_{2,2}$ denotes the momentum/winding lattice corresponding
to a two--dimensional sublattice $\Lambda^{(2)}$ of the
compactification lattice $\Lambda$, which we assume
to have a decomposition $\Lambda = \Lambda^{(2)} \oplus
\Lambda^{(4)}$, as this happens for example for the ${\bf Z}_{4}$
twist. If Wilson lines are switched on, $\Gamma_{p+2,2}$ does
not factorize any more but can still be described in terms of
$\Gamma_{p}$ and $\Gamma_{2,2}$.

The matrix $\phi = \phi^{(-1)}$
\be
\phi = \frac{1}{2} \left( \begin{array}{c}
- A_{Ii}E^{i}_{\;\nu} \\
D_{ik}E^{k}_{\;\nu} \\
E^{i}_{\;\nu}\\
\end{array} \right) \in M(p+2,2,{\bf R})
\eq
is the analog of $\Phi$, because it
satisfies the constraint equation of a $SO(p+2,2)$ coset
with the standard metric replaced by the lattice metric
$H_{p+2,2}$ of ${\Gamma}_{p+2,2}$
\be
\phi^{T} H_{p+2,2} \, \phi = -I_{2}
\eq
where
\be
H_{p+2,2} = \left( \begin{array}{ccc}
C_{(p)}^{-1} & 0 & 0 \\
0 & 0 & I_{2} \\
0 & I_{2}& 0 \\
\end{array} \right)
\label{SOcosetLB}
\eq
and $C_{(p)}$ is the lattice metric of $\Gamma_{p}$.

In order to introduce complex coordinates and to
make explicit the K\"ahler structure of the
$SO(p+2,2)$ coset, we will now transform equation
(\ref{SOcosetLB}) into
its standard form.
This can be done in two steps.  The first step
consists in
converting the lattice indices
$I,J,\ldots = 1,\ldots,p$
which refer to a lattice basis of $\Gamma_{p}$, into
orthonormal indices $M,N,\ldots$ as well as converting
the lattice indices $i,j,\ldots = 1,2$, which refer to
lattice bases of $\Lambda_{2}$ (as lower indices) and
of $\Lambda^{*}_{2}$ (as upper indices),
into orthonormal indices $\mu, \nu$.
This is done in a similar way to the one
discussed in the
last subsection, and one arrives at the coordinate
$\wh{\phi} = \wh{\phi}^{(-1)}$ introduced already there.

The second step for bringing equation
(\ref{SOcosetLB}) into its canonical form consists in
replacing the metric $\wh{H}_{p+2,2}$ by the standard metric
$\eta_{p+2,2}$. This is achieved by
\be
\wh{\bf v}^{T} \wh{\phi} = \wt{\bf v}^{T} \wt{\phi}
\eq
with
\be
\wt{\phi} = \left( \begin{array}{ccc}
I_{p} & 0 & 0 \\
0 & \frac{1}{\sqrt{2}} I_{2} & \frac{1}{\sqrt{2}} I_{2} \\
0 & \frac{1}{\sqrt{2}} I_{2} & - \frac{1}{\sqrt{2}} I_{2} \\
\end{array} \right) \wh{\phi} \;\;\;\mbox{      ,      }
\;\;\; \wt{\bf v}^{T} = \wh{\bf v}^{T}
\left( \begin{array}{ccc}
I_{p} & 0 & 0 \\
0 & \frac{1}{\sqrt{2}} I_{2} & \frac{1}{\sqrt{2}} I_{2}  \\
0 & \frac{1}{\sqrt{2}} I_{2} & - \frac{1}{\sqrt{2}} I_{2} \\
\end{array} \right)
\eq
In terms of the real moduli we have that
\be
\wt{\phi} = \frac{1}{2} \left( \begin{array}{c}
- A \\ \frac{1}{\sqrt{2}} (SD + S^{T,-1}) \\
\frac{1}{\sqrt{2}} (SD - S^{T,-1}) \\
\end{array} \right)
\eq
with $A = A^{(-1)}$, etc.

The new homogenous coset coordinate $\wt{\phi}$
satisfies the standard coset relation
\be
\wt{\phi}^{T} \eta_{p+2,2} \, \wt{\phi} = - I_{2}.
\eq
Recall that $\eta_{p+2,2}$ is the standard
metric of type $(+)^{p+2} (-)^{2}$.

We can now introduce complex coset coordinates by \cite{Gil}
\be
\wt{\phi} = \left( \begin{array}{cc}
\phi_{(1)}^{1} &  \phi_{(2)}^{1} \\
   \vdots      & \vdots  \\
\phi_{(1)}^{p+4} & \phi_{(2)}^{p+4} \\
\end{array} \right) \in M(p+4,2,{\bf R})
\longrightarrow
\phi_{c} = \left( \begin{array}{c}
\phi_{(1)}^{1} + i \,  \phi_{(2)}^{1} \\
   \vdots        \\
\phi_{(1)}^{p+4} + i \, \phi_{(2)}^{p+4} \\
\end{array} \right)
\in M(p+4,1,{\bf C})
\eq
The complex coordinate $\phi_{c}$ satisfies the
equations
\be
\phi_{c}^{+} \eta_{p+2,2}\, \phi_{c} = -2,\;\;\;
\phi_{c}^{T} \eta_{p+2,2} \, \phi_{c} = 0
\eq
and is therefore the standard complex homogenous coordinate
on the $SO(p+2,2)$ coset \cite{Gil}. Note that we can replace
$\wt{\bf v}^{T} {\tilde \phi}$ by $\wt{\bf v}^{T} \phi_{c}$
in the mass formula, because
\be
{\bf p}^2_R
= \wt{\bf v}^{T} {\tilde \phi} {\tilde \phi}^{T} \wt{\bf v} =
\wt{\bf v}^{T} \phi_{c} \phi^{+}_{c} \wt{\bf v} =
|\wt{\bf v}^{T} \phi_{c}|^{2}
\eq
Thus the moduli dependent part of the mass is proportional
to the norm squared of a complex number.

Our next step towards the derivation of the mass formula is
to solve the complex constraint equations
in terms of unconstrained complex coordinates and to make explicit
the K\"ahler structure of the moduli space and  the K\"ahler
potential.  This procedure is well known both in the mathematical
\cite{Gil} and in the physics \cite{FerKouLuZw}
literature. We will use first the
physicists approach and then explain the relation to
the results in the mathematical literature.

\subsubsection*{A bounded realization}

One way of solving the constraint equations is to first extract
a scale factor $\sqrt{Y}$ from the coordinates, where $Y$
is a positive, real analytic functions of the moduli.
It turns out that this function is closely related to the
K\"ahler potential \cite{FerKouLuZw}. In terms of the new variables
$y \in {\bf C}^{p+4}$
\be
y = \sqrt{Y} \phi_{c}
\eq
the constraints read
\be
\sum_{i=1}^{p+2} |y_{i}|^{2} - |y_{p+3}|^{2} - |y_{p+4}|^{2}
= - 2 Y
\label{sumybary}
\eq
and
\be
\sum_{i=1}^{p+2} y_{i}^{2} - y_{p+3}^{2} - y_{p+4}^{2} = 0
\label{sumyy}
\eq
One can now take the first $p+2$ variables as the independent
ones and express the other two in terms of them \cite{FerKouLuZw}
\be
y_{i} = z_{i},\;i=1,\ldots,p+2,\;\;\;
y_{p+3} = \frac{1}{2} \left(
1 + \sum_{i=1}^{p+2} z_{i}^{2} \right),\;\;\;
y_{p+4} = \frac{i}{2} \left(
1 - \sum_{i=1}^{p+2} z_{i}^{2} \right)
\eq
This solves equation (\ref{sumyy}).
Defining ${\bf z} = (z_{i}) \in {\bf C}^{p+2}$
we see that equation (\ref{sumybary}) yields that
\be
Y = Y_{b} = \frac{1}{4} \left( 1 + (\ov{\bf z}^{T} \ov{\bf z})
({\bf z}^{T} {\bf z})
- 2 \ov{\bf z}^{T}  {\bf z}
\right).
\label{yb}
\eq
Solution (\ref{yb}) is $SO(p+2)$ symmetric.

The domain defined by $Y>0$ has two connected components,
which is readily seen from
$|{\bf z}^{T} {\bf z}| = 1 \Rightarrow Y= 0$. Choosing
$|{\bf z}^{T} {\bf z}| < 1$ defines a bounded open domain
in ${\bf C}^{p+2}$, called a complex polydisc,
\be
PD_{p+2} = \{ {\bf z} \in {\bf C}^{p+2} |\; |{\bf z}^{T} {\bf z}| <1
\mbox{   and   }
1 + (\ov{\bf z}^{T} \ov{\bf z})
({\bf z}^{T} {\bf z})
- 2 \ov{\bf z}^{T}  {\bf z}
> 0 \}
\eq
which provides the standard bounded realization of the
$SO(p+2,2)$ coset \cite{Gil}. The domain possesses a K\"ahler
metric, with K\"ahler potential \cite{CalVes}
\be
K = - \log Y_{b}.
\eq


\subsubsection*{An unbounded realization}

Another useful parametrization is defined by taking ${\bf y}=$
$(y_{1},\ldots,,y_{p+1},y_{p+3})$ as independent variables.
Setting \cite{FerKouLuZw,CLM}
\be
y_{p+2} = -i \left( 1 - \frac{1}{4} \left( -\sum_{i=1}^{p+1} y_{i}^{2}
   + y_{p+3}^{2} \right) \right),\;\;\;
y_{p+4} = i \left( 1 + \frac{1}{4} \left( -\sum_{i=1}^{p+1} y_{i}^{2}
   + y_{p+3}^{2} \right) \right)
\eq
solves the rescaled constraint equations with
\be
Y = Y_{u} = \frac{1}{4} \left( (y_{p+3} + \ov{y_{p+3}} )^{2} -
\sum_{i=1}^{p+1}  (y_{i} + \ov{y_{i}} )^{2} \right).
\eq
This solution is $SO(p+1,1)$ symmetric.

Again, $Y>0$ has two connected components, because
$y_{p+3} + \ov{y}_{p+3} \rightarrow 0$ implies $Y \rightarrow 0$. Taking for
definiteness $y_{p+3} + \ov{y}_{p+3} > 0$ we get the unbounded
open domain
\be
L_{+}^{p+1,1} + i\; {\bf R}^{p+2} := \{
{\bf y} \in {\bf C}^{p+2} | \;
(y_{p+3} + \ov{y}_{p+3} )^{2} - \sum_{i=1}^{p+1}
(y_{i} + \ov{y}_{i} )^{2} >0,\;
y_{p+3} + \ov{y}_{p+3} > 0 \}
\eq
which differs by a factor of $i$ from the one used in
the mathematical literature \cite{Gil}. Note that the
imaginary part of ${\bf y}$ is unconstrained whereas the real
part lives in the forward light cone of a $p+2$ dimensional
Minkowski space.

Using the holomorphic transformation
between the two parametrizations given in \cite{FerKouPorZwi}
we know that
\be
K = - \log Y_{u}
\eq
also is a K\"ahler potential.

\subsubsection*{Another unbounded realization}

For applications one prefers to rearrange the complex moduli
${\bf y}$ in terms of a $T$ and an $U$ modulus
and additional complex Wilson line moduli $B_{k}, C_{k}$.
Then, for vanishing Wilson lines, one is left with an $SO(2,2)$ coset
which factorizes into two $SU(1,1)$ cosets parametrised by the $T$ and
the $U$ modulus.
To do this one has to set (generalizing the treatment of
$SO(4,2)$ cosets \cite{FerKouLuZw,CLM})
\be
T = y_{p+1} + y_{p+3},\;\;\;
2U = y_{p+3} - y_{p+1},\;\;\;
B_{k} = y_{2k-1} - i y_{2k},\;\;\;
C_{k} = y_{2k-1} + i y_{2k}
\eq
with $k=1,\ldots, r$.
If $p$ is even, then $r = \frac{p}{2}$.
If $p$ is odd, $r = \frac{p-1}{2}$ and then
there is one additional unpaired complex coordinate
\be
A = y_{p}
\eq
In terms of the new moduli, the $Y_{u}$ function now reads
\be
Y_{u} = \frac{1}{2} (T+\ov{T}) (U+\ov{U})
- \frac{1}{4} \sum_{k=1}^{r}
(B_{k}+\ov{C_{k}}) (C_{k}+\ov{B_{k}})
- \frac{1}{4}(A + \ov{A})^{2}
\eq
For completeness, let us display how the $y_{i}$ look in terms
of the new moduli
\beqa
y_{1} &=& \frac{1}{2} (B_{1} + C_{1}),\;\;\;
y_{2} =  \frac{i}{2} (B_{1} - C_{1}), \ldots,
\mbox{   and  }
y_{p} = A,\mbox{  if $p$ is odd}\nonumber\\
y_{p+1} &=& \frac{1}{2}(T-2U),\;\;\;
y_{p+2} = -i \left( 1 - \frac{1}{4}(2TU - \sum_{k} B_{k} C_{k}
- A^{2}) \right) \nonumber\\
y_{p+3} &=& \frac{1}{2}(T+2U),\;\;\;
y_{p+4} = i \left( 1 + \frac{1}{4} (2 TU - \sum_{k} B_{k}C_{k}
 -  A^{2}) \right)
\label{y1p4}
\eeqa

Let us now comment on several special cases. First, by setting
all of the Wilson moduli to zero, $B_{k} = C_{k} = A =0$,
one obtains the K\"ahler potential for
a $SO(2,2)$ coset, which factorizes into two $SU(1,1)$ cosets
parametrised by the $U$ and the $T$ modulus, respectively.
Again, $Y > 0$ has two connected components, and
$U$ and $T$ have been defined in such a way, that the condition
$y_{p+3} + \ov{y}_{p+3} > 0$ implies that they both have
a positive real part, as usual.

Next, by inspection of the K\"ahler potential, one sees that,
for a fixed value of $k$,
$T,U,B_{k},C_{k}$ parametrize
a subspace which is a $SO(4,2)$ coset. Likewise
$T,U,A$ parametrize a $SO(3,2)$ coset. Note also that by setting
$B_{k} = C_{k}$ one can truncate a $SO(4,2)$ coset to a
$SO(3,2)$ coset. Moreover, for even $p=2m$ we can
eliminate half of the moduli by setting
$U + \ov{U} = r_{1}$ and
$(\ov{B}_{k} + C_{k})(B_{k} + \ov{C}_{k})$
$ = r_{2}\ov{A}_{k} A_{k}$,
with real positive constants $r_{1,2}$ and thus
truncate the $SO(p+2,2)$ coset to a $SU(m+1,1)$ coset. This is
obvious from the fact that $Y$ then takes the form
\be
Y = \frac{1}{2} r_{1}(T + \ov{T}) - \frac{1}{4} r_{2}
\sum_{k} \ov{A}_{k} A_{k}
\eq
which leads to the K\"ahler potential of a $SU(m+1,1)$ coset.

\subsubsection*{The mass formula}

Having found various parametrizations $y = \sqrt{Y} \phi_{c}$
with $y =y({\bf z})$, $y({\bf y})$ or
$y = y(T,U,B_{k},C_{K},A)$ of the coset, we can substitute
each of them into the mass formula with the result that
\be
{\bf p}_{R}^{2} = \frac{ | \wt{\bf v}^{T} y |^{2} }{Y}.
\label{pr2}
\eq
Therefore the mass squared can be written as the ratio of
the square of the norm of
a chiral mass ${\cal M} =\wt{\bf v}^{T} y$, which is
a holomorphic function of the moduli and contains the dependence
on the quantum numbers, and of a real analytic positive function
$Y$, which is related to a K\"ahler potential by $K = -\log Y$.

Let us recall that this expression is invariant under
the group of target space modular transformations. To be
precise we will  consider here the modular group of
the sublattice $\Gamma_{p+2,2}$ only. The connection of this group
with the full modular group was discussed in the extended version
of our paper \cite{CLM}. A general lattice vector
${\bf P} \in \Gamma_{p+2,2}$ can be decomposed as
\be
{\bf P} = v^{A} {\cal E}_{A}^{\; M} (m)  e_{M}, \;\;\;
A,M = 1,\ldots,p+4.
\eq
Here ${\bf v} = (v^{A})$ denotes the set of all components
with respect to a lattice basis of $\Gamma_{p+2,2}$ and
$e_{M}$ an orthonormal basis. The $p+4$ bein
${\cal E}_{A}^{\; M} (m)$
connecting the two bases
contains the dependence on the moduli $m=(T,U,B_{k},C_{k},A)$.
Note that
in \cite{CLM} the full $p+4$ bein $({\cal E}_{A}^{\;M})$
was denoted by $\phi$,
a symbol that we now use for the right moving part
$({\cal E}_{A}^{\;i})$ of it.
Note also that we discussed the full Narain lattice there,
but these considerations also apply to sublattices.
Now a modular transformation consists of acting
with a matrix $\Omega^{-1} \in SO(p+2,2,{\bf Z})$ on
the quantum numbers,
\be
{\bf v} \rightarrow \Omega^{-1} {\bf v},
\eq
while simultanously transforming the moduli $m \rightarrow m'$
such that ${\bf P}$ is invariant \cite{Spa,CLM}.

As we discussed in some detail in \cite{CLM}
the modular group can be
interpreted as a discrete subgroup of the deformation group
$SO(p+2,2)$ and therefore, the transformation of the moduli
is given by the left action of this discrete subgroup
on the $SO(p+2,2)$ coset. Starting from this observation
we then showed how the concrete transformation laws of the
moduli can be worked out.

We can now combine this with the results of this section
to deduce the general form of the transformation of
$Y$ and ${\cal M}$. First we know that $Y$ only depends
on the moduli, but not on the quantum numbers, and that it is
related to the K\"ahler potential by $K = -\log Y$.
Since the moduli are holomorphic coset coordinates,
and modular transformations result from the left action
of $SO(p+2,2)$ on its coset, which is a symmetric
K\"ahler manifold, $K$ must transform by a
K\"ahler transformation
\be
K \rightarrow K + F + \ov{F},
\eq
where $F$ is a holomorphic function of the moduli.
Therefore $Y \rightarrow  e^{-F-\ov{F}} Y$.
But since by construction ${\bf P} = ({\bf p}_{L};{\bf p}_{R})$ and
therefore ${\bf p}_{R}^{2}$ are invariant, this implies
\be
{\cal M} \rightarrow e^{-F} {\cal M}
\eq

\subsubsection*{Examples: The $SO(4,2)$, $SO(3,2)$ and
$SO(2,2)$ mass formulae}

Let us illustrate the formalism with a concrete example, namely
a 2--torus with 2 two--component Wilson lines turned on.
This
leads to a $SO(4,2)$ coset. The reference compactification
lattice is chosen to be the $A_{1} \oplus A_{1}$ root
lattice. We also have to chose a two--dimensional
sublattice of the $E_{8} \oplus E_{8}$ lattice and, again, we
take it to be $A_{1} \oplus A_{1}$. Let us work out, step by step,
the transformations of the quantum numbers ${\bf v}$.
To implement the first step ${\bf v}$ $\rightarrow$ $\wh{\bf v}$,
we need the matrices
\be
({\cal E}_{I}^{\;M}) = (T_{i}^{\;\mu}) =
\left( \begin{array}{cc}
\sqrt{2} & 0 \\ 0 & \sqrt{2} \\
\end{array} \right),\;\;\;
(T^{i}_{\;\mu}) =
\left( \begin{array}{cc}
\frac{1}{\sqrt{2}} & 0 \\ 0 & \frac{1}{\sqrt{2}} \\
\end{array} \right)
\eq
in order to convert the lattice basis of our reference
lattice to an orthonormal one (with respect to the
Euclidean scalar product)
\be
\wh{\bf v}^T = \left( \wh{q}_{1}, \wh{q}_{2},\wh{n}_{1}, \wh{n}_{2},
\wh{m}_{1}, \wh{m}_{2} \right) =
\left( \sqrt{2} q_{1},\sqrt{2} q_{2},\sqrt{2} n_{1},\sqrt{2} n_{2},
\frac{1}{\sqrt{2}} m_{1},\frac{1}{\sqrt{2}} m_{1} \right)
\label{vhat}
\eq
Note that we have written all of the indices as lower ones
for simplicity.
In a second step, we have to switch to a basis which is
also orthonormal with respect to the pseudo--euclidean
metric. This gives
\be
\wt{\bf v}^T=
\left( \wt{q}_{1}, \wt{q}_{2},\wt{n}_{1}, \wt{n}_{2},
\wt{m}_{1}, \wt{m}_{2} \right) =
\left(\sqrt{2} q_{1},\sqrt{2} q_{2}, n_{1} + \frac{1}{2} m_{1},
n_{2} + \frac{1}{2} m_{2}, n_{1} - \frac{1}{2} m_{1},
n_{2} - \frac{1}{2} m_{2} \right)
\eq
Next, let us use the solution (\ref{y1p4})
of the coset equations for the case
$p=2$ with $U,T,B,C$ as independent variables.
Then the chiral mass is given by
\beqa
{\cal M} &=& \sum_{i=1}^{6} \wt{v}_{i} y_{i} =
-i \left(
m_{2} -i m_{1} U + i n_{1} T -n_{2} (TU - \frac{1}{2} BC)
\right. \nonumber\\
&+& \frac{i}{\sqrt{2}} q_{1} (B+C) - \frac{1}{\sqrt{2}} q_{2} (B-C)
\left. \right)
\label{massutbc}
\eeqa
Using the mass formula (\ref{pr2}),
with the K\"ahler potential for $p=2$ inserted in it, then
gives the following mass formula for an $SO(4,2)$ coset
\be
{\bf p}_{R}^{2} = \frac{
| m_{2} -i m_{1} U + i n_{1} T -n_{2} (TU - \frac{1}{2} BC)
+ \frac{i}{\sqrt{2}} q_{1} (B+C)
- \frac{1}{\sqrt{2}} q_{2} (B-C) |^{2}}
{\frac{1}{2} (T + \ov{T})(U + \ov{U})
- \frac{1}{4} (B + \ov{C})(\ov{B} + C)}
\label{p42utbc}
\eq


The mass formula for a $SO(3,2)$-coset
can now be gotten from (\ref{massutbc})
in a straightforward way.
Namely, this time we have to choose a one--dimensional sublattice
of $\Gamma_{16}$ which we take to be the first of the two
$A_{1}$ lattices
of the former example.
Now, the two real Wilson lines will have one component only.
Setting the second components to zero $A_{2i} = 0$,
$i=1,2$ implies $B=C$. Note that this is a consistent
truncation, because $B-C$ is the coefficent of $q_2$ in (\ref{p42utbc}).
The mass formula then reads
\beqa
i \, {\cal M} = m_2 - i m_1 U + i  n_1 T + n_2 ( -U T + \frac{1}{2} \, B^2)
+  i\sqrt{2}\,q_1 \,B
\label{mass32}\eeqa
The mass formula for a $SO(2,2)$-coset is obtained by setting $B=C=0$,
yielding \cite{FerKouLuZw}
\beqa
i \, {\cal M} = m_2 - i m_1 U + i  n_1 T - n_2 U T
\label{mass22}
\eeqa

\subsubsection{Complex eigenvalues and the $SU(m+1,1)$ coset
\label{SUmP11} }

We will now discuss the moduli subspace
corresponding to complex eigenvalues.  We will, however, restrict
ourselves
to the case of a non--degenerate
rightmoving complex eigenvalue, $n_{i}= 1$. In geometrical terms this
means that we again focus on a two torus with Wilson line moduli switched
on,  but
this time with the twist on this subsector acting as a rotation and not as
a reflection.
As in the case of the $SO(p+2,2)$ coset, one can step by step repeat the
procedure given there for introducing a new homogenous coset coordinate
${\tilde \phi}$ satisfying
\be
\wt{\phi}^{T} \eta_{2m +2, 2} {\tilde \phi} = -I_{2}.
\label{SUcosetReal}
\eq
In this case, however, one has to take into account that only
half of the components of ${\tilde \phi}$ are independent.  This
can be done in the following way.
After a suitable reordering of the components
\be \phi' = P \wt{\phi},\;\;\; {\bf v}'^{T} = \wt{\bf v}^T P^{-1}
\eq
via
\be
P = \left( \begin{array}{cccccc}
I_{m} & 0 & 0 & 0 & 0 & 0 \\
0 & 0 & I_{2} & 0 & 0 & 0 \\
0 & 0 & 0 & 0 & I_{2} & 0 \\
0 & I_{m}& 0 & 0 & 0 & 0 \\
0 & 0 & 0 & I_{2} & 0 & 0 \\
0 & 0 & 0 & 0 & 0 & I_{2} \\
\end{array} \right)
\eq
one finds that $\phi'$ has the form
\be
\phi' =
\left( \begin{array}{cc}
                    \phi'_{1} &\phi'_{2} \\
                   -\phi'_{2} & \phi'_{1} \\
              \end{array} \right)
\eq
This effectively leads to the replacement
$\eta_{2m+2,2} \rightarrow \eta_{m+1,1} \oplus \eta_{m+1,1}$
in (\ref{SUcosetReal}).
Then, by introducing the complex coordinate
\be
\phi_{c} = \phi'_{1} + i\, \phi'_{2} \in
M(m + 2,1,{\bf C})
\eq
one finds that $\phi_{c}$ satisfies the relation
\be
\phi_{c}^{+} \eta_{m+1,1}
\phi_{c}  = - 1
\label{SUcosetCompl}
\eq
characterizing the coset $SU(m + 1,1)$/
$(SU(m+1) \otimes U(1))$ \cite{Gil}.
Finally, one also has to introduce complex quantum numbers
by
\be
v_{(c)}^{i} = v'_{i} + i v'_{m+2+i},\;\;\;
i = 1,\ldots, m+2
\eq
in order to be able to rewrite the mass formula as
\be
{\bf p}_{R}^{2} = {\bf v}'^{T} \phi' \phi'^{T} {\bf v}'
= |{\bf v}_{(c)}^{T} \phi_{c}|^{2}
\eq

\subsubsection*{A bounded realization}

Since the homogenous coset coordinate $\phi_{c}$ is
again a complex vector (and not a matrix) we can proceed
as in the last subsection. First we introduce rescaled
coordinates $y_{i} = \sqrt{Y} \phi_{c}^{i}$,
$i=1,\ldots, m+2$ and obtain the equation
\be
\sum_{i=1}^{m+1} \ov{y}_{i} y_{i} - \ov{y}_{m+2} y_{m+2}
= - Y.
\eq
Let us introduce unconstrained coordinates ${\bf z} = (z_{i},z)$,
$i=1,\ldots, m$ by
\be
y_{i} = z_{i},\;i=1,\ldots,m,\;\;\;
y_{m+1} = z,\;\;\;
y_{m+2} = 1.
\label{SUyz}
\eq
This solves (\ref{SUcosetCompl}) with
\be
Y = Y_{b} = 1 - \ov{\bf z}^{T} {\bf z}.
\eq
Note that $Y>0$ implies that $1 - \ov{\bf z}^{T} {\bf z} > 0$.
Therefore we have found a realization of the
coset by the bounded open domain \cite{Gil}
\be
D_{m+1} =
\{ {\bf z} \in {\bf C}^{m+1} | \; 1 - \ov{\bf z}^{T} {\bf z} > 0
\}
\eq
and the standard K\"ahler potential for this realization is \cite{CalVes}
\be
K = - \log Y_{b}.
\eq

Comparing to standard projective coordinates (which are also
known to provide a solution to the constraints \cite{Gil})
\be
Z_{i} := \frac{\phi_{c}^{i}}{\phi_{c}^{m+2}}
=\frac{y_{i}}{y_{m+2}} = z_{i},\;i=1,\ldots,m,\;\;\;
Z_{m+1} := \frac{\phi_{c}^{m+1}}{\phi_{c}^{m+2}}
=\frac{y_{m+1}}{y_{m+2}} = z
\eq
we see that those are identical to the $z_{i},z$.

\subsubsection*{An unbounded realization}

Again, we would like to have another, unbounded representation
in terms of a $T$ modulus parametrizing a $SU(1,1)$ coset
and additional
complex Wilson line moduli
$A_{i}, i =1,\ldots,m$.
The $y_i = \sqrt{Y} \phi^i_c$ are now given by
\be
y_{i} = A_{i},\; i =1,\ldots,m,\;\;\;
y_{m+1} = \frac{1}{2}(T-1),\;\;\;
y_{m+2} = \frac{1}{2}(T+1).
\label{yAT}
\eq
This solves (\ref{SUcosetCompl}) with
\be
Y = Y_{u} = \frac{1}{2} (T + \ov{T}) -
\sum_{i=1}^{m} \ov{A}_{i} A_{i}
\eq
Clearly we have found an unbounded realization, because
the imaginary part of $T$, for example, is not constrained at all,
whereas the real part can be arbitrarily large.

There is a second way of connecting the bounded and the
unbounded realization. Namely one can start with the
bounded realization and then introduce $T$ and $A_{i}$
by the map
\be
z \rightarrow T = \frac{1-z}{1+z},\;\;\;
z_{i} \rightarrow A_{i} := \frac{z_{i}}{1 + z}
\label{DisktoHS}
\eq
For vanishing Wilson lines this reduces to the standard
map from the open unit disc onto the right half plane
\be
D_{1} = \{ z \in {\bf C} | \, |z| < 1 \} \rightarrow
H = \{ T \in {\bf C} | T + \ov{T} > 0  \}.
\eq
Substituting the transformation (\ref{DisktoHS})
into $K' = -\log Y_{u}$ yields
\be
K' = -\log \left(
1 - \ov{z}z - \sum_{i=1}^{m} \ov{z}_{i} z_{i}
\right) + \log \left( |1+z|^{2} \right)
\eq
which differs from $K = - \log Y_{b}$ by a K\"ahler
transformation.
Note that, when relating $z,z_i$ and $T,A_i$ by equating
(\ref{yAT}) and (\ref{SUyz}), this is equivalent to
relating them by (\ref{DisktoHS}) modulo this
K\"ahler transformation.

\subsubsection*{The mass formula}

Again, the mass formula is given by the ratio of the square of the chiral
mass and the function $Y$
\be
{\bf p}_{R}^{2} = \frac{|{\bf v}_{(c)}^{T} y|^{2}}{Y}
\eq
where $y$ and $Y$ are functions of the complex moduli
$T, A_{i}, i=1,\ldots,m$.

Again this expression is invariant under the group of modular
transformations. The relevant group $SU(m+1,1,{\bf Z})$
is a subgroup of the group $SO(p+2,2,{\bf Z})$,
namely the normalizer with respect to the ${\bf Z}_{N}$
group generated by the twist \cite{Spa}.
Recall that these groups explictly depend on
the reference lattice, so we did not make this explicit
in our notation. For the same reasons as discussed before
in the case of $SO(p+2,2)$ cosets, $Y$ and the chiral
mass ${\cal M}$ transform as
\be
Y \rightarrow e^{ -F -\ov{F} } Y,\;\;\;
{\cal M} \rightarrow e^{-F} {\cal M},
\eq
where $F$ is a holomorphic function of the moduli.

\subsubsection*{Example: The $SU(2,1)$ and $SU(1,1)$
mass formulae}

We will again illustrate the general procedure with a
concrete example, a two dimensional ${\bf Z}_{3}$ orbifold
with one independent two--component Wilson line.
This time we take both the reference compactification lattice
and the sublattice of $\Gamma_{16}$ to be $A_{2}$ root lattices.
Both the internal and the gauge twist are taken to be
the $A_{2}$ Coxeter twist.

Now the transformation matrices from the lattice to the
orthonormal basis are given by
\be
({\cal E}_{I}^{\;M}) = (T_{i}^{\;\mu}) =
\left( \begin{array}{cc}
\sqrt{2} & 0 \\ - \frac{1}{\sqrt{2}} & \sqrt{\frac{3}{2}} \\
\end{array} \right),\;\;\;
(E^{i}_{\;\mu}) =
\left( \begin{array}{cc}
\frac{1}{\sqrt{2}} & \frac{1}{6} \sqrt{6} \\
0 & \frac{1}{3} \sqrt{6} \\
\end{array} \right)
\eq
The transformed quantum numbers are given by
\beqa
\wh{\bf v}^T &=&
\left( \wh{q}_{1}, \wh{q}_{2},\wh{n}_{1}, \wh{n}_{2},
\wh{m}_{1}, \wh{m}_{2} \right) =
\left( \sqrt{2} q_{1} - \frac{1}{\sqrt{2}} q_{2},
\sqrt{\frac{3}{2}} q_{2},
\sqrt{2} n_{1} - \frac{1}{\sqrt{2}} n_{2},
\sqrt{\frac{3}{2}} n_{2},
\frac{1}{\sqrt{2}} m_{1} , \right. \nonumber\\
&& \frac{1}{6} \sqrt{6} m_{1} +
\frac{1}{3} \sqrt{6} m_{2} \left. \right)
\eeqa
Diagonalizing the pseudo--euclidean lattice metric
yields
\beqa
\wt{\bf v}^{T} &=&
( \sqrt{2} q_{1} - \frac{1}{\sqrt{2}} q_{2},
\sqrt{\frac{3}{2}} q_{2},
n_{1} - \frac{1}{2} n_{2} + \frac{1}{2} m_{1},
\frac{1}{2} \sqrt{3} n_{2} + \frac{1}{6} \sqrt{3} m_{1}
+ \frac{1}{3} \sqrt{3} m_{2}, \nonumber\\
&& n_{1} - \frac{1}{2} n_{2} - \frac{1}{2} m_{1},
\frac{1}{2} \sqrt{3} n_{2} - \frac{1}{6} \sqrt{3} m_{1}
- \frac{1}{3} \sqrt{3} m_{2} )
\eeqa
Next, we have to reorder the components $\wt{\bf v} \rightarrow$
${\bf v}'$ and finally complexify them, ${\bf v}' \rightarrow$
${\bf v}_{c}$. Introducing the complex quantum numbers
\[
q_{c} = \sqrt{2} q_{1} - \frac{1}{\sqrt{2}} q_{2} + i
\sqrt{\frac{3}{2}} q_{2},
\]
\[
n_{c} = n_{1} - \frac{1}{2} n_{2} + \frac{1}{2} m_{1}
+ i \frac{\sqrt{3}}{2} ( n_{2} + \frac{1}{3} m_{1} + \frac{2}{3} m_{2} )
\]
\be
m_{c} = n_{1} - \frac{1}{2} n_{2} - \frac{1}{2} m_{1}
+ i \frac{\sqrt{3}}{2} ( n_{2} - \frac{1}{3} m_{1} - \frac{2}{3} m_{2} )
\eq
yields that
\be
{\bf v}_{(c)}^T = (q_{c},n_{c},m_{c})
\eq
Therefore, the chiral mass (setting $y_{1} = y$) is given by
\be
{\cal M} = \sum_{i=1}^{3} v_{(c)}^{i} y_{i}
= q_{c} y + n_{c} \frac{1}{2} (T-1) + m_{c} \frac{1}{2} (T+1)
= q_{c} y + \frac{1}{2}(n_{c} + m_{c}) T + \frac{1}{2} (m_{c} - n_{c})
\eq
giving raise to the mass formula
\be
{\bf p}_{R}^{2} = \frac{ |
q_{c} y + \frac{1}{2}(n_{c} + m_{c}) T + \frac{1}{2} (m_{c} - n_{c})
|^{2} }{\frac{1}{2} (T + \ov{T}) - y \ov{y} }
\label{MFSUd}
\eq

We now proceed to show that one can get the
mass formula (\ref{MFSUd})
of an $SU(2,1)$ coset by a suitable truncation of the one
for the $SO(4,2)$ coset
 given in (\ref{p42utbc}). To do this truncation correctly, one has
to take two things into account.
First, the lattice $\Lambda$ must be proportional to the
$A_{2}$ root lattice in order to have the $A_{2}$ Coxeter
twist as a lattice automorphism. This freezes the
$U$ modulus to the value $U=  \frac{1}{2}(\sqrt{3} + i)$,
while $T$ is still arbitrary.
Secondly, one has to
choose inside the $E_{8} \oplus E_{8}$ lattice a
sublattice with the appropriate symmetry. Taking an $A_{2}$
sublattice amounts to setting
\be
(\wh{q}_{1}, \wh{q}_{2}) =
( \sqrt{2} q_{1} - \frac{1}{\sqrt{2}} q_{2},
\sqrt{\frac{3}{2}} q_{2} )
\eq
in (\ref{vhat}). Also note that the Wilson line moduli have to be fixed
to $B = \sqrt{3} {\cal A}$, $C=0$ \cite{CLM}.

Specializing in this way, the chiral mass given in (\ref{massutbc})
turns into
\be
{\cal M} = \frac{\sqrt{3}}{2} q_{c}
{\cal A} + \frac{1}{2} (m_{c} + n_{c})
T + \frac{\sqrt{3}}{2} (m_{c} - n_{c})
\eq
with the complex quantum numbers as defined above. The mass formula
(\ref{p42utbc}) then turns into
\be
{\bf p}_{R}^{2} =
\frac{ |\frac{\sqrt{3}}{2} q_{c} A + \frac{1}{2} (m_{c} + n_{c})
T + \frac{ \sqrt{3}}{2} (m_{c} - n_{c}) |^{2} }{ \sqrt{3}
(\frac{1}{2} (T + \ov{T}) - \frac{\sqrt{3}}{4} {\cal A} \ov{{\cal A}} )}
\label{p21at}
\eq
Equation (\ref{MFSUd}) is then indeed obtained from (\ref{p21at})
by rescaling
$T$ by a factor 1/$\sqrt{3}$ and
by setting $y=\frac{{\cal A}}{2}$.

And finally,
when switching off the complex Wilson line $\cal A$, ${\cal A}=0$,
one arrives at the wellknown mass formula for a $SU(1,1)$-coset
\beqa
{\cal M}= \frac{i}{2}\left(
  (m_c + n_c) T + \sqrt{3} (m_c - n_c) \right)
\eeqa

\section{Target space modular invariant orbits of massive untwisted states
\label{ModOrbUnt}}

\setcounter{equation}{0}

Massive
untwisted states play an important role in the context of 1-loop
corrections to gauge and gravitational couplings
\cite{Vkap}-\cite{AntNaTay}, \cite{DFKZ}-\cite{AntTay},
\cite{VafaOog,FerKouLuZw,MunLu}
in ${\bf Z}_N$-orbifold models.
  These states give
rise to moduli dependent
 threshold corrections, which are given in terms of automorphic
functions of the modular group under consideration
\cite{Louis,May1,VafaOog,FerKouLuZw}.
We will thus
focus on massive untwisted states in the following.

We will assume that the internal 6-torus factorises into $T_6 = T_2
\oplus T_4$ and that the lattice twist $\theta$ acts on the 2-torus
$T_2$ as a ${\bf Z}_2$-twist.  We will then focus on the $SO(p+2,2)$-coset
space associated with the $T_2$ and discuss its mass formula.  That is,
we will consider those massive untwisted states which have
non-vanishing quantum numbers
${\bf v}^T=(q^1,...,q^p,n^1,n^2,m_1,m_2)$
in the Narain sublattice $\Gamma_{p+2,2} \subset \Gamma_{22,6}$,
as discussed in section \ref{SOpP22}.  We will, in
addition, also allow
these massive untwisted states to carry non-vanishing quantum
numbers in an orthogonal sublattice $\Gamma_{20-p,4}$ with
$\Gamma_{p+2,2} \oplus \Gamma_{20-p,4}
\subset \Gamma_{22,6}$.

Recall that
the level matching condition for physical states in the heterotic string
reads
\beqa
{\bf p}^2_L - {\bf p}^2_R = 2(N_R + 1 - N_L) + {\bf P}^2_R
- {\bf P}^2_L
\label{level}
\eeqa
where $({\bf p}_L;{\bf p}_R) \in \Gamma_{p+2,2}$ and
$({\bf P}_L;{\bf P}_R) \in \Gamma_{20-p,4}$.
The mass formula for physical states
can then be written as
\beqa
\frac{\alpha'}{2}  M^2=
 {\bf p}^2_R + {\bf P}^2_R + 2 N_R
\eeqa
The untwisted states associated with the Narain sublattice $\Gamma_{p+2,2}$
do, on
the other hand, satisfy
\beqa
{\bf p}^2_L - {\bf p}^2_R = 2n^T m + q^T {\cal C} q
\label{int}
\eeqa
where ${\cal C}$ denotes the lattice metric of the sublattice $\Gamma_p$
of the $E_8 \oplus E_8$ lattice $\Gamma_{16}$, as explained in
section \ref{SOpP22}.
Then, equating (\ref{level}) and (\ref{int}) yields
\beqa
{\bf p}^2_L - {\bf p}^2_R
= 2(N_R + \frac{1}{2} {\bf P}^2_R + 1 - N_L - \frac{1}{2} {\bf P}^2_L)
= 2n^T m + q^T {\cal C} q
\label{compat}
\eeqa

We have shown in section \ref{SOpP22} that for a $SO(p+2,2)$-coset
${\bf p}^2_R$ can be written as
\beqa
{\bf p}^2_R =\frac{|{\cal M}|^2}{Y}
\eeqa
where $Y$ is related to the K\"{a}hler potential by $K=-\log Y$, and
where ${\cal M}$ is a holomorphic function of the complex coordinates
for the $SO(p+2,2)$-coset.  {From} the study of a few examples
in the past \cite{May1,VafaOog,FerKouLuZw},
it is expected that a prominent role in threshold corrections
to gauge and gravitational couplings is going to be played by those
massive untwisted
states which satisfy $2N_R + {\bf P}^2_R=0$, so that
\beqa
\frac{\alpha'}{2} M^2 = {\bf p}^2_R =  \frac{|{\cal M}|^2}{Y}
\label{massso}
\eeqa
Thus, we will in the following
only consider untwisted states for which $2N_R + {\bf P}^2_R=0$.
Then, (\ref{compat}) turns into
\beqa
{\bf p}^2_L - \frac{\alpha'}{2} M^2
= 2(1 - N_L - \frac{1}{2} {\bf P}^2_L) = 2n^T m + q^T {\cal C} q
\label{pdiff}
\eeqa
Note that, for any given $2N_L + {\bf P}^2_L$,
the orbit
$ 2n^T m + q^T {\cal C} q=2(1 - N_L - \frac{1}{2} {\bf P}^2_L)$
is invariant under modular $SO(p+2,2,{\bf Z})$-transformations.
Let us now consider the following cases:

a) first, consider untwisted states for which $N_L=0, {\bf P}^2_L=0$.
This defines an
orbit for which
\beqa
2n^T m + q^T {\cal C} q =2
\label{orb0}
\eeqa
This is the orbit which is relevant for the discussion of the
stringy Higgs effect.
It can namely happen \cite{IbaLuLer} that
at certain finite
points in the fundamental region of the
moduli space certain (otherwise) massive states precisely
become massless, $M^2=0$.
Then, at these special points in moduli space one has that
${\bf p}^2_L=2$, and, hence, one has additional massless gauge bosons in
the spectrum as well as additional massless scalar fields.  These additional
massless gauge bosons enhance the gauge group of the underlying
two-dimensional ${\bf Z}_2$-orbifold.  This enhancement can give rise to an
additional $SU(2)$, as discussed in section \ref{ExGGCompS}.  Thus, it it
crucial to take into account orbit (\ref{orb0}) when discussing
1-loop corrections to gauge couplings associated with the enhanced
gauge group of the compactification sector
of the orbifold.  Similarly, orbit (\ref{orb0}) must
also be taken into account when discussing threshold corrections
to gravitational couplings.

b) now consider untwisted states for which
$2N_L+ {\bf P}^2_L=2$.  This defines
an orbit for which ${\bf p}^2_L={\bf p}^2_R$ and, hence,
\beqa
2n^T m + q^T {\cal C} q =0
\label{orb1}
\eeqa
An example is provided by setting $N_L = 0, {\bf P}^2_L =2$.
Then, take ${\bf P}^2_L=2$
to be a vector in the root lattice of the
gauge group $G'$ which is not affected by the moduli associated
with $\Gamma_{p+2,2}$.  For instance, $G'$ can be taken to be
the
$E'_8$ in the hidden sector left unbroken
when turning on Wilson lines.
Then, the orbit given in
(\ref{orb1}) is the relevant one for the discussion
of threshold corrections for this type of gauge couplings
in the presence of Wilson lines.

For untwisted states with quantum numbers $q=0$, (\ref{orb1}) obviously
reduces down to $n^T m = 0$.  This reduced orbit is not invariant under
general modular $SO(p+2,2,{\bf Z})$-transformations anymore, but it is still
invariant under modular transformations belonging to the subgroup
$SL(2,{\bf Z})_T\otimes SL(2,{\bf Z})_U \subset SO(p+2,2,{\bf Z})$.
This reduced orbit is the
one which has been discussed quite extensively
in the literature \cite{VafaOog,FerKouLuZw,May1}
in the context of $(2,2) {\bf Z}_N$-orbifold theories.  In this context,
it is well known that there are no finite
points in the fundamental region of the $(T,U)$-moduli space where
(otherwise) massive states might
become massless.  Indeed, by setting \cite{FerKouLuZw}
$n_1=r_1 s_2, n_2 = s_1 s_2 , m_1 = - r_2 s_1$ and $m_2=r_1 r_2$,
it follows that ${\bf p}^2_R = 2 \left(\frac{|r_1 + i s_1 U|^2}
{U + \bar{U}}\right)
\left(\frac{|r_2 + i s_2 T|^2}{T + \bar{T}} \right)$.
This shows that
it is only in the large radius limit, namely
at $T=\infty$ with $U$ fixed
 (and vice-versa), that states
(Kaluza-Klein states) become massless.
This is
in manifest contrast to what happens for the orbit discussed in a).

c) finally, consider untwisted states for which
$2N_L + {\bf P}^2_L \ge 4$.
Inspection of (\ref{pdiff}) shows that there aren't any points,
finite or otherwise,
in moduli space for which $M^2=0$, since then ${\bf p}^2_L < 0$, which
isn't allowed.

Hence, it is only for the orbits a) and b) that it can happen that
massive states become massless at some special points in the
fundamental region of moduli space.  Threshold corrections to
couplings should then exhibit a singular behaviour at precisely those
points in moduli space.  Thus, it appears that the interesting
physics contained in threshold corrections is associated with orbits a)
and b).  This is then why we will be focussing on orbits a) and b)
in the following.

It was shown \cite{May1,FerKouLuZw} in the context of threshold corrections
to the $E_6$ and $E'_8$
gauge couplings in $(2,2) {\bf Z}_N$-orbifold theories, that
it is of relevance to consider the quantity
 $\sum_{orbit} \log {\cal M}$.
There, the relevant orbit is the reduced orbit discussed in case b),
given by $n^T m =0 ,  q = 0$.
Indeed, when suitably regularised \cite{FerKouLuZw},
$\sum_{n^T m=0,q=0} \log {\cal M}|_{reg} $
$= \log\left( \eta^{-2}(T) \eta^{-2}(U) \right)$,
this quantity
is precisely the object appearing in these threshold corrections.
In the context of threshold corrections to
gravitational couplings,
on the other hand, one should apriori consider both the orbits a) and b).
We will thus introduce
the following quantities for cases a) and b), respectively
\beqa
\Delta_{0} &=& \frac{1}{L_0}
\sum_{2n^T m + q^T {\cal C} q=2} \log {\cal M} \nonumber\\
\Delta_{1} &=& \frac{1}{L_1}
\sum_{2n^T m + q^T {\cal C} q=0} \log {\cal M}
\label{w012}
\eeqa
It is implied in (\ref{w012})
that a regularisation procedure should exist for turning
these formal expressions into meaningful ones.  This regularisation
procedure should be compatible with the transformation property of
$\log {\cal M}$ under modular $SO(p+2,2,{\bf Z})$-transformations.
Thus, each
 of the these $\Delta_i$, $i=0,1$,  is expected to be expressed
in terms of automorphic functions of the modular group $SO(p+2,2,{\bf Z})$.
The constants $L_i$ are such so as to ensure that,
under the subgroup of $SL(2,{\bf Z})_{T,U}$-transformations,
each of the $\exp{\Delta_i}$ has a modular weight of $-1$.

Now consider the case when all Wilson lines have been turned off.
As will be discussed in detail in the next section,
$\Delta_0$ should contain a term of the form
\beqa
\Delta_0 \propto \log \;\{ (j(T) - j(U))^r \;\eta(T)^{-2}
\;\eta (U)^{-2} \} + \dots
\label{wtu0}
\eeqa
where $r$ denotes some non-zero integer and where $j(T)$ denotes the
absolute modular invariant function.  We will explain that
such a term should arise when summing over all the states lying on the
orbit $n^T m =1, q=0$.  The singular behaviour
 of this term at the points
in the fundamental region where $T=U$ reflects the fact
that (otherwise) massive
states become massless at these points in moduli space, as discussed
in section \ref{ExGGCompS}.
  The role of the
eta-terms in (\ref{wtu0}) is to ensure the correct transformation
property of $\exp{\Delta}_0$
under $SL(2,{\bf Z})_{T,U}$-transformations.\footnote{\label{foot4}
As stated
above, we have made the important assumption that the regularisation
procedure used in (\ref{w012}) respects both holomorphicity and
modular covariance.  Otherwise additional terms
transforming inhomogenously under modular transformations
might appear
on the r.h.s. of (\ref{wtu0}).  Such additional terms do not
exhibit singular behaviour in the fundamental region, and so their
appearance would not
change the singular behaviour of $\Delta_0$.}

$\Delta_1$, on the other hand, will contain a term of the form
\beqa
\Delta_1 \propto \log \;\{ \eta (T)^{-2}\; \eta (U)^{-2} \} + \dots
\label{wtu1}
\eeqa
As shown in \cite{FerKouLuZw},
this term arises when summing over all massive states
for which $ n^T m =0 , q=0$.

As discussed above, all the physically interesting information
relevant to the threshold
corrections to gravitational couplings should be contained in
$\Delta_0$ and in $\Delta_1$.
These gravitational threshold corrections should
then, for the case of vanishing Wilson moduli, be proportional to
\beqa
\Delta_0 + \Delta_1 \propto
\alpha \; \log  (j(T) - j(U))  + \beta \; \log \{\eta(T)^{-2}
\;\eta (U)^{-2} \}
\label{w}
\eeqa
where the constants $\alpha$ and $\beta$
have to be determined
from an appropriate string scattering amplitude calculation.
We will, in the next sections, study $\Delta_0$ and $\Delta_1$ in
detail.  We will, in particular, also study the effect of non-vanishing
Wilson moduli on (\ref{w}).  The last section will be devoted
to a discussion of threshold corrections to gauge and gravitational
couplings.

Note that
the analogous result to (\ref{w}) for a $SU(m+1,1)$-coset can
be obtained by an appropriate truncation of the mass formula (\ref{massso})
for a $SO(2m+2,2)$-coset, as discussed in section \ref{SUmP11}.

\section{Automorphic functions for the orbit $2 n^T m + q^T {\cal C} q = 2$
\label{AutFunNulOrb} }

\setcounter{equation}{0}

In this section we will consider massive untwisted states possessing
$N_L=0, {\bf P}^2_L=0$.  Then
the associated modular invariant
orbit is defined by
\beqa
2n^T m + q^T {\cal C} q =2
\label{orbzero}
\eeqa
It can happen that
certain  massive states, characterised by a set of quantum numbers satisfying
(\ref{orbzero}), become massless at special points in moduli space.
Then, at these special points one has that
${\bf p}^2_L = 2$, and, hence, there are some additional
massless states corresponding to massless gauge bosons.  Thus, orbit
(\ref{orbzero}) is of big relevance for the stringy Higgs-effect.

We will in the following try to construct automorphic functions
for the orbit (\ref{orbzero}).
We will, for concretenesss, first consider
a $SO(4,2)$-coset associated with an internal 2-torus $T_2$.
At the end of this section, by making use
of the truncation procedure given in
section \ref{SUmP11}, we will also discuss
automorphic functions for a $SU(2,1)$-coset.

Consider the quantity $\Delta_0$ introduced in
(\ref{w012}).  Taking into account that the Cartan matrix $\cal C$
for the Narain sublattice $\Gamma_{4;2}$ under consideration
reads ${\cal C} = 2 diag (1,1)$, yields (\ref{w012}) as
\beqa
\Delta_{0} &\propto& \sum_{n^T m + q^2 =1} \log {\cal M} \nonumber\\
&=&
\sum_{n^T m =1 \;,\;  q=0} \log {\cal M} +
\sum_{n^T m =0 \;,\;  q^2 = 1} \log {\cal M} +
\sum_{n^T m = -1 \;,\;  q^2 = 2} \log {\cal M}
+ \dots
\label{wzero}
\eeqa
where we have rewritten the sum over all the states laying on the
$SO(4,2,{\bf Z})$-invariant
orbit $n^T m + q^2 = 1$ into a sum over $SL(2,{\bf Z})_{T,U}$-invariant
orbits $n^T m =constant$.
Then, (\ref{wzero})
can be conveniently written as
\beqa
\Delta_{0}  \propto \sum_{q^2 \ge 0} \Delta_{0,q^2} =
\Delta_{0,0} + \Delta_{0,1}
+ \dots
\eeqa
where
\beqa
\Delta_{0,0} =
\sum_{n^T m =1, q=0} \log {\cal M} \;,\;\;
\Delta_{0,1} =
 \sum_{n^T m =0, q^2=1} \log {\cal M}
\label{www012}
\eeqa
Each of the $\exp{\Delta_{0,q^2}}$
 should carry a definite modular weight
under
$SL(2,{\bf Z})_{T,U}$-transformations.
The complete $\exp{\Delta}_0$ should then
have a modular weight of $-1$.

We begin by studying the case where the Wilson lines $B$ and $C$
have been switched off.  We will later generalise our results
to the case of non-vanishing Wilson lines.

As already mentioned, there is an enhancement of the gauge group
of the model at some special points in the $(T,U)$-moduli space.
The relevant untwisted states are the ones
for which $q_1=q_2=0, n^T m = 1$.  For these states
the mass formula (\ref{massutbc}) reduces to
\beqa
{\cal M} = m_2 - i m_1 U + i  n_1 T - n_2 U T
\label{masstu}
\eeqa
Note that we have redefined the chiral mass ${\cal M}$
by an irrelevant overall phase factor for convenience.
Let us first discuss the issue of enhancement of the gauge group
in the context of a two-dimensional toroidal model.  We will then, in
a second step, discuss it in the context of the two-dimensional
${\bf Z}_2$-orbifold.
At generic values for the moduli $T$ and $U$ the gauge group
of the two-dimensional torus $T_2$ under consideration is given by
$G=U(1) \otimes U(1)$.  At special points in the $(T,U)$-moduli space,
namely for i) $T=U$, ii) $T=U=1$ and iii) $T=U=\rho=e^{\frac{i\pi}{6}}$,
this generic gauge group is enhanced to
i) $G=SU(2)\otimes U(1)$, ii) $SU(2) \otimes SU(2)$ and iii)
$G=SU(3)$ \cite{IbaLuLer}, as we discussed in some detail
in section \ref{ExGGCompS}.

Let us consider case i) first. Setting $T=U$ turns (\ref{masstu}) into
\beqa
{\cal M} = m_2 + i ( n_1 - m_1) T - n_2 T^2
\label{masstt}
\eeqa
Then, the holomorphic mass (\ref{masstt}) vanishes for
the following 2 untwisted states
\beqa
m_1=n_1=\pm 1 \;\;,\;\; m_2=n_2=0
\label{st1}
\eeqa
Note that these states satisfy $n^T m =1$.  Thus, they
are precisely the 2 states corresponding to the 2 additional gauge bosons
which enhance the gauge group to $G=SU(2) \otimes U(1)$ at the points
where $T=U$.

Next, consider case ii).  Setting $T=U=1$ turns (\ref{masstt}) into
\beqa
{\cal M} = m_2 - n_2 + i (n_1 -  m_1)
\label{masstu1}
\eeqa
In addition to the 2 states (\ref{st1}) there are now 2 additional
ones
\beqa
m_2 = n_2 = \pm 1 \;\;,\;\; m_1=n_1=0
\label{st2}
\eeqa
satisfying $n^T m =1$, for which the holomorphic mass (\ref{masstu1})
vanishes.  These 2 states correspond to the 2 additional massless
gauge bosons which further enhance the gauge group from
$G=SU(2)\otimes U(1)$ to $G=SU(2) \otimes SU(2)$.

Finally, consider case iii).  Inserting $T=U=e^{\frac{i \pi}{6}}$ into
(\ref{masstt}) yields
\beqa
{\cal M} = \frac{1}{2} \left(
2 m_2 - n_2 - n_1 + m_1
 + i \sqrt{3} ( n_1 - m_1  - n_2)\right)
\label{massturo}
\eeqa
The holomorphic mass (\ref{massturo}) vanishes not only
for the states (\ref{st1}), but also for the following
4 additional states satisfying $n^T m =1$
\beqa
m_2 &=& n_2 = 1 \;:\;\;\; m_1= - 1\;,\; n_1=0 \;\; or
\; \;m_1=0 \;,\; n_1=1
\nonumber\\
m_2 &=& n_2 = - 1 \;:\;\;\; m_1=0 \;,\;  n_1= -1 \;\;
or \;\; m_1=1 \;,\; n_1=0
\label{st3}
\eeqa
These 4 additional massless states correspond again to 4 additional
massless gauge bosons which enhance the gauge group to $G=SU(3)$.
Thus, the ratio of additional massless gauge bosons for the cases
i)-iii) is given by 1:2:3.  Note that there is an infinite set of states
with dual transformed quantum numbers, whose masses vanish at
the transformed critical points.

The enhancement of the gauge group for the
${\bf Z}_2$-orbifold  occurs at the same points in moduli space as
in the toroidal case.  The associated gauge groups are, however,
smaller,
since one has to project
out all the
non-twist invariant states.  The resulting gauge groups
are then, for the cases i)-iii), given by
i)$U(1)$, ii)$U(1) \otimes U(1)$ iii) $SU(2)$ at level
$k=4$, as discussed in section \ref{ExGGCompS}.
Thus, the ratio of additional massless gauge bosons is, as in the
toroidal case, given by 1:2:3.

Now, what do the individual $\Delta_{0,q^2}$ given in (\ref{www012})
look like? First consider $\Delta_{0,0}$.  It
is defined in terms of a sum
over all the states laying on the orbit $n^T m=1, q=0$.  As shown above,
this is precisely the orbit for which some of the massive states
become massless at the special points i)-iii)
in the fundamental region of the
$(T,U)$-moduli space.  Thus, at each of these special points $\Delta_{0,0}$
has to exhibit a logarithmic singularity.  The order of this
singularity should be determined by the number of states
which become massless at these special points.  As shown above,
the ratio of additional massless states at the special points
i)-iii) is 1:2:3.  Thus, the order of the zeros of
$\exp{\Delta_{0,0}}$ should be 1:2:3 for the points i)-iii).
According to well-known theorems of modular functions \cite{Kobl}, the order
of the zeros and of the poles, together with the modular weights
(possibly leading to non-trivial multiplier systems), determine a
modular function in a unique way.  Applying this to the above case
yields that
\beqa
\exp{\Delta_{0,0}} \propto (j(T) - j(U))^r\; \eta(T)^{-2} \;\eta(U)^{-2}
\label{w00jj}
\eeqa
where $r$ denotes some non-zero integer.  $j(T)$ denotes the
absolute modular invariant function.\footnote{In (\ref{w00jj}) we
have ignored possible additional multiplicative terms of the
type $\exp f(T,U)$ with $f(T,U)$ exhibiting regular behaviour
inside the fundamental region.  Such terms might be necessary
in order to reproduce
the correct asymptotic behaviour of $\Delta_{0,0}$
as $T,U \rightarrow \infty$ \cite{WitLouLu}.\label{footreg} }
The formal definition (\ref{www012}) of $\Delta_{0,0}$ implies that
$\Delta_{0,0}$ has to carry a modular weight of $-1$ under
$SL(2,{\bf Z})_{T,U}$-transformations.  This explains the presence of the
eta-terms in
(\ref{w00jj}), which indeed ensure that $\exp{\Delta_{0,0}}$
transforms with modular weight of $-1$ under
$SL(2,{\bf Z})_{T,U}$-transformations.
Furthermore, $\exp{\Delta_{0,0}}$, as
given in (\ref{w00jj}), obviously vanishes
for $T=U$.
For $T$ in the vicinity of $U$
one has that $j(T) - j(U) = j'(U) ( T - U)$.  For $T \sim U =1$
one finds that $j(T)-j(U)= \frac{1}{2} j''(1) (T-1)^2$,
because of $j'(1) =0$.  And, finally, for
$T \sim U = \rho$ one
 has that
$ j(T) - j(U) = \frac{1}{3!} j'''(\rho) (T- \rho)^3 $,
because $j'(\rho)=j''(\rho)=0$.
Thus, the ansatz $j(T) - j(U)$
does indeed
reflect the ratio 1:2:3 in the order of the zeros of $\exp{\Delta_{0,0}}$.
This observation is just a reflection of the fact that the order of
the zeros of the $j$-function is determined by the order of  the
fixed points of the duality group; this order is proportional to
the number of massless states at the fixed points, as proven in
(\ref{orfixp}).

Comparing (\ref{www012}) with (\ref{w00jj}),
on the other hand, gives that
\beqa
\Delta_{0,0} &=&
\sum_{n^T m =1} \log ( m_2 - i m_1 U + i  n_1 T - n_2 U T ) \nonumber\\
&=& \log \{(j(T) - j(U))^r\; \eta(T)^{-2} \;\eta(U)^{-2}\} + \dots
\label{repj}
\eeqa
for some non-zero integer $r$.
In (\ref{repj}) a regularisation of the infinite sum
is implied.  Thus, (\ref{repj})
yields a representation of $j(T)^r \; \eta(T)^{-2}$ as a regularised
sum (of chiral masses)
over all the states on the orbit $n^T m = 1, q=0$.  It would be very
interesting
to prove this directly by explicitly performing the sum.
To our knowledge such a representation
is unknown in the literature.  The dots in (\ref{repj}) stand for
additional terms which are regular inside the fundamental region
of the $T,U$-moduli space (see footnote \ref{footreg}).

Let us now switch on
the Wilson lines $B$ and $C$ and discuss the resulting modifications to
(\ref{repj}).  We will, unfortunately, have to restrict ourselves to the
case where $B$ and $C$ are small, that is, we will work to lowest non-trivial
order in $B$ and $C$, only. When switching on Wilson lines, $\Delta_{0,0}$
becomes equal to
\beqa
\Delta_{0,0} =
\sum_{n^T m =1} \log \{ m_2 - i m_1 U + i  n_1 T + n_2 ( -U T
+ \frac{1}{2} B C )\}
\label{delzero}
\eeqa
as can be
seen from (\ref{massutbc}).
Then, a suitable ansatz for
the modification to (\ref{repj}) is, to lowest order in $BC$, given by
\beqa
\Delta_{0,0} &=&
\log \left( j(T) - j(U) +\frac{1}{2} B C \;X (T,U) \right)^r
+ \log \left(\eta(T)^{-2} \;\eta(U)^{-2} + \frac{1}{2} B C \;Y (T,U)
\right) \nonumber\\
&+& \dots
\label{w00bc}
\eeqa
where $X(T,U)$ and $Y(T,U)$ will be determined in the following.
 Let us first discuss the modification to the $(j(T) -j(U))$-term
given by $X(T,U)$.  Under $SL(2,{\bf Z})_U$-transformations
\cite{AntGaNarTay,CLM}
\beqa
U &\rightarrow& \frac{\alpha U -i \beta}{i \gamma U + \delta}
\;,\;\;\; T \rightarrow T - \frac{i \gamma}{2} \frac{BC}{i \gamma U + \delta}
\;,\;\;\; \alpha \delta - \beta \gamma = 1 \nonumber\\
B &\rightarrow& \frac{B}{i \gamma U + \delta} \;\;,\;\;
C \rightarrow \frac{C}{i \gamma U + \delta}
\label{utbctransf}
\eeqa
it follows that, to lowest order in $BC$,
\beqa
j(T) - j(U) \rightarrow \;\;
j(T) -j(U) \; -\;  \frac{i \gamma}{2} \; \frac{BC}{i \gamma U + \delta}
\;\; j'(T)
\eeqa
Then, if one requires that
$j(T) - j(U)
 + \frac{1}{2} BC X(T,U)$ should be invariant under
$SL(2,{\bf Z})_{T,U}$-transformations (\ref{utbctransf}), one obtains that
$X(T,U)$ has to transform as
\beqa
X(T,U) \rightarrow (i \gamma U + \delta )^2 \; X(T,U)
+ i \gamma ( i \gamma U + \delta) \; j'(T)
\eeqa
Similarly it follows that, under
$SL(2,{\bf Z})_T$-transformations, $X(T,U)$ has to transform to lowest order
in $BC$ as
\beqa
X(T,U) \rightarrow  (i \gamma T + \delta )^2 \; X(T,U)
- i \gamma ( i \gamma T + \delta) \; j'(U)
\label{xtutransf}
\eeqa
Note that, in addition,
$X(T,U)$ has to vanish at $T=U$.
Only then does $\Delta_{0,0}$
again become singular at $T=U$.  This has to be the case because, as
discussed in section \ref{ExGGCompS},
the enhancement of the gauge group
still occurs at $T=U$ when
Wilson lines are switched on.  Indeed, when inserting (\ref{st1})
into (\ref{delzero}) it follows that $\Delta_{0,0}$ has a
singularity at $T=U$.  Similarly,
when inserting (\ref{st2}) into (\ref{delzero}) shows that another singularity
of $\Delta_{0,0}$ occurs at
$T=U=\sqrt{1+\frac{BC}{2}}$.  And finally,
when inserting (\ref{st3}) into (\ref{delzero}) yields yet another singularity
of $\Delta_{0,0}$ at $T=U=\frac{i}{2} + \sqrt{\frac{3}{4}
+ \frac{BC}{2}}$, thus substanciating our claims in section 2.
All these requirements do not uniquely specify $X(T,U)$, however.  The
following $X(T,U)$ satisfies all the above requirements
\beqa
X(T,U) &=&
\partial_U
\log \eta^2(U) \; j'(T) -
\partial_T
\log \eta^2(T) \; j'(U)   +
a \left( j(T) - j(U) \right) \eta^4(T) \eta^4(U) \nonumber\\
&+&
{\cal O}((BC)^2)
\label{x0}
\label{ytu12}
\eeqa
where $a$ is an arbitrary constant which cannot be determined alone
by symmetry arguments.  Note that the first term in (\ref{ytu12})
transforms inhomogenously as in (\ref{xtutransf}), whereas the second
term in (\ref{ytu12}) transforms homogenously with modular weight $2$
under
$SL(2,{\bf Z})_{T,U}$-transformations.\footnote{This is, strictly speaking,
not quite correct.  We have ignored the issue of field-independent
phases $\phi=\phi(\alpha,\beta,\gamma,\delta)$, which arise when
performing modular transformations.  These phases, also called multiplier
systems, show up in the transformation laws of the Wilson line moduli $B$
and $C$ as well as in the transformation law of the eta-function.
Thus, in order to really construct the appropriate
function $X(T,U)$, care needs to be taken of all the different
multiplier systems appearing in the transformation laws.}

One can now try to determine
the function $Y(T,U)$ in a similar manner \cite{AntGaNarTay}
by demanding that
$\eta(T)^{-2} \;\eta(U)^{-2} + \frac{1}{2} B C \;Y (T,U)$
should transform with modular weight of $-1$ under
$SL(2,{\bf Z})_{T,U}$-transformations.  Then, one has to require $Y(T,U)$
to transform as \cite{AntGaNarTay}
\beqa
Y(T,U) \rightarrow ( i \gamma U + \delta) Y(T,U)
+ i \gamma \eta^{-2}(U) \partial_T \eta^{-2}(T)
\eeqa
under $SL(2,{\bf Z})_U$-transformations as well as \cite{AntGaNarTay}
\beqa
Y(T,U) \rightarrow ( i \gamma T + \delta) Y(T,U)
+ i \gamma \eta^{-2}(T) \partial_U \eta^{-2}(U)
\eeqa
under $SL(2,{\bf Z})_T$-transformations.  These transformation laws
do, however, not
uniquely
specify $Y(T,U)$.  The following $Y(T,U)$ transforms appropriately
up to corrections of order $(BC)^2$
\beqa
Y(T,U) = -
\eta(T)^{-2} \;\eta(U)^{-2} \;\; \partial_T \log \eta^2(T)
 \;\;\partial_U \log \eta^2(U) + b  \; \eta^2(T)\; \eta^2(U)
+ \cdots
\label{y0}
\eeqa
where $b$ denotes an arbitrary constant which cannot be determined
alone by symmetry arguments.
Note that the first term \cite{AntGaNarTay} in (\ref{y0})
transforms inhomogenously under
$SL(2,{\bf Z})_{T,U}$-transformations, whereas the second
term in (\ref{y0}) transforms homogenously with modular
weight $1$.\footnote{We have, again, ignored the issue of multiplier systems
appearing in the transformation laws given above.}
Hence, it follows from (\ref{w00bc}) that the $BC$-corrected
$\exp{\Delta_{0,0}}$
is given by
\beqa
\exp{\Delta_{0,0}} &\propto& (j(T) - j(U))^r\; \eta(T)^{-2} \;\eta(U)^{-2}
\nonumber\\
&+& \frac{1}{2} BC \;r X(T,U) (j(T) - j(U))^{r-1}
\eta(T)^{-2} \;\eta(U)^{-2} \nonumber\\
&+& \frac{1}{2} BC \; Y(T,U) (j(T) - j(U))^r
\eeqa
with $X(T,U)$ and $Y(T,U)$ given by (\ref{x0}) and (\ref{y0}),
respectively.

Next, let us discuss $\Delta_{0,1}$ in the presence of Wilson lines.
The sum now runs over all the states for which $n^T m =0, q^2 = 1$.
This set of states can be divided into two subsets.  The first
subset consists of the 4 states $n=m=0, q^2=1$, whereas the
second subset consists of all the states for which $(n,m) \neq (0,0),
q^2 =1$. Then, it follows from (\ref{massutbc}) that
\beqa
\Delta_{0,1} = 2 \log (B+C) + 2 \log (B-C) + \;\; \sum_{ (n,m) \neq (0,0),
q^2 =1} \log {\cal M} + const
\label{w01bc}
\eeqa
The first two terms in (\ref{w01bc}) are the contribution arising from
the 4 states $n=m=0, q^2=1$.  When turning off the Wilson lines,
these 4 states become massless, thus giving rise to logarithmic
singularities in $\Delta_{0,1}$, as required.
The last term in (\ref{w01bc}),
$\sum_{ (n,m) \neq (0,0),q^2 =1} \log {\cal M}$,
can be written as
\beqa
\sum_{ (n,m) \neq (0,0),q^2 =1} \log {\cal M}
= \sum_{ (n,m) \neq (0,0)} \log \{{\tilde {\cal M}} + \frac{i}{\sqrt{2}}
(B+C) \} \nonumber\\
 \;\; +
\sum_{ (n,m) \neq (0,0)} \log \{{\tilde {\cal M}} - \frac{i}{\sqrt{2}}
(B+C) \}   \;\;
+
\sum_{ (n,m) \neq (0,0)} \log \{{\tilde {\cal M}} + \frac{1}{\sqrt{2}}
(B-C) \} \nonumber\\
 \;\; + \sum_{ (n,m) \neq (0,0)} \log \{{\tilde {\cal M}}- \frac{1}{\sqrt{2}}
(B-C) \}
\label{w01sec}
\eeqa
where
\beqa
{\tilde {\cal M}} =
m_2 - i m_1 U + i  n_1 T + n_2 ( -U T
+ \frac{1}{2} B C )
\eeqa

Then, expanding (\ref{w01sec}) to lowest order in $B$ and $C$ yields
that
\beqa
 \sum_{ (n,m) \neq (0,0),q^2 =1} \log {\cal M}
&=& 4 \left(
\sum_{ (n,m) \neq (0,0)} \log (m_2 -i m_1 U + i n_1 T - n_2 UT)
\right. \nonumber\\
&+& \left.  \frac{1}{2} B\,C
\sum_{ (n,m) \neq (0,0)} \frac{n_2}{(
m_2 - i m_1 U + i  n_1 T - n_2 U T )} \right) \nonumber\\
\;\; &+& \;  2 \; B\; C \sum_{ (n,m) \neq (0,0)} \frac{1}{(
m_2 - i m_1 U + i  n_1 T - n_2 U T )^2}
\label{l00}
\eeqa
As we will show in the next section,
the terms in the big bracket in
(\ref{l00})
give, when suitably regularised, rise to
\beqa
&&\left(
\sum_{ (n,m) \neq (0,0)} \{ \log (m_2 -i m_1 U + i n_1 T - n_2 UT)
+ \frac{BC}{2}
\frac{n_2}{(
m_2 - i m_1 U + i  n_1 T - n_2 U T )} \}\right)|_{reg}  \nonumber\\
&&= \log \left( \eta(T)^{-2} \;\eta(U)^{-2} - \frac{1}{2} B C \;
\eta(T)^{-2} \;\eta(U)^{-2} \;\; \partial_T \log \eta^2(T)
 \;\;\partial_U \log \eta^2(U) \right)
\label{orbnm00}
\eeqa
Furthermore, (\ref{l00}) should not exhibit any singular behaviour
at finite points in the fundamental region of moduli space,
and it should transform appropriately
under $SL(2,{\bf Z})_{T,U}$-transformations.
Together
with (\ref{orbnm00}) this then restricts
(\ref{l00}), to lowest order in $B$ and $C$, to be
given by\footnote{Holomorphic regularisation of the
last term in (\ref{l00}) yields
$\sum_{ (n,m) \neq (0,0)} \frac{1}{(
m_2 - i m_1 U + i  n_1 T - n_2 U T )^2}
\propto G_2 (T) G_2 (U)$. The holomorphically regularised Eisenstein
function $G_2(T)$, however, transforms inhomogenously
under $SL(2,{\bf Z})_{T}$-transformations.  We thus suspect that the
procedure given in (\ref{l00}) for computing $\Delta_{0,1}$ is too naive.}
\beqa
&&\sum_{ (n,m) \neq (0,0),q^2 =1} \log {\cal M}
= 4 \log \left( \eta(T)^{-2} \;\eta(U)^{-2} \right. \nonumber\\
&&- \left. \frac{1}{2} B C \;
\eta(T)^{-2} \;\eta(U)^{-2} \;\; \partial_T \log \eta^2(T)
 \;\;\partial_U \log \eta^2(U) \right)
+ c BC \eta(T)^{4} \;\eta(U)^{4}
\label{zz}
\eeqa
where $c$ is an arbitrary constant which cannot be fixed
by symmetry considerations alone.
Note that the last term in (\ref{zz}) is invariant under
$SL(2,{\bf Z})_{T,U}$-transformations.\footnote{Ignoring again complications
due to the appearance of multiplier systems.}
Equation (\ref{zz}) can also be written as
\beqa
\sum_{ (n,m) \neq (0,0),q^2 =1} \log {\cal M}
= 4 \log \left( \eta(T)^{-2} \;\eta(U)^{-2} + \frac{1}{2} B C \;Y (T,U)
\right)
\label{zzz}
\eeqa
with $Y$ given in (\ref{y0}).
Inserting (\ref{zzz})
into (\ref{w01bc}) yields that
\beqa
\Delta_{0,1} &=& p\; \log (B+C) + p\; \log (B-C) \nonumber\\
&+& t\;
\log \left( \eta(T)^{-2} \;\eta(U)^{-2} + \frac{1}{2} B C \;Y (T,U)
\right) + const
\label{w1}
\eeqa
with appropriate constants $p$ and $t$.
Note that each of the factors in $\exp{\Delta_{0,1}}$ transforms
with modular weight of $-1$ under $SL(2,{\bf Z})_{T,U}$-transformations.

Inserting (\ref{w00bc}) and (\ref{w1}) into (\ref{wzero})
yields that
\beqa
\Delta_0 &\propto& r
\log \left( j(T) - j(U) +\frac{1}{2} B C \;
\left( \partial_U
\log \eta^2(U) \; j'(T) -
\partial_T
\log \eta^2(T) \; j'(U) \right)
\right) \nonumber\\
&+& {\tilde t} \;  \log \left(\eta(T)^{-2} \; \eta(U)^{-2} -
\frac{1}{2} B C
\eta(T)^{-2} \; \eta(U)^{-2} \;\; \partial_T \log \eta^2(T)
\;\;\partial_U \log \eta^2(U) \right) \nonumber\\
&+& {\tilde b} \; B C\; \eta^4(T)\; \eta^4(U)
+ p \;\log (B+C) + p\; \log (B-C)
 + \dots
\label{w0fin}
\eeqa
where, again, $p,r,{\tilde t}$ and ${\tilde b}$
denote some appropriate constants.
Note that each of the terms in (\ref{w0fin}) is either
invariant or the logarithm of a term
transforming with weight of $-1$ under
modular $SL(2,{\bf Z})$-transformations.  Also note that, up to
quadratic order in $B$ and $C$, the terms given in (\ref{w0fin})
are the only ones one can write down which can exhibit
singular behaviour at finite points in the moduli space.
At these points, otherwise massive states become massless, thus inducing
this singular behaviour. The terms in (\ref{w0fin})
represent possible threshold corrections terms which
one should find in calculations
of the running of gauge and gravitational couplings in orbifolds with
${\bf Z}_2$-sectors.
In that context, the coefficients
$p,r,{\tilde t}$ and ${\tilde b}$ are related to
beta-function coefficients as well as to K\"{a}hler and $\sigma$-model
anomaly coefficients.  This will be discussed in the last section
of this paper.
Finally, note that the terms given in (\ref{w0fin}) transform
appropriately under modular $SL(2,{\bf Z})_{T,U}$-transformations, but not
under additional transformations belonging to the
full modular
$SO(4,2,{\bf Z})$-group.  The dots in (\ref{w0fin}) stand for additional
contributions, which we haven't computed, whose role is to restore
the proper transformation behaviour of $\Delta_0$ under
full $SO(4,2,{\bf Z})$-transformations.

Finally, let as look at the $SU(2,1)$-coset discussed in
section \ref{SUmP11}, which was based on an $A_2$-root lattice.  Applying
the truncation $U=\frac{1}{2}(\sqrt{3} + i), C=0, B= \sqrt{3} {\cal A}$
to (\ref{w0fin}) yields that
\beqa
\Delta_0 &\propto& r
\log  j(T)
+ {\tilde t}\; \log \eta(T)^{-2}
+ {\tilde p} \; \log {\cal A} + const
 + \dots
\label{w0fin21}
\eeqa
where we have used that $j(U=\rho)=0$.
The dots in (\ref{w0fin21}) stand for additional terms whose role is
to restore the proper transformation behaviour of
 (\ref{w0fin21}) under transformations belonging to the full
modular $SU(2,1,{\bf Z})$-group.

\section{Automorphic functions for the orbit $2 n^T m + q^T {\cal C} q = 0$
\label{AutFun2Orb}}

\setcounter{equation}{0}

In this section we will consider massive untwisted states possessing
$2N_L + {\bf P}^2_L=2$.  Then,
the associated modular invariant
orbit is defined by
\beqa
2n^T m + q^T {\cal C} q =0
\eeqa

Let us again look at the example of an
$SO(4,2)$-coset, for concreteness.
As in the previous section,
by
taking into account that the Cartan matrix $\cal C$
for the Narain sublattice $\Gamma_{4;2}$ under consideration
reads ${\cal C} = 2 diag (1,1)$, then yields (\ref{w012}) as
\beqa
\Delta_{1} &\propto& \sum_{n^T m + q^2 =0} \log {\cal M}
=
\sum_{n^T m =0 \;,\;  q=0} \log {\cal M} +
\sum_{n^T m =-1 \;,\;  q^2=1} \log {\cal M} +
 \dots
\label{wone}
\eeqa
where we have rewritten the sum over all the states laying on the
$SO(4,2,{\bf Z})$-invariant
orbit $n^T m + q^2 = 0$ into a sum over $SL(2,{\bf Z})_{T,U}$-invariant
orbits $n^T m =constant$.

Let us consider the term $\sum_{n^T m =0 \;,\;  q=0} \log {\cal M}$
in (\ref{wone}).  That is, let us
consider massive untwisted states for which $q_1=q_2=0, n^T m =0$
and $(n,m) \neq 0$.
For these states the mass formula (\ref{massutbc})
reduces
to
\beqa
{\cal M} = m_2 - i m_1 U + i  n_1 T + n_2 ( -U T + \frac{1}{2} \, B C)
\eeqa
Expanding to lowest order in $B$ and $C$ yields
\beqa
\sum_{n^T m =0 \;,\;  q=0} \log {\cal M} &=&
\sum_{(n,m) \neq 0 } \log
(m_2 - i m_1 U + i  n_1 T - n_2 U T ) \nonumber\\
&+& \frac{1}{2} BC
\sum_{(n,m)\neq 0} \frac{n_2}{
(m_2 - i m_1 U + i  n_1 T - n_2 U T )} + {\cal O}((BC)^2)
\label{nmzero}
\eeqa
Introducing a
  set of integers
$r_1, r_2, s_1$ and $s_2$ \cite{FerKouLuZw}
subject to $n^T m =0, (n,m) \neq 0,$
\beqa
m_2 = r_1 \, r_2 \;\;,\;\;
n_2 = s_1 \, s_2 \;\;,\;\;
m_1 = - r_2 \, s_1 \;\;,\;\;
n_1 = r_1 \, s_2
\eeqa
allows one to rewrite (\ref{nmzero})
as
\beqa
\sum_{n^T m =0 \;,\;  q=0} \log {\cal M} &=&
\sum_{(r_1,s_1) \neq (0,0)}
\log (r_1 + i s_1 U) \;\; +
\sum_{(r_2,s_2) \neq (0,0)}
\log(r_2 + i s_2 T) \nonumber\\
&+& \frac{1}{2} BC \;
(\sum_{(r_1,s_1) \neq (0,0)} \frac{s_1}{r_1 + i s_1 U})
(\sum_{(r_2,s_2) \neq (0,0)} \frac{s_2}{r_2 + i s_2 T})\nonumber\\
&+& {\cal O}((BC)^2)
\eeqa
and, hence, as
\beqa
\sum_{n^T m =0 \;,\;  q=0} \log {\cal M} &=&
\sum_{(r_1,s_1) \neq (0,0)}
\log (r_1 + i s_1 U) \;\; +
\sum_{(r_2,s_2) \neq (0,0)}
\log(r_2 + i s_2 T) \nonumber\\
&-& \frac{1}{2} BC \;
(\partial_U \sum_{(r_1,s_1) \neq (0,0)}
 \log(r_1 + i s_1 U))
\;\;(\partial_T \sum_{(r_2,s_2) \neq (0,0)}
\log(r_2 + i s_2 T))\nonumber\\
&+& {\cal O}((BC)^2)
\label{pptu}
\eeqa
Using that upon regularisation, $\sum_{(r_1,s_1) \neq (0,0)}
\log (r_1 + i s_1 U) = \log \eta^{-2} (U)$ \cite{VafaOog,FerKouLuZw},
allows one to rewrite
(\ref{pptu}) as
\beqa
\sum_{n^T m =0 \;,\;  q=0} \log {\cal M} &=&
\log \left(\eta^{-2} (U)
\eta^{-2} (T) \right) - \frac{1}{2} BC \;\partial_U \log \eta^{-2} (U)
 \; \partial_T \log \eta^{-2} (T) \nonumber\\
&+& {\cal O}((BC)^2)
\label{anto}
\eeqa
This can also be written as
\beqa
\sum_{n^T m =0 \;,\;  q=0} \log {\cal M} &=&
\log \left( \eta^{-2} (U)
\eta^{-2} (T)  (1 - \frac{1}{2} BC \;\partial_U \log \eta^{2} (U)
 \; \partial_T \log \eta^{2} (T)  )\right) \nonumber\\
&+& {\cal O}((BC)^2)
\label{antoniad}
\eeqa
Inserting (\ref{antoniad}) into (\ref{wone})
yields that
\beqa
\Delta_1 \propto
\log \left( \eta^{-2} (U)
\eta^{-2} (T)  (1 - \frac{1}{2} BC \;\partial_U \log \eta^{2} (U)
 \; \partial_T \log \eta^{2} (T)  ) \right) + \dots
\label{w1exp}
\eeqa
This result agrees with the one given in \cite{AntGaNarTay}, which was
obtained
by requiring $\exp{\Delta_1}$ to
transform with weight $-1$
under $SL(2,{\bf Z})_{T,U}$-transformations \cite{AntGaNarTay}.
Note that $\Delta_1$ also has to transform
appropriately under additional $SO(4,2,{\bf Z})$-transformations.
The dots in (\ref{w1exp}) stand for additional contributions,
which we haven't computed, whose role is to restore the proper
transformation behaviour of $\Delta_1$
under transformations belonging to the
full modular
$SO(4,2,{\bf Z})$-group.

\section{Threshold corrections
\label{ThreshCor}}

\setcounter{equation}{0}

In this section, we will discuss 1-loop corrections to gauge
and gravitational couplings in the context of $(0,2) {\bf Z}_N$-orbifold
compactifications.  We will begin by reviewing some well-known
facts about effective gauge couplings in locally
$N=1$ supersymmetric
effective quantum field theories (EQFT).

Consider first the case of a locally supersymmetric EQFT
with gauge group $G=\otimes G_a$
where the
light charged particles
are exactly massless, and where the massive charged
fields
decouple at some scale, say $M_{X}$.  Then, at energy
scales $p^2 \ll M^2_{X}$, the 1-loop corrected low energy
gauge couplings are given by \cite{Vkap,DKL,Louis,DFKZ,KapLu}
\beqa
\frac{1}{g_a^2(p^2)} = \frac{1}{g_a^2(M^2_{X})}
+ \frac{b_a}{16 \pi^2} \log \frac{M^2_{X}}{p^2} +
\frac{{\tilde \Delta}_a}{16 \pi^2}
\label{grun}
\eeqa
where $b_a$ is the coefficient
of the 1-loop $N=1$ $\beta$-function, $\beta_a=\frac{b_a g_a^3}{16 \pi^2}$,
computed from the massless charged spectrum of the theory.  $b_a$ describes
the running between $M^2_{X}$ and $p^2 \ll M^2_{X}$ and is given by
$b_a = - 3 c(G_a) + \sum\limits_C  T_a(r_C)$.  Here, $c(G_a)$ denotes
the quadratic Casimir of the gauge group and
the sum is over chiral matter superfields transforming under some
representation $r$ of the gauge group $G_a$.
Also, $T_a(r) = Tr_r ( T_{(a)}^2)$, where
$T_{(a)}$ denotes a generator of $G_a$.
${\tilde \Delta}_a$, on the other hand, determines the boundary conditions
for the running gauge couplings at $M^2_{X}$ and is given by
\beqa
{\tilde \Delta}_a = \left(c_a K - 2 \sum \limits_r
T_a(r)\log \det g_r\right) +\Delta_a
\label{tildel}
\eeqa
The massive charged fields, which have been integrated out, contribute
a finite threshold correction $\Delta_a$ to the low energy gauge coupling.
There are, however, also contributions to ${\tilde \Delta}_a$
from the massless modes in
the theory.
They too
need to be taken into account when
 discussing effective couplings.
These massless contributions arise due to non-vanishing
K\"{a}hler and $\sigma$-model anomalies
\cite{Louis,DFKZ,IbLu,CO2,KapLu}, present
in generic supergravity-matter
systems, and they are given by the term $c_a K - 2 \sum \limits_r
 T_a(r)\log \det g_r$
in (\ref{tildel}).  Here, $c_a = - c(G_a) + \sum \limits_C  T_a(r_C)$,
and $g_r$ denotes the $\sigma$-model
metric of the massless subsector of the matter
fields in the representation $r$.

Next consider the case where some gauge or matter particles do have small
masses of the order $M_I \ll M_{X}$.  In a regime where
$M^2_I \ll p^2 \ll M^2_{X}$, all interactions can be described in terms
of a massless EQFT, whereas for $p^2 \ll M^2_I$ there is another EQFT
given
in terms of the truly massless fields, only \cite{KapLu}.
We will assume that, at the
threshold scale $M_I$, supersymmetry remains unbroken whereas the
gauge group $G=\otimes G_a$ is spontaneoulsy broken
down to $\hat{G}$.  Let us consider one such factor $G_a$ and assume
that it gets spontaneously broken down to $G_a \rightarrow
\hat{G}_a = \otimes_i \hat{G}_{a,i}$.
Let us then discuss the running of the coupling $\hat{g}_{a,i}(p^2)$
of one such subgroup $\hat{G}_{a,i}$.
In order to simplify the notation,
we will, in the following, simply denote this subgroup $\hat{G}_{a,i}$
by $\hat{G}_a$ and its associated coupling constant
$\hat{g}_{a,i}$ by
$\hat{g}_{a}$.
At low energies, $p^2 \ll M^2_I$, the effective gauge coupling
$\hat{g}_a^2(p^2)$ is given as \cite{MunLu,KapLu}
\beqa
\frac{1}{\hat{g}_a^2(p^2)} = \frac{1}{\hat{g}_a^2(M^2_I)}
+ \frac{\hat{b}_a}{16 \pi^2} \log \frac{M^2_I}{p^2} +
\frac{1}{16 \pi^2}
\left(
\hat{c}_a K - 2 \sum \limits_{\hat{r}}
{\hat T}_a(\hat{r})\log \det g_{\hat{r}} \right)
\label{hatga2p2}
\eeqa
where the coefficients $\hat{b}_a = -3 c (\hat{G}_a) +
\sum\limits_{\hat{C}} {\hat T}_a({\hat{r}_{\hat{C}}})
$ and $\hat{c}_a =-c(\hat{G}_a) + \sum \limits_{\hat{C}}
{\hat T}_a(\hat{r}_{\hat{C}})$ are now
determined only in terms of the truly massless fields transforming
under $\hat{G}_{a}$.  Here, ${\hat C}$ denotes a truly
massless chiral superfield transforming under some representation
${\hat r}_{\hat C}$ of the unbroken gauge group ${\hat G}_{a}$.
Above the threshold $M_I$, the running of the gauge coupling is
determined by the gauge group $G_a$.  Hence, $\hat{g}_a (M^2_I)$
is given by
\beqa
\frac{1}{\hat{g}_a^2(M^2_I)} = \frac{1}{g^2_a(M^2_{X})}
+ \frac{b_a}{16 \pi^2} \log \frac{M^2_{X}}{M^2_I}
+ \frac{\Delta_a}{16 \pi^2}
\label{hatga2}
\eeqa
Here, $b_a$ describes the running between $M_{X}$ and $M_I$ and is
given by $b_a = -3 c(G_a) + \sum\limits_C T_a(r_C)$, where the sum
runs over all the light chiral matter superfields charged under $G_a$.
$\Delta_a$ denotes the contribution from all the massive charged
states which decouple at $M_{X}$.

It is useful to note that the running of the gauge coupling between
$M_X$ and $M_I$ can also be described in terms of the light fields
which are charged under the gauge group ${\hat G}_{a}$.  Above the
threshold $M_I$ all of these light fields are effectively massless.
Thus, they all contribute to the running
and therefore, at least for regular embeddings,
$b_a$ can also be written as $b_a = -3 c(\hat{G}_a) - 3 \sum\limits_V
{\hat T}_a({\hat r}_V) + \sum\limits_C {\hat T}_a({\hat r}_C)$.
Here, ${\hat r}_V$ and ${\hat r}_C$
denote the representation of a light vector
multiplet and of a light chiral multiplet, respectively.
As an example, consider the breaking of $G_a=SU(5)$ down to
$SU(3) \otimes SU(2) \otimes U(1)$.  Decomposing
the adjoint representation of $SU(5)$ into representations of
$SU(3) \otimes SU(2)$, ${\bf 24}={\bf (8,1)}+
{\bf (1,3)}+
{\bf (1,1)}+
{\bf (3,2)}+
{\bf ({\bar 3},2)}$, yields that $c(SU(5))=c(SU(3)) +
4 {\hat T}_{SU(3)} ({\bf 3})
= 10$.  Decomposing ${\bf 5} ={\bf (3,1)} + {\bf (1,2)}$ yields
that
$T_{SU(5)}({\bf 5}) =
{\hat T}_{SU(3)}({\bf 3}) = 1$.  More generally,
$\sum \limits_C T_{SU(5)}(r_C) =
\sum \limits_C {\hat T}_{SU(3)}({\hat r}_C)$, and it follows then indeed that
$b_{SU(5)} = -3 c(SU(3)) -3
 \sum\limits_V
{\hat T}_{SU(3)}({\hat r}_V) + \sum\limits_C {\hat T}_{SU(3)}({\hat r}_C)$.

We now turn to string theory and consider orbifolds with $N=2$ spacetime
sectors. The reason for this is as follows.
Explicit string scattering amplitude calculations of
threshold corrections to gauge
\cite{DKL,AntNaTay,May1,Bailin1,AntGaNarTay,AntTay}
and gravitational \cite{AntGaNa}
couplings in the context of $(2,2) {\bf Z}_N$-orbifold compactifications
show that
a non-trivial moduli dependence of these thresholds
only arises
if the orbifold point twist group ${\cal P}$ contains a subgroup
$\tilde{\cal P}$
that, by itself,
would produce an orbifold with $N=2$ spacetime supersymmetry.
Furthermore, if
the underlying $T_6$ torus factorises into
$T_6=T_2 \oplus T_4$, where the $T_2$ remains untwisted under the
action of $\tilde{\cal P}$, then
the moduli dependent
threshold corrections associated with this $T_2$
are invariant under $\Gamma = SL(2,{\bf Z})_{T,U}$.
We will, in the following, stick to those $(0,2) {\bf Z}_N$-orbifolds for which
$T_6=T_2 \oplus T_4$ with the untwisted plane lying in $T_2$, and we will
derive formulae for
the gauge and gravitational threshold corrections
associated with this untwisted plane $T_2$.
As an example one can think of
a ${\bf Z}_4$-orbifold, for which
$\tilde{\cal P}$ is the ${\bf Z}_2$ generated by
$\tilde{\theta} = \theta^2= (\Omega^2,
\Omega,\Omega)$, where $\Omega = e^{\pi i}$.

Let us consider the case where the gauge group $G=\otimes
G_a$ in the observable sector
of the $(2,2) {\bf Z}_N$-orbifold gets broken to
a subgroup $\hat{G}$ by turning on Wilson moduli.  Thus, the
Wilson moduli act as Higgs fields.
Let us assume
that this breaking takes place when turning on Wilson moduli $B$ and $C$
associated with the $T_2$ (in the decomposition
$T_6=T_2 \oplus T_4$).  We will now determine the running of the
gauge couplings in both the hidden and the observable sectors from
equations (\ref{grun}),
(\ref{hatga2p2}) and (\ref{hatga2}).
We will throughout this section be working to lowest order in the
Wilson moduli $B$ and $C$.  As a consequence, the results given
below will only be invariant under the subgroup $SO(2,2,{\bf Z})$
of the modular group $SO(4,2,{\bf Z})$.  Also, for notational simplicity
we will not
include the Kac-Moody level $k_a$ into the equations below.
The tree-level gauge couplings in the observable and hidden sectors
are given by $g_a^{-2}(M^2_{string})=g_{E'_8}^{-2}(M^2_{string})=
\frac{S+ {\bar S}}{2}$.
In the case of vanishing Wilson lines $B$ and $C$,
the
$\sigma$-model
metric $g_{\hat{C}}$ of a charged matter field/
twisted modulus $\hat{C}$
exhibits the following dependence on the moduli $T,U$,
namely $g_{\hat{C}}=\left( (T+{\bar T})(U+{\bar U}) \right)^{n_{\hat C}}$,
where $n_{\hat C}$ denotes the modular weight of $\hat C$.  In the
presence of Wilson lines $B$ and $C$ one then expects
the $\sigma$-model metric $g_{\hat{C}}$ to be given by
\beqa
g_{\hat C}
= \left( (T+{\bar T})(U+{\bar U}) - \frac{1}{2}
(B+{\bar C})(C+{\bar B}) \right)^{n_{\hat C}}
\label{gcmetric}
\eeqa
Let us first discuss the running, in the
presence of Wilson lines $B$ and $C$, of the
gauge couplings associated with the part of the gauge group
which is not affected by turning on $B$ and $C$.  For concreteness,
take this to be the case for the $E_8'$ in the hidden sector.
Of the orbits discussed in section
\ref{ModOrbUnt}, there is one orbit which is the relevant one for the
discussion of threshold corrections to
$g_{E'_8}$, namely the orbit $2 n^T m + q^T {\cal C} q =0$.
As also discussed in section \ref{ModOrbUnt},
there are no finite points in the fundamental region of moduli space
for which the massive states on this orbit might become massless
and, hence, no additional threshold scales.  Thus, for this case the
only relevant threshold scale is given by $M_X=M_{string}$.
Inspection of (\ref{w1exp}) shows that the threshold corrections
$\Delta_{E'_8}$ should be given by
\beqa
\Delta_{E'_8} = - \alpha_{E'_8}
\log \left(| \eta (U)
\eta (T)|^4  |1 - \frac{1}{2} BC \;\partial_U \log \eta^{2} (U)
 \; \partial_T \log \eta^{2} (T)  |^{-2} \right)
\label{deltae8}
\eeqa
where $\alpha_{E'_8} = - c(E'_8)$ \cite{DFKZ}.
Inserting $K = - \log \left((T+{\bar T})(U+{\bar U})-
\frac{1}{2}(B+{\bar C})(C+{\bar B}) \right)$ and (\ref{deltae8})
into (\ref{grun}) and (\ref{tildel}) yields that
\beqa
\frac{1}{g_{E'_8}^2(p^2)} &=&
\frac{S+{\bar S}}{2} +  \frac{b_{E'_8}}{16 \pi^2}
\log \frac{M^2_{string}}{p^2}  \nonumber\\
&-&
\frac{\alpha_{E'_8}}{16 \pi^2}
\log\left((T+{\bar T})(U+{\bar U})-
\frac{1}{2}(B+{\bar C}^2)(C+{\bar B}) \right)
\nonumber\\
&-&
\frac{\alpha_{E'_8}}{16 \pi^2}
\log \left(| \eta (U)
\eta (T)|^4  |1 - \frac{1}{2} BC \;\partial_U \log \eta^{2} (U)
 \; \partial_T \log \eta^{2} (T)  |^{-2} \right)
\eeqa
where $b_{E'_8}= -3c(E'_8)$.
Note that
we have not yet taken into account
the Green-Schwarz mechanism.  The Green-Schwarz mechanism can remove
an amount $\delta_{GS}$ from the above
\cite{CO1,DFKZ}.  It might, in addition,
also remove a
modular invariant function \cite{KapPriv}, yielding
\beqa
\frac{1}{g_{E'_8}^2(p^2)} &=&
\frac{Y}{2} +  \frac{b_{E'_8}}{16 \pi^2}
\log \frac{M^2_{string}}{p^2}  \nonumber\\
&-&
\frac{(\alpha_{E'_8} - \delta_{GS})}{16 \pi^2}
\log\left((T+{\bar T})(U+{\bar U})-
\frac{1}{2}(B+{\bar C}^2)(C+{\bar B}) \right)
\nonumber\\
&-&
\frac{(\alpha_{E'_8}-\delta_{GS})}{16 \pi^2}
\log \left(| \eta (U)
\eta (T)|^4  |1 - \frac{1}{2} BC \;\partial_U \log \eta^{2} (U)
 \; \partial_T \log \eta^{2} (T)  |^{-2} \right) \nonumber\\
\label{runE8}
\eeqa
where $Y= S+{\bar S} - \frac{\delta_{GS}}{8 \pi^2}
\log\left((T+{\bar T})(U+{\bar U})-
\frac{1}{2}(B+{\bar C})(C+{\bar B}) \right)
+ (modular\, inv. \, function)$.

The effective gauge coupling (\ref{runE8}) has to be invariant under
modular $SL(2,{\bf Z})_{T,U}$ transformations.
The quantity $Y$ is invariant, when taking into account that the
dilaton acquires a non-trivial transformation behaviour
\cite{DFKZ,MunLu}
at the 1-loop level under $SL(2,{\bf Z})_{T,U}$-transformations, that is
$S \rightarrow
 S - \frac{1}{8 \pi^2} \delta_{GS} \log (i \gamma U + \delta)$
under (\ref{utbctransf}), etc.  Then it indeed follows that
(\ref{runE8}) is invariant under $SL(2,{\bf Z})_{T,U}$-transformations.

Next, let us
look at the running of the gauge coupling constants
in the observable sector.  Consider a
non-abelian factor $G_a$ and assume that
it gets broken down to
$\hat{G}_a = \otimes_i {\hat G}_{a,i}$
when turning on the Wilson moduli $B$ and $C$.
Let us then discuss the running of the coupling $\hat{g}_{a,i}(p^2)$
of one such subgroup $\hat{G}_{a,i}$.
In order to simplify the notation,
we will, in the following, simply denote this subgroup $\hat{G}_{a,i}$
by $\hat{G}_a$ and its associated coupling constant
$\hat{g}_{a,i}$ by
$\hat{g}_{a}$.  This time, the relevant orbit is the one for
which $n^T m = 0  , q^2 = 1$.
Inspection of (\ref{w1})
shows that there are now 3 threshold
scales in the presence of non-vanishing Wilson lines $B$ and $C$, namely
$M_X=M_{string}, M_I=|B+C|M_{string}$ and $M'_I=|B-C|M_{string}$.\footnote{
All moduli are taken to be dimensionless.}
We will in the following set $B=C$ for simplicity.  Then the discussion
simplifies and the remaining 2 threshold scales are given by
$M_X=M_{string}, M_I=|B+C|M_{string}$.
Furthermore, it follows from (\ref{gcmetric}) that for $B=C$
\beqa
 \hat{c}_a K - 2 \sum \limits_{\hat{r}}
T_a(\hat{r})\log \det g_{\hat{r}} = -\hat{\alpha}_{a}
\log \left( (T+{\bar T})
(U+ {\bar U}) - \frac{1}{2} (B+{\bar B})^2\right)
\eeqa
where $\hat{\alpha}_{a} = - c(\hat{G}_a) + \sum \limits_{\hat{C}}
 {\hat T}_a({\hat r}_{\hat{C}})(1+2n_{\hat{C}})$.
Inspection of (\ref{w1}), on the other hand,
 shows that the threshold corrections $\Delta_a$ should be given by
\beqa
\Delta_a = - \alpha_{a}
\log\left(|\eta(T)\eta(U)|^{4}|1-\frac{1}{2} B^2 \partial_U \log \eta^2(U)
\partial_T \log \eta^2(T) + r B^2 \eta^4(T) \eta^4 (U)
|^{-2} \right) \nonumber\\
\eeqa
where $r$ denotes an unknown coefficient which cannot be determined
by symmetry considerations alone.  $\alpha_a$ can either be written in
terms of the gauge group $G_a$ as $\alpha_a =
- c(G_a) + \sum \limits_C T_a(r_C)(1+ 2 n_C)$, or in terms of
the unbroken gauge group ${\hat G}_a$ as
$\alpha_{a}=-c(\hat{G}_a) -
\sum\limits_V
{\hat T}_a({\hat r}_V) + \sum\limits_C {\hat T}_a({\hat r}_C)(1+2n_C)$.
Here the $n_C$ denote the modular weights of the light chiral
superfields.  Then, it follows from (\ref{hatga2p2}) and
(\ref{hatga2}) that
the effective gauge coupling associated with the
unbroken subgroup $\hat{G}_a$ is given by
\beqa
\frac{1}{\hat{g}_a^2(p^2)} &=&
\frac{S+{\bar S}}{2} +  \frac{\hat{b}_a}{16 \pi^2}
\log \frac{M^2_{string}}{p^2}
+ \frac{(\hat{b}_a-b_a)}{16 \pi^2} \log |B|^2\nonumber\\
&-&
\frac{\hat{\alpha}_{a}}{16 \pi^2}
\log\left((T+{\bar T})(U+{\bar U})-\frac{1}{2}(B+{\bar B}^2) \right)
\nonumber\\
&-&
\frac{\alpha_{a}}{16 \pi^2}
\log\left(|\eta(T) \eta(U)|^{4}
|1-\frac{1}{2} B^2 \partial_U \log \eta^2(U)
\partial_T \log \eta^2(T) \right. \nonumber\\
&+& \left. r B^2 \eta^4(T) \eta^4 (U)
|^{-2}\right)
\label{rustr}
\eeqa
$r$ denotes a constant which cannot be determined by symmetry.
Note that
we have not yet taken into account
the Green-Schwarz mechanism.  The Green-Schwarz mechanism can, again, remove
an amount $\delta_{GS}$ from the above
\cite{CO1,DFKZ} and possibly
also a
modular invariant function \cite{KapPriv}, yielding
\beqa
\frac{1}{\hat{g}_a^2(p^2)} &=&
\frac{Y}{2} +  \frac{\hat{b}_a}{16 \pi^2}
\log \frac{M^2_{string}}{p^2}
+ \frac{(\hat{b}_a-b_a)}{16 \pi^2} \log |B|^2\nonumber\\
&-&
\frac{(\hat{\alpha}_{a} - \delta_{GS})}{16 \pi^2}
\log\left((T+{\bar T})(U+{\bar U})-\frac{1}{2}(B+{\bar B}^2) \right)
\nonumber\\
&-&
\frac{(\alpha_{a}    - \delta_{GS})}{16 \pi^2}
\log\left(|\eta(T) \eta(U)|^{4}
|1-\frac{1}{2} B^2 \partial_U \log \eta^2(U)
\partial_T \log \eta^2(T) \right. \nonumber\\
&+& \left. r B^2 \eta^4(T) \eta^4 (U)
|^{-2}\right)
\label{run}
\eeqa
where $Y= S+{\bar S} - \frac{\delta_{GS}}{8 \pi^2}
\log\left((T+{\bar T})(U+{\bar U})-\frac{1}{2}(B+{\bar B}^2) \right)
+ (modular\, inv. \, function)$.
The quantity $Y$ is invariant
under $SL(2,{\bf Z})_{T,U}$-transformations, as
discussed above.  Requiring
(\ref{run}) also to be invariant
under $SL(2,{\bf Z})_{T,U}$-transformations
 (\ref{utbctransf}) leads to
the following restriction on the spectrum
\beqa
\hat{b}_a - b_a = \hat{\alpha}_{a}
- \alpha_{a}
\eeqa
which can be rewritten as
\beqa
\sum_V {\hat T}_a (r_V) = - \sum_C n_C {\hat T}_a (r_C) +
\sum_{\hat C} n_{\hat C} {\hat T}_a ({\hat r}_{\hat C})
\eeqa
Restrictions of a similar type on the spectrum where already considered
in \cite{MunLu}.

For completeness, let us consider the case of vanishing Wilson lines
$B=C=0$.  Then, equation (\ref{run}) turns into
\beqa
\frac{1}{g_a^2(p^2)} &=&
\frac{Y}{2} +  \frac{b_a}{16 \pi^2}
\log \frac{M^2_{string}}{p^2}  \nonumber\\
&-&
\frac{(\alpha_{a} - \delta_{GS})}{16 \pi^2}
\log\left((T+{\bar T})(U+{\bar U}) \right)
\nonumber\\
&-&
\frac{(\alpha_{a}    - \delta_{GS})}{16 \pi^2}
\log\left(|\eta(T) \eta(U)|^{4}\right)
\eeqa
where $Y= S+{\bar S} - \frac{\delta_{GS}}{8 \pi^2}
\log\left((T+{\bar T})(U+{\bar U}) \right)
+ (modular\, inv. \, function)$.  This describes the running of
the gauge coupling associated with the unbroken group $G_a$
\cite{DKL,Louis,AntNaTay,DFKZ}.

An example resembling the above situation
is provided by a ${\bf Z}_4$-orbifold
of the type $T_6 = T_2 \oplus T_4$.  In the $(2,2)$ case, the gauge
group in the observable sector is given by $G=E_6 \otimes SU(2)
\otimes U(1)$.  Now consider turning on complex Wilson lines $B=C$
associated with the $T_2$.  Then, the $E_6 \otimes U(1)$ gets
broken down to an $SO(8) \otimes U(1)'$, whereas the $SU(2)$ remains
intact.  At energies below $M^2_I$, $p^2 \ll M^2_I$, the
running of the gauge coupling $g_{SO(8)}$ is determined in terms of the
truly massless representations of $SO(8)$.  At energies above $M^2_I$,
on the other hand, the running is determined in terms of the light
representations of $E_6$.

Let us now consider a $(2,2) {\bf Z}_N$-orbifold
model with $N=2$ spacetime supersymmetry.  Let us again assume that
the internal torus factorises as $T_6=T_2 \oplus T_4$, and that
the $T_2$ is not twisted under the action of the internal twist,
thus leading to a model with $N=2$ spacetime supersymmetry.  Let us
discuss the
threshold corrections to the gauge couplings associated with
the gauge group of the internal $T_2$.
As discussed extensively in section \ref{AutFunNulOrb}, the
orbit relevant to
the discussion of the threshold corrections to these gauge
couplings is given by $n^T m =1, q^2 =0$.  Then, inspection of (\ref{repj})
shows that a different threshold scale $M_I =|j(T) - j(U)|M_{string}$
arises in this context besides $M_X = M_{string}$.  For points where
$T=U$ in the $(T,U)$-moduli space, the
generic gauge group $U(1) \otimes U(1)$ gets enlarged to
$SU(2) \otimes U(1)$, because 2 additional $N=2$ vector multiplets
become massless at these points.  At generic values in the $(T,U)$-moduli
space, the running of one such effective $U(1)$-gauge coupling should,
in analogy to (\ref{rustr}),
be given by
\beqa
\frac{1}{\hat{g}_{U(1)}^2(p^2)} &=&\frac{1}{\hat{g}_{tree}^2}
+  \frac{\hat{b}_{U(1)}}{16 \pi^2}
\log \frac{M^2_{string}}{p^2}
+
 \frac{(\hat{b}_{U(1)} - b_{U(1)})}{16 \pi^2} \log |j(T)-j(U)|^2 \nonumber\\
&-&
\frac{\hat{\alpha}_{U(1)}}{16 \pi^2}
\log(T+{\bar T})(U+{\bar U})|\eta(T) \eta(U)|^{4}
\label{gau1}
\eeqa
Here, the $N=2$ $\beta$-function coefficient $\hat{b}_{U(1)}$ vanishes,
 $\hat{b}_{U(1)}=0$, since there are no
massless hyper multiplets
 in the theory, which
are charged under the $U(1)$.  The running above the threshold
scale $M_I$ is determined in terms of the $N=2$ $\beta$-function
coefficient $b_{U(1)}$ associated with the light charged multiplets
in the theory.  Then,
$b_{U(1)}$ is given
by $b_{U(1)} = \hat{b}_{U(1)} + 2 b_{vec}^{N=2}=2 b_{vec}^{N=2}$, where
$b_{vec}^{N=2}$ denotes the $N=2$ $\beta$-function coefficient
for one $N=2$ vector multiplet.  This is so, because at energies
above the threshold $M_I$ the 2 additional $N=2$ vector multiplets
are effectively massless.
Finally, the coefficient $\hat{\alpha}_{U(1)}$ is related to the
$N=2$ $\beta$-function coefficient \cite{DKL} as $\hat{\alpha}_{U(1)}
= \hat{b}_{U(1)}=0$.
Note that (\ref{gau1})
 is manifestly invariant under modular
$SL(2,{\bf Z})_{T,U}$-transformations\footnote{As discussed in footnote
(\ref{foot4}), we have ignored the issue of appearance of additional
non covariant terms.
This will be discussed in \cite{WitLouLu}.
Note that such additional terms do not change the singular behaviour
of (\ref{gau1}).}
and
also that we have ignored a possible
removal due to the
Green-Schwarz mechanism.

Finally, at points where $T=U\ne 1,\rho$
there is no threshold scale $M_I$ anymore
and
the running of the effective $SU(2)$
coupling is simply given by
\beqa
\frac{1}{g_{SU(2)}^2(p^2)} &=&\frac{1}{g_{tree}^2}
+  \frac{b_{SU(2)}}{16 \pi^2}
\log \frac{M^2_{string}}{p^2}
-
\frac{{\alpha}_{SU(2)}}{16 \pi^2}
\log(T+{\bar T})^2|\eta(T)|^{8}
\label{gaut1}
\eeqa
Here, $b_{SU(2)}=-2c(SU(2))$ denotes the $N=2$ $\beta$-function coefficient
associated with the massless
$N=2$ $SU(2)$-vector multiplet in the theory.  $\alpha_{SU(2)}$
is again related to $b_{SU(2)}$ \cite{DKL} by
$\alpha_{SU(2)} = b_{SU(2)}$.

Let us now turn to the discussion of moduli dependent threshold
corrections to gravitational couplings in the context of $(0,2) {\bf Z}_N$-
orbifold theories.  As in the gauge case, we will stick to those orbifolds
for which the underlying
$T_6$ torus factorises into
$T_6=T_2 \oplus T_4$
and we will discuss
the moduli dependent
threshold corrections associated with this $T_2$.
The gravitational coupling we will be considering in the following
is the one associated with a ${\cal C}^2$ in the low energy
effective action of $(0,2) {\bf Z}_N$-compactifications
of the heterotic string,
${\cal L} = - \frac{1}{2} {\cal R} + \frac{1}{4}
\frac{1}{g^2_{grav}}{\cal C}^2 + \frac{1}{4} \Theta_{grav}
{\cal R}_{mnpq}{\tilde {\cal R}^{mnpq}}
+ \frac{1}{\rho^2}{\cal R}^2_{mn} + \frac{1}{\sigma^2} {\cal R}^2$.
Here, ${\cal C}^2$ denotes the square of the Weyl tensor
${\cal C}_{mnpq}$.
The conventional choice \cite{ZWIEB,GROSS,TSEYT} for the tree-level couplings
to quadratic gravitational curvature terms is taken to be the one
where $\frac{1}{g^2_{grav}}=
- \frac{1}{2} \frac{1}{\rho^2} =
 \frac{3}{2} \frac{1}{\sigma^2} =
\frac{S+{\bar S}}{2}$.  Then,
in the gravitational sector, the dilaton only couples
to the Gauss-Bonnet combination $GB = {\cal C}^2 -2 {\cal R}^2_{mn}
+ \frac{2}{3} {\cal R}^2$ at the tree-level, ${\cal L}=
- \frac{1}{2} {\cal R} +
\frac{1}{4}
\Re S \;GB + \frac{1}{4} \Im S \;
{\cal R}_{mnpq}{\tilde {\cal R}^{mnpq}}$.

Let us first consider
the running of the gravitational coupling $g^2_{grav}$
in the context of $(2,2) {\bf Z}_N$-orbifold models with
gauge group $G$ ($B=C=0$).
In analogy
to the gauge case (\ref{rustr}),
and ignoring a possible removal by the Green-Schwarz mechanism for the
time being,
the running is, in the presence of
a threshold $p^2 \ll M^2_I \ll M^2_X=M^2_{string}$, given by
\beqa
\frac{1}{g_{grav}^2(p^2)} &=&
\frac{S+{\bar S}}{2}
+  \frac{\hat{b}_{grav}}{16 \pi^2}
\log \frac{M^2_I}{p^2}
- \frac{b_{grav}}{16 \pi^2} \log \frac{M^2_I}{M^2_{string}}\nonumber\\
&-&
\frac{\hat{\alpha}_{grav}}{16 \pi^2}
\log(T+{\bar T})(U+{\bar U})-
\frac{\alpha_{grav}}{16 \pi^2}
\log|\eta(T) \eta(U)|^{4}
\label{rugra}
\eeqa
As already discussed,
at the points in the $(T,U)$-moduli space where
$T=U$, the gauge group occuring in the compactification sector
of the model becomes enhanced to $U(1)$
because
an additional $N=2$ vector multiplet becomes
massless.  Thus, the threshold scale $M_I$ is to be identified
with $M_I = |j(T)-j(U)|M_{string}$, as
discussed above.  Equation (\ref{rugra})
describes the running of $g^2_{grav}$ at generic points in the
$(T,U)$-moduli space.  The 1-loop $\beta$-function coefficient
$\hat{b}_{grav}$ describes the running at low momenta $p^2 \ll
M^2_I$ and is thus determined in terms of the truly massless modes
in the theory.  The 1-loop $\beta$-function coefficient
$b_{grav}$ describes the running between $M_I$ and $M_{string}$
and is thus determined in terms of all the light modes
in the theory.
Above the threshold $M_I$ there is only one additional light
multiplet around, namely the additional $N=2$ vector multiplet.
Decomposing this $N=2$ vector multiplet into one $N=1$ vector multiplet
and one $N=1$ chiral multiplet,
it follows that $b_{grav}-\hat{b}_{grav}=\delta^{V}_{grav}+\delta^C_{grav}$,
where $\delta^{V}_{grav}$ and $\delta^{C}_{grav}$
denote the gravitational 1-loop
$\beta$-function coefficient for an $N=1$ vector
and chiral multiplet, respectively.
$b_{grav} - \hat{b}_{grav}$ is also proportional to the trace anomaly
coefficient of this
additional $N=2$ vector multiplet.
The coefficient $\hat{\alpha}_{grav}$ denotes
the contribution from the truly massless modes due to
non-vanishing K\"{a}hler and $\sigma$-model anomalies,
and is given by \cite{IbLu,COLu} $\hat{\alpha}_{grav} = \frac{1}{24}
(21 + 1 - \dim G + \gamma_M + \sum\limits_{\hat C}
(1+2n_{\hat C}))$, where the sum now runs over all
the massless
chiral matter/twisted moduli fields $\hat{C}$ with modular weight
$n_{\hat C}$, see equation (\ref{gcmetric}).  $\gamma_M$ denotes the
contribution from the untwisted modulinos.
For the case that there is no Green-Schwarz
mechanism for the $T_2$ under consideration, $\hat{\alpha}_{grav}$ is
also
computed \cite{AntGaNa}
to be $\hat{\alpha}_{grav}= {\tilde b}_{grav}^{N=2}$,
where ${\tilde b}_{grav}^{N=2}$ denotes
the trace anomaly contribution to ${\cal C}^2$
of all the truly massless $N=2$ multiplets in the associated $N=2$-orbifold
with gauge group $\tilde{G}$
generated by ${\tilde{\cal P}}$, $T^m_m= -
\frac{1}{(4 \pi)^2} {\tilde b}_{grav}^{N=2} {\cal C}^2$.
This was checked explicitly for the example of a ${\bf Z}_4$-orbifold in
\cite{COLu}.
Finally, the coefficient $\alpha_{grav}$ denotes the contribution
from all the massive states which decouple at $M_{string}$ and
is also given by $\alpha_{grav} = {\tilde b}_{grav}^{N=2}$ \cite{AntGaNa}.

Inserting all of this into (\ref{rugra}) yields
\beqa
\frac{1}{g_{grav}^2(p^2)} &=& \frac{S+{\bar S}}{2}
+  \frac{\hat{b}_{grav}}{16 \pi^2}
\log \frac{M^2_{string}}{p^2}
-
 \frac{(\delta^{V}_{grav}
+ \delta^{C}_{grav})}{16 \pi^2} \log |j(T)-j(U)|^2 \nonumber\\
&-&
\frac{\hat{\alpha}_{grav}}{16 \pi^2}
\log(T+{\bar T})(U+{\bar U})|\eta(T) \eta(U)|^{4}
\label{rugrad}
\eeqa
which is manifestly invariant under modular
$SL(2,{\bf Z})_{T,U}$-transformations.
If one also takes the Green-Schwarz mechanism into account, then
(\ref{rugrad}) gets, in analogy to the gauge case, modified to
\beqa
\frac{1}{g_{grav}^2(p^2)} &=& \frac{Y}{2}
+  \frac{\hat{b}_{grav}}{16 \pi^2}
\log \frac{M^2_{string}}{p^2}
-
 \frac{(\delta^{V}_{grav} + \delta^{C}_{grav})}
{16 \pi^2} \log |j(T)-j(U)|^2 \nonumber\\
&-&
\frac{(\hat{\alpha}_{grav} - \delta_{GS})}{16 \pi^2}
\log(T+{\bar T})(U+{\bar U})|\eta(T) \eta(U)|^{4}
\label{greenru}
\eeqa
where $Y = S + {\bar S} - \frac{\delta_{GS}}{8 \pi^2} \log ((T + {\bar T})
(U + {\bar U}) -\frac{1}{2}(B+{\bar B})^2) + (modular \, inv. \,
function)$, as pointed out by Kaplunovsky \cite{KapPriv}.  Note that
(\ref{greenru}) is again invariant under modular
$SL(2,{\bf Z})$-transformations.

The running of the effective gravitational coupling $g_{grav}$ at the
points where $T=U \ne 1,\rho$
is, on the other hand, given as follows.  Since
$T=U$,
there is no additional threshold scale $M_I$, and thus
it follows from (\ref{greenru}) that
\beqa
\frac{1}{g_{grav}^2(p^2)} &=& \frac{Y}{2}
+  \frac{b_{grav}}{16 \pi^2}
\log \frac{M^2_{string}}{p^2}  -
\frac{(\hat{\alpha}_{grav}+ \tilde{\alpha}-\delta_{GS})}{16 \pi^2}
\log(T+{\bar T})^2|\eta(T)|^{8}\nonumber\\
\label{rugratu}
\eeqa
where now
$Y = S + {\bar S} - \frac{\delta_{GS}}{8 \pi^2} \log ((T + {\bar T})^2
 -\frac{1}{2}(B+{\bar B})^2) + (modular \, inv. \,
function)$.
Here, $\tilde{\alpha}$ denotes the contribution to the K\"{a}hler and
$\sigma$-model anomalies due to the additional $N=2$ vector multiplet
which has become massless at $T=U \ne 1,\rho$.
Since a $N=2$ vector multiplet can be decomposed into a
$N=1$ vector multiplet and into a $N=1$ chiral multiplet $C$, it follows
that
$\tilde{\alpha}=\frac{1}{24}(-1+1+2n_c)=-\frac{2}{24}$, which is
precisely twice
the trace anomaly of one $N=2$ vector multiplet.
Note that, in the associated $N=2$ orbifold generated by ${\tilde P}$,
the gauge group associated with the internal $T_2$ gets enhanced from
$U(1)\otimes U(1)$ to $SU(2) \otimes U(1)$ at $T=U$, resulting
in 2 additional $N=2$ vector multiplets.  Thus, $\tilde{\alpha}$ is indeed
equal to the additional trace anomaly contribution in the associated
$N=2$ orbifold generated by $\tilde{P}$.
The coefficients
$b_{grav}$ and $\hat{\alpha}_{grav}$ are the same as in the previous case
of generic $T$ and $U$ values.

Next, consider turning on
Wilson lines $B$ and $C$.  Let us again set $B=C$, for simplicity.
This introduces a new threshold scale $M_I'=|B|M_{string}$ into the
game, at which
the gauge group $G$ of the model gets broken down
to some subgroup $H$, thus changing the spectrum of the massless multiplets
in the theory.  Consider first the generic case where $T \ne U$.
The resulting changes to (\ref{rugrad}) read
\beqa
\frac{1}{g_{grav}^2(p^2)} &=& \frac{S+{\bar S}}{2}
+  \frac{\hat{b}_{grav,H}}{16 \pi^2}
\log \frac{M^2_{string}}{p^2}
+  \frac{(\hat{b}_{grav,H}  - b_{grav,G})}{16 \pi^2}
\log |B|^2 \nonumber\\
&-& \frac{(\delta^{V}_{grav} + \delta^{C}_{grav})}{16 \pi^2} \log |j(T)-j(U)
\nonumber\\
&+&\frac{1}{2} B^2 \;
\left( \partial_U
\log \eta^2(U) \; j'(T) -
\partial_T
\log \eta^2(T) \; j'(U) \right)
|^2 \nonumber\\
&-&
\frac{\hat{\alpha}_{grav,H}}{16 \pi^2}
\log\left((T+{\bar T})(U+{\bar U})-\frac{1}{2}(B+{\bar B})^2\right)\nonumber\\
&-& \frac{\alpha_{grav,G}}{16 \pi^2} \;
\log | \eta (U)
\eta (T)|^4  | 1 - \frac{1}{2} B^2 \;\partial_U \log \eta^{2} (U)
 \; \partial_T \log \eta^{2} (T) \nonumber\\
&+& r \; B^2\; \eta^4(T)\; \eta^4(U) |^{-2}
\label{rutub}
\eeqa
The coefficient $\hat{b}_{grav,H}$ describes the running of the coupling
constant at low energies and is thus determined entirely by the truly
massless modes of the theory with gauge group $H$.  Above the threshold
$M'_I$ the running is described by $b_{grav,G}$, which is now
determined by all the light modes in the theory.
$\hat{\alpha}_{grav,H}$ denotes the contribution from the truly massless
fields to the K\"{a}hler and $\sigma$-model anomalies, and should
thus be given by $\hat{\alpha}_{grav,H}= \frac{1}{24}(21+1-\dim H
+ \gamma_M + \sum_{\hat C}(1+2n_{\hat C}))$, where the sum goes over all
the truly massless chiral matter/twisted moduli multiplets.
Finally, $\alpha_{grav,G}$ describes the contribution of all the massive
states which decouple at $M_{string}$ and should thus still be
given as before by $\alpha_{grav,G}={\tilde b}^{N=2}_{grav}$.
As explained in section \ref{AutFunNulOrb}, $r$
denotes a
constant which cannot be determined by symmetry arguments.
Note that all of the orbits discussed in section \ref{ModOrbUnt} are relevant
for the gravitational case.

If a possible Green-Schwarz removal is taken into account, then
(\ref{rutub}) turns into
\beqa
\frac{1}{g_{grav}^2(p^2)} &=& \frac{Y}{2}
+  \frac{\hat{b}_{grav,H}}{16 \pi^2}
\log \frac{M^2_{string}}{p^2}
+  \frac{(\hat{b}_{grav,H}  - b_{grav,G})}{16 \pi^2}
\log |B|^2 \nonumber\\
&-& \frac{(\delta^{V}_{grav} + \delta^{C}_{grav})}{16 \pi^2} \log |j(T)-j(U)
\nonumber\\
&+&\frac{1}{2} B^2 \;
\left( \partial_U
\log \eta^2(U) \; j'(T) -
\partial_T
\log \eta^2(T) \; j'(U) \right)
|^2 \nonumber\\
&-&
\frac{(\hat{\alpha}_{grav,H}- \delta_{GS})}{16 \pi^2}
\log\left((T+{\bar T})(U+{\bar U})-\frac{1}{2}(B+{\bar B})^2\right)\nonumber\\
&-& \frac{(\alpha_{grav,G}-\delta_{GS})}{16 \pi^2} \;
\log | \eta (U)
\eta (T)|^4  | 1 - \frac{1}{2} B^2 \;\partial_U \log \eta^{2} (U)
 \; \partial_T \log \eta^{2} (T) \nonumber\\
&+& r \; B^2\; \eta^4(T)\; \eta^4(U) |^{-2}
\label{greenrutub}
\eeqa
where $Y = S + {\bar S} - \frac{\delta_{GS}}{8 \pi^2} \log ((T + {\bar T})
(U + {\bar U}) -\frac{1}{2}(B+{\bar B})^2) + (modular \, inv. \,
function)$ \cite{KapPriv}.

The effective coupling (\ref{greenrutub}) should be invariant under modular
$SL(2,{\bf Z})_{T,U}$ transformations, which leads to the requirement that
\beqa
\hat{b}_{grav,H}-b_{grav,G}=\hat{\alpha}_{grav,H}-\alpha_{grav,G}
\eeqa

Finally, let us discuss the running in the presence of a Wilson line $B$
for the case where
$T=U \ne 1,\rho$. When taking into account the Green-Schwarz mechanism,
the running (\ref{rugratu}) is
 modified to
\beqa
\frac{1}{g_{grav}^2(p^2)} &=& \frac{Y}{2}
+  \frac{(\hat{b}_{grav,H} + \delta^{V}_{grav} + \delta^{C}_{grav})}
{16 \pi^2}
\log \frac{M^2_{string}}{p^2}
+  \frac{(\hat{b}_{grav,H}  - b_{grav,G})}{16 \pi^2}
\log |B|^2 \nonumber\\
&-&
\frac{(\hat{\alpha}_{grav,H}+ \tilde{\alpha}-\delta_{GS})}{16 \pi^2}
\log\left((T+{\bar T})^2-\frac{1}{2}(B+{\bar B})^2\right)\nonumber\\
&-& \frac{(\alpha_{grav,G}+ \tilde{\alpha}-\delta_{GS})}{16 \pi^2} \;
\log |\eta (T)|^8  | 1 - \frac{1}{2} B^2
 \;( \partial_T \log \eta^{2} (T)  )^2
 + r \; B^2\; \eta^8(T) |^{-2}\nonumber\\
\label{rutb}
\eeqa
where $Y = S + {\bar S} - \frac{\delta_{GS}}{8 \pi^2} \log  ((T + {\bar T})^2
-\frac{1}{2}(B+{\bar B})^2) + (modular \, inv. \,
function)$ \cite{KapPriv}.
The coefficients are all given in the above discussion.

\section{Conclusions}

In this paper we have computed
string threshold corrections from the massive string spectrum for Abelian
orbifold compactifications. Our discussion contains two main results which were
not obtained before in the literature. First, we derive from the massive
spectrum the
dependence of the threshold functions on the continuous Wilson line moduli
at least for small values of these fields. Second, we derive the
threshold functions also for those gauge groups whose massless charged spectrum
is enlarged at certain points in the orbifold moduli spaces.
In terms of target space free energies these thresholds
correspond to summation orbits which were neglected in previous
considerations. This
discussion includes the stringy Higgs effect, i.e. that a gauge
group gets spontaneously broken to some subgroup when the moduli take
values away from the critical points in the moduli space. Then  the
corresponding
threshold functions possess logarithmic singularities at the critical
points in ${\cal M}$. This also happens   for the  gravitational threshold
corrections since the whole spectrum contributes to the loop
corrections of the gravitational coupling constants. We have shown that
if the appearance of the additional massless states is related to
the six-dimensional orbifold, the correct singularity structure of
the threshold functions is provided by the absolute modular invariant
function $j$. (This result was already anticipated in \cite{CveLu} in quite
general terms.)
For the case that the enlargement of the massive spectrum is associated
with the Wilson line fields, one encounters logarithmic singularities in
these fields. (The leading Wilson line singularities
in the threshold functions
were discussed in \cite{MunLu,CaCaMu} by using renormalization group
arguments.)

Our results can be applied in various ways. First, the complete moduli
dependence of the threshold functions is relevant for several
soft supersymmetry breaking parameters after breaking
of local $N=1$ supersymmetry. In particular, the
Wilson line dependence is important for the $\mu$-problem
\cite{AntGaNarTay,GM,CaMu}.
Second one can use the threshold corrections for the non-perturbative
gaugino condensation mechanism in
the hidden gauge sector. Specifically, if the
hidden gauge group is spontaneosly broken by Wilson line moduli, the
effective superpotential depends on these fields; thus one can determine
the non-perturbatively fixed  vacuum expectation values of these fields.
Even more interesting effects may happen if the discontinuities
in the spectrum related to the six-dimensional orbifold enter the
non-perturbative superpotential. This might be the case when considering
non-perturbative superpotential by gravitational instantons
\cite{PSrStr}, since in this
case the gravitational threshold corrections are most likely to be relevant
for the non-perturbative dynamics.
Then the non-perturbative superpotential will contain also the absolute
modular invariant function $j$, and the vacuum structure of the
moduli depends on the singularities of
the threshold functions in an interesting way \cite{CveLu}.
Finally, these types of superpotentials are also important for the formation
of domain walls and other topological objects, as discussed in
\cite{CvQue,CvGrSo}.

\vskip .1in

{\bf Acknowledgement}

We would like to thank V. Kaplunovsky, J. Louis and C. Kounnas
for fruitful discussions.  We understand that V. Kaplunovsky, C. Kounnas
and E. Kiritsis have also been considering the influence of the
enhanced symmetry points on threshold corrections.

\begin{table}[h]
\[
\begin{array}{|c|c|c|c|c|c|} \hline
\mbox{Class} & \mbox{Twist} & \mbox{Order} & \mbox{Eigenvalues} &
\mbox{Minimal GG} &
\mbox{Maximal GG} \\ \hline \hline
1 & \emptyset & 1 & [0] & E_{8} & E_{8} \\ \hline
2 & A_{1} & 4^{*} & [2] & E_{7} & E_{7} U_{1} \\ \hline
3 & A_{1}^{2} & 4^{*} & [2,2] & D_{6} & D_{7} U_{1} \\ \hline
4 & A_{2} & 3 & [1,2] & E_{6} & E_{7} U_{1} \\ \hline
5 & A_{1}^{3} & 4^{*} & [2,2,2]& D_{4} A_{1}
  & E_{6} A_{1} U_{1} \\ \hline
6 & A_{2} A_{1} & 12^{*} & [4,8,6]
  & A_{5} & D_{6} U_{1}^{2} \\ \hline
7 & A_{3} & 8^{*} & [2,6,4]
  & D_{5} & D_{6} U_{1}^{2} \\ \hline
8 & A_{1}^{4,I} & 2 & [1,1,1,1]
  & D_{4} & E_{7} A_{1} \\ \hline
9 & A_{1}^{4,II} & 4^{*} & [2,2,2,2]
  & A_{1}^{4} & A_{7} U_{1} \\ \hline
10 & A_{2} A_{1}^{2} & 12^{*} & [4,8,6,6]
   & A_{3} U_{1} & D_{5} A_{1} U_{1}^{2} \\ \hline
11 & A_{2} A_{2} & 3 & [1,2,1,2]
   & A_{2} A_{2} & D_{7} U_{1} \\ \hline
12 & A_{3} A_{1} & 8^{*} & [2,6,4,4]
   & A_{3} A_{1} & D_{4} A_{2} U_{1}^{2} \\
             \hline
14 & D_{4} & 6 & [1,5,3,3]
   & D_{4} & E_{6}U_{1}^{2} \\ \hline
15 & D_{4}(a_{1}) & 4 & [1,3,1,3]
   & D_{4} & E_{6} A_{1} U_{1} \\ \hline
16 & A_{1}^{5} & 4^{*} & [2,2,2,2,2]
   & A_{1}^{3} & D_{6}A_{1}U_{1} \\ \hline
17 & A_{2} A_{1}^{3} & 12^{*} & [4,8,6,6,6]
   & A_{1} U_{1}^{2} & A_{5} A_{1} U_{1}^{2} \\ \hline
18 & A_{2}^{2} A_{1} & 12^{*} & [4,8,4,8,6]
   & A_{2} U_{1} & D_{5} A_{1} U_{1}^{2}
   \\ \hline
19 & (A_{3}A_{1}^{2})^{I} & 8^{*} & [2,6,4,4,4]
   & A_{3} & A_{4} A_{1}^{2} U_{1}^{2} \\ \hline
20 & (A_{3}A_{1}^{2})^{II} & 8^{*} & [2,6,4,4,4]
   & A_{1}^{2} U_{1} & D_{5}A_{1}^{2}U_{1}  \\ \hline
23 & A_{5} & 12^{*} & [2,10,4,8,6]
   & A_{2} A_{1} & D_{4} A_{1} U_{1}^{3} \\ \hline
\end{array}
\]
\caption{Table of unbroken gauge groups for vanishing and for
generic Wilson lines. The notation is explained in the text.
The data are taken from references [38,39,40].}
\end{table}

\newpage

\begin{table}[h]
\[
\begin{array}{|c|c|c|c|c|c|} \hline
\mbox{Class} & \mbox{Twist} & \mbox{Order} & \mbox{Eigenvalues} &
\mbox{Minimal GG} &
\mbox{Maximal GG} \\ \hline \hline
24 & D_{4} A_{1} & 12^{*} & [2,10,6,6,6]
   & A_{1}^{3} & A_{5} U_{1}^{3} \\ \hline
25 & D_{4}(a_{1})A_{1} & 4 & [1,3,1,3,2]
   & A_{1}^{3} & A_{7} U_{1} \\ \hline
26 & D_{5} & 8 & [1,7,3,5,4]
   & A_{3} & D_{5} U_{1}^{3} \\ \hline
28 & A_{1}^{6} & 4^{*} & [2,2,2,2,2,2]
   & A_{1}^{2} & D_{5}A_{3} \\ \hline
29 & A_{2} A_{1}^{4} & 6 & [2,4,3,3,3,3]
   & U_{1}^{2}  & D_{6} A_{1} U_{1} \\ \hline
30 & A_{2}^{2} A_{1}^{2} & 12^{*}  & [4,8,4,8,6,6]
   & U_{1}^{2}  &A_{3}^{2} U_{1}^{2} \\ \hline
31 & A_{2}^{3} & 3 & [1,2,1,2,1,2]
   & A_{2} & E_{6} A_{2} \\ \hline
32 & A_{3} A_{1}^{3} & 8^{*} & [2,6,4,4,4,4]
   & A_{1} U_{1} & A_{3} A_{2} A_{1}^{2} U_{1} \\ \hline
34 & (A_{3}^{2})^{I} & 4 & [1,3,1,3,2,2]
   & A_{1}^{2} &D_{6} A_{1} U_{1} \\ \hline
35 & (A_{3}^{2})^{II} & 8^{*} & [2,6,2,6,4,4]
   & U_{1}^{2} & A_{3}^{2} U_{1}^{2} \\ \hline
38 &(A_{5}A_{1})^{I} & 6 & [1,5,2,4,3,3]
   & A_{2} & D_{5} A_{1}^{2} U_{1} \\ \hline
39 &(A_{5}A_{1})^{II} & 12^{*} & [2,10,4,8,6,6]
   & A_{1} U_{1} & A_{3}A_{1}^{2} U_{1}^{3} \\ \hline
40 & A_{6} & 7 & [1,6,2,5,3,4]
   & A_{1}U_{1} & D_{4} A_{2} U_{1}^{2} \\ \hline
41 & D_{4} A_{1}^{2} & 12^{*} & [2,10,6,6,6,6]
   & A_{1}^{2} & D_{4} A_{1}^{2} U_{1}^{2} \\ \hline
42 & D_{4} A_{2} & 6 & [1,5,3,3,2,4]
   & U_{1}^{2} & A_{6} U_{1}^{2} \\ \hline
43 & D_{4}(a_{1}) A_{2} & 12 & [3,9,3,9,4,8]
   & U_{1}^{2} & A_{5} A_{1} U_{1}^{2} \\ \hline
44 & D_{5} A_{1} & 8 & [1,7,3,5,4,4]
   & A_{1} U_{1} & A_{4} A_{2} U_{1}^{2} \\ \hline
47 & D_{6}(a_{1}) & 8 & [1,7,2,6,3,5]
   & A_{1}^{2} & A_{5} U_{1}^{3} \\ \hline
48 & D_{6}(a_{2}) & 12^{*} & [2,10,2,10,6,6]
   & A_{1}^{2} & A_{3} A_{1}^{2} U_{1}^{3} \\ \hline
49 & E_{6} & 12 & [1,11,4,8,5,7]
   & A_{2} & D_{4} U_{1}^{4} \\ \hline
\end{array}
\]
\caption{Table of unbroken gauge groups for vanishing and for
generic Wilson lines (continued).}
\end{table}

\newpage

\begin{table}[h]
\[
\begin{array}{|c|c|c|c|c|c|} \hline
\mbox{Class} & \mbox{Twist} & \mbox{Order} & \mbox{Eigenvalues} &
\mbox{Minimal GG} &
\mbox{Maximal GG} \\ \hline \hline
51 & E_{6}(a_{2}) & 6 & [1,5,1,5,2,4]
   & A_{2} & D_{5} A_{1} U_{1}^{2} \\ \hline
52 & A_{1}^{7} & 4^{*} & [2,2,2,2,2,2,2]
   & A_{1} & A_{7} A_{1} \\ \hline
53 & A_{2}^{3} A_{1} & 12^{*} & [4,8,4,8,4,8,6]
   & U_{1} & A_{5} A_{2} U_{1} \\ \hline
54 & A_{3} A_{1}^{4} & 8^{*} & [2,6,4,4,4,4,4]
   & U_{1} & D_{4} A_{3} U_{1} \\  \hline
56 & A_{3}^{2} A_{1} & 4 & [1,3,1,3,2,2,2]
   & A_{1} & D_{5} A_{3} \\ \hline
59 & A_{5} A_{1}^{2}  & 12^{*} & [2,10,4,8,6,6,6]
   & U_{1} & A_{3} A_{1}^{3} U_{1}^{2} \\ \hline
60 & A_{5} A_{2} & 12^{*} & [2,10,4,8,6,4,8]
   & A_{1} & A_{2}^{3} A_{1}U_{1} \\ \hline
62 & (A_{7})^{I} & 8 & [1,7,2,6,3,5,4]
   & A_{1} & D_{4} A_{1}^{2} U_{1}^{2} \\ \hline
64 & D_{4} A_{1}^{3} & 12^{*} & [2,10,6,6,6,6,6]
   & A_{1} & A_{3}^{2} A_{1} U_{1} \\ \hline
66 & D_{4}(a_{1}) A_{3} & 8^{*} & [2,6,2,6,2,6,4]
   & U_{1} & A_{3} A_{2} A_{1}^{2} U_{1} \\ \hline
67 & D_{5} A_{1}^{2} & 8 & [1,7,3,5,4,4,4]
   & U_{1} & A_{5} A_{1} U_{1}^{2} \\ \hline
71 & D_{6}(a_{2})A_{1} & 12^{*} & [2,10,2,10,6,6,6]
   & A_{1} & A_{3}A_{1}^{3}U_{1}^{2} \\ \hline
72 & E_{6}A_{1} & 12 & [1,11,4,8,5,7,6]
   & U_{1} & A_{2}^{2}A_{1}U_{1}^{3} \\ \hline
74 & E_{6}(a_{2}) A_{1} & 12^{*} & [2,10,2,10,4,8,6]
   & U_{1} & A_{3}A_{1}^{2}U_{1}^{3} \\ \hline
80 & E_{7}(a_{2}) &12 & [1,11,2,10,5,7,6]
   & A_{1} & A_{3}A_{1}U_{1}^{4} \\ \hline
83 & A_{1}^{8} & 2 & [1,1,1,1,1,1,1,1]
   & - & D_{8} \\ \hline
84 & A_{2}^{4} & 3 & [1,2,1,2,1,2,1,2]
   & - &A_{8} \\ \hline
85 & A_{3}^{2}A_{1}^{2} &4 & [1,3,1,3,2,2,2,2]
   & - &A_{7}A_{1} \\ \hline
87 & A_{5} A_{2} A_{1} & 6 & [1,5,2,4,3,2,4,3]
   & - & A_{5}A_{2} A_{1} \\ \hline
88 & A_{7}A_{1} & 8 & [1,7,2,6,3,5,4,4]
   & - & A_{3}^{2} A_{1} U_{1} \\ \hline
\end{array}
\]
\caption{Table of unbroken gauge groups for vanishing and for
generic Wilson lines (continued).}
\end{table}

\newpage

\begin{table}[h]
\[
\begin{array}{|c|c|c|c|c|c|} \hline
\mbox{Class} & \mbox{Twist} & \mbox{Order} & \mbox{Eigenvalues} &
\mbox{Minimal GG} &
\mbox{Maximal GG} \\ \hline \hline
90 & D_{4} A_{1}^{4} & 6 & [1,5,3,3,3,3,3,3]
   & - & A_{7} U_{1} \\ \hline
91 & D_{4}^{2} & 6 & [1,5,3,3,1,5,3,3]
   & - & D_{4}A_{3}U_{1} \\ \hline
92 & D_{4}(a_{1})^{2} & 4 & [1,3,1,3,1,3,1,3]
   & - & D_{5} A_{3} \\ \hline
93 & D_{5}(a_{1}) A_{3} & 12 &[2,10,3,9,6,3,9,6]
   & - & A_{3}^{2} A_{1} U_{1} \\ \hline
96 & D_{8}(a_{1}) & 12 & [1,11,3,9,3,9,5,7]
   & - & A_{2}A_{1}^{4}U_{1}^{2} \\ \hline
98 & D_{8}(a_{3}) &8 & [1,11,3,9,3,9,5,7]
   & - & A_{3}A_{2}A_{1}^{2}U_{1} \\ \hline
99 & E_{6} A_{2} & 12 & [1,11,4,8,5,7,4,8]
   & - & A_{2}^{3} U_{1}^{2} \\ \hline
100& E_{6}(a_{2})A_{2} & 6 & [1,5,1,5,2,4,2,4]
   & - & A_{5}A_{2}U_{1} \\ \hline
102&E_{7}(a_{2})A_{1} &12 & [1,11,2,10,5,7,6,6]
   & - & A_{3}A_{1}^{2}U_{1}^{3} \\ \hline
103&E_{7}(a_{4})A_{1} & 6 & [1,5,1,5,1,5,3,3]
   & - & A_{5}A_{1}^{2}U_{1} \\ \hline
107&E_{8}(a_{3}) & 12 & [1,11,1,11,5,7,5,7]
   & - & A_{2} A_{1}^{3} U_{1}^{3} \\ \hline
111&E_{8}(a_{7}) & 12 & [1,11,2,10,2,10,5,7]
   & - & A_{2}^{2}A_{1}U_{1}^{3} \\ \hline
112& E_{8}(a_{8}) & 6 & [1,5,1,5,1,5,1,5]
   & - & A_{4} A_{3} U_{1} \\ \hline
\end{array}
\]
\caption{Table of unbroken gauge groups for vanishing and for
generic Wilson lines (continued).}
\end{table}

\end{document}